\documentclass[]{aa}

\usepackage[varg]{txfonts}
\usepackage{placeins}
\usepackage{lscape}
\usepackage{textcomp}
\usepackage{gensymb}
\usepackage[urlcolor=cyan, colorlinks=true, citecolor=blue, linkcolor=blue]{hyperref}
\bibpunct{(}{)}{;}{a}{}{,}

\begin{document}

\title{UV-irradiated outflows from low-mass protostars in Ophiuchus with JWST/MIRI}
\author{I.~M.~Skretas\inst{1}, A. Karska\inst{1,2}, L. Francis\inst{3}, W. R. M. Rocha\inst{3}, M. L. van Gelder\inst{3}, Ł. Tychoniec\inst{3}, M. Figueira\inst{1,4}, M. Sewi{\l}o\inst{5,6,7}, F. Wyrowski\inst{1}, and P. Schilke\inst{8}}

\institute{$^{1}$ Max-Planck-Institut für Radioastronomie, Auf dem Hügel 69, 53121, Bonn, Germany \\
$^{2}$ Centre for Modern Interdisciplinary Technologies, Nicolaus Copernicus University in Toruń, Wileńska 4, 87-100 Toruń, Poland\\
$^{3}$ Leiden Observatory, Leiden University, PO Box 9513, 2300 RA Leiden, The Netherlands\\
$^{4}$ National Centre for Nuclear Research, Pasteura 7, 02-093, Warszawa, Poland\\
$^{5}$ Exoplanets and Stellar Astrophysics Laboratory, NASA Goddard Space Flight Center, Greenbelt, MD 20771, USA\\
$^{6}$ Center for Research and Exploration in Space Science and Technology, NASA Goddard Space Flight Center, Greenbelt, MD 20771, USA\\
$^{7}$ Department of Astronomy, University of Maryland, College Park, MD 20742, USA\\
$^{8}$ Physikalisches Institut der Universität zu K{\"o}ln, Z{\"u}lpicher Str. 77, D-50937 K{\"o}ln, Germany
}

\date{Received April 1, 2025; accepted September 9 2025}
\titlerunning {UV-irradiated outflows in Ophiuchus}
\authorrunning{I.~Skretas et al.~2025}

\abstract
{The main accretion phase of protostars is characterized by the ejection of material in the form of bipolar jets/outflows. In addition, external UV irradiation can potentially have a significant impact on the excitation conditions within these outflows. High-resolution observations in the mid-infrared (IR) allow us to investigate the details of those energetic processes  through the emission of shock-excited H$_2$ .}
{Our aim is to spatially resolve H$_2$ and ionic/atomic emission within the outflows of low-mass protostars, and investigate its origin in connection to shocks influenced by external ultraviolet irradiation.}
{We analyze spectral maps of 5 Class I protostars in the Ophiuchus molecular cloud from the James Webb Space Telescope (JWST) Medium Resolution Spectrometer (MIRI/MRS). The MIRI/MRS field-of-view covers an area between $\sim$3.2$\arcsec$ $\times$ 3.7$\arcsec$ at 6 $\mu$m and 6.6$\arcsec$ $\times$ 7.7$\arcsec$ at 25 $\mu$m and  with a resolution of $\sim$0.3 to 1$\arcsec$, corresponding to spatial scales of a few 100 AUs.}
{Four out of five protostars in our sample show strong H$_2$, [\ion{Ne}{II}], and [\ion{Fe}{II}] emission associated with outflows/jets. Pure rotational H$_2$ transitions from S(1) to S(8) are found and show two distinct temperature components on Boltzmann diagrams with rotational temperatures of $\sim$500-600 K and $\sim$1000-3000 K respectively. Both $C$-type shocks propagating at high pre-shock densities (n$_\text{H} \ge$10$^4$ cm$^{-3}$) and $J$-type shocks at low pre-shock densities (n$_\text{H} \le$10$^3$ cm$^{-3}$) reproduce the observed line ratios. However, only $C$-type shocks produce sufficiently high column densities of H$_2$, whereas predictions from a single $J$-type shock reproduce the observed rotational temperatures of the gas better. 
A combination of various types of shocks could play a role in protostellar outflows as long as UV irradiation is included in the models. The origin of this radiation is likely internal, since no significant differences in the excitation conditions of outflows are seen at various locations in the cloud.} 
{Observations with MIRI offer an unprecedented view of protostellar outflows, allowing us to determine the properties of outflowing gas even at very close distances to the driving source. Further constraints on the physical conditions within outflows can be placed thanks to the possibility of direct comparisons of such observations with state-of-the-art shock models.}


\maketitle
\section{Introduction}

Bipolar outflows are the most prominent sign of ongoing star formation. They are considered an essential part of the star formation process as they remove material and excess angular momentum from the circumstellar disk -- protostar system, thus enabling accretion, along the disk, onto the protostar \citep{Pudritz07,Frank14,Bal16,RayFer21}. Furthermore, protostellar outflows might limit the star formation efficiency by injecting large amounts of energy and momentum into the surrounding interstellar medium (ISM) and dispersing the natal envelope \citep[e.g.,][]{Fall10, Frank14}. Since protostellar outflows transfer material from the protoplanetary disk, they also have a significant impact on the early disk evolution and planet formation \citep{tych20}. In addition to wide-angle outflows, narrow well-collimated jets are launched at distances close to the protostar \citep[e.g.,][]{Pudritz1983, Ferreira97, Frank14, Bai16,Podio21}.    
  
\begin{figure*}[ht!]
\centering 
\includegraphics[width=0.8\linewidth]{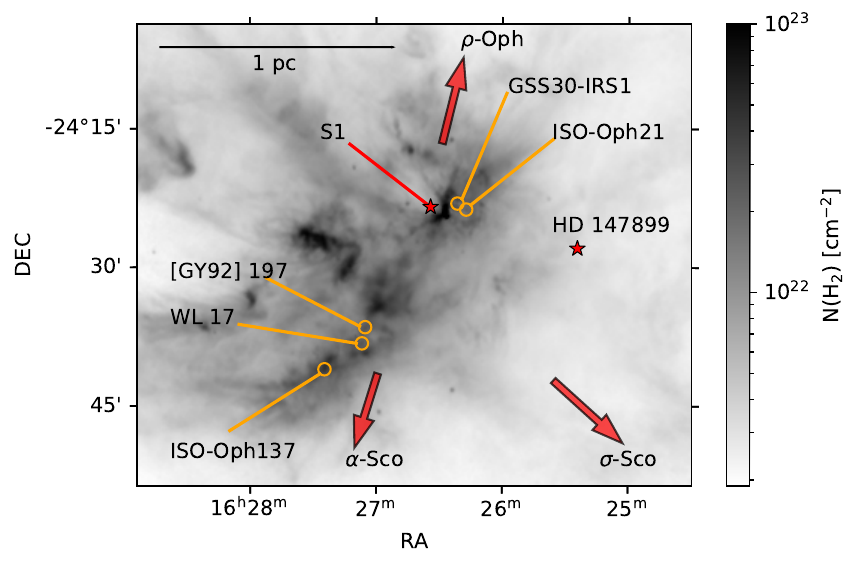} 
\caption{\label{fig:region} Overview of the studied region. H$_2$ column density map from \cite{lad20} derived from \textit{Herschel} data. Orange circles mark the location of the targeted sources and red stars mark the nearby massive stars HD147899 and S1. The red arrows point to the direction of the surrounding massive stars $\sigma$-Sco, $\alpha$-Sco and $\rho$-Oph located at a distance of $\sim4.2$-4.3, $\sim$4.3-5.2 and $\sim$2.3-3.1 pc from the protostars, respectively.}
\end{figure*}

To investigate the origin and formation of protostellar outflows and jets, it is vital to observe them at their origin, close to the protostars. Powerful interferometers such as the Atacama Large Millimeter/submillimeter Array (ALMA) and the Northern Extended Millimeter Array (NOEMA) are well fit to study energetic processes deep inside dusty molecular envelopes of protostars. Low-$J$ transitions of CO have been commonly employed to trace the large scale protostellar outflows \citep[e.g.,][]{Arce13,tych21,skretas23}. Observations in additional tracers, for example, SiO and SO, characterized the cold gas component of the jets \citep[e.g.,][]{Lee20,Podio21}. However, as the jet and outflow propagate through the surrounding ISM, they compress and heat the gas, thus creating a significant warm gas component, which is best traced in the infrared (IR) regime \citep{kn96,fp10}. 

Space observatories such as Infrared Space Observatory (ISO), \textit{Spitzer Space Telescope} and \textit{Herschel Space Observatory} detected bright emission from molecular hydrogen (H$_2$) \citep[e.g.,][]{Nisini03,Maret09,Neufeld2009} and H$_2$O coming from shocked material in the outflows from low-mass protostars \citep{Kristensen12,kri17co,karska13,karska14a,vd2021}. In addition, surprisingly bright OH emission was detected \citep[e.g.,][]{wamp13}, suggesting a possible role of ultraviolet (UV) radiation in setting the abundance of H$_2$O \citep{karska14b}. The inclusion of external UV irradiation of shock models by \cite{MK15} allowed to reproduce the observed ratios of OH/H$_2$O lines toward outflow shocks with relatively low UV fields of about 0.1-10 times the the average interstellar radiation field \citep{karska18}. Follow-up observations of CN/HCN ratio sensitive to photodissociation suggested that even higher UV fields may be at play in clusters of low-mass protostars \citep{mir21}. The detection of irradiation tracers, such as C$_2$H and c-C$_3$H$_2$ at the walls of outflow cavities further reinforces the prevalence of UV emission in protostellar outflows \citep{tych21,Legouellec23}. Shock models for a broad parameter space have been recently developed and allow us to estimate gas conditions and UV fields in various environments using, in particular, mid-IR lines of H$_2$ \citep{Godard19,lehmann20,lehmann22,Kristensen23}.

With the advent of James Webb Space Telescope (JWST), protostellar outflows can now be directly probed at spatial scales that reveal the origin of H$_2$ as well as atomic emission arising from shocks. Early JWST results showcased a rich molecular emission toward the outflow of the Class 0 protostar HH 211 and a surprisingly weak atomic and ionized emission 
\citep{ray23}, while another young protostar IRAS16253 showed a clear ionized jet in absence of molecular gas \citep{narang24}. Toward the more evolved Class I protostars, TMC 1 and TMC 1A, a highly collimated, ionized jet was detected in various [\ion{Fe}{II}] transitions and a broader cavity was seen in H$_2$ lines \citep[][]{harsono23,assani24,tych24}. The zoom-in view toward the famous outflow from HH46 presented a complicated outflow structure, including possible UV production at the location of its bow-shock \citep{nisini24}. Finally, the detection of rotational, suprathermal OH lines toward an intermediate-mass protostar HOPS 370 provided the evidence for the photodissociation of water by fast shocks along the jet \citep{neu24}.

\begin{table*} 
\caption{Source properties and integration times used for the  observations with MIRI/MRS \label{table:cat}} 
\centering 
\begin{tabular}{l l l l l l l c}
\hline \hline 
No. & Source &  \begin{tabular}[c]{@{}c@{}}R.A. (J2000)\\  $\left(  \text{hh:mm:ss} \right)$ \end{tabular}&\begin{tabular}[c]{@{}c@{}}Dec. (J2000)\\  $\left(  \text{dd:mm:ss} \right)$ \end{tabular}& \begin{tabular}[c]{@{}c@{}}$T_\text{bol}$\\  $\left(  \text{K} \right)$ \end{tabular} &  \begin{tabular}[c]{@{}c@{}}$L_\text{bol}$\\  $\left(  \text{L}_\odot \right)$ \end{tabular}& Other names &\begin{tabular}[c]{@{}c@{}}Integration time\\  $\left(  \text{s} \right)$ \end{tabular}\\
\hline    
    \multicolumn{7}{c}{\textbf{$\rho$ Oph A}}\\
    1 & GSS30-IRS 1 & 16:26:21.34& -24:23:05.05& $172$ & 12.72  & Elias 21, Oph-emb 8 & 55.5\\
    2 & ISO-Oph 21  &16:26:17.23& -24:23:45.98& $490$ & $0.83$ & CRBR12, CRBR 2315.8-1700 & 277.5\\ 
    \hline
\multicolumn{7}{c}{\textbf{$\rho$ Oph E}}\\
	3 & [GY92] 197 & 16:27:05.24& -24:36:30.27& $110$  & $1.53$ & LFAM26 & 499.5\\
	4 & WL 17 & 16:27:06.76& -24:38:15.62& $310$ & $0.73$ & [GY92]~205 & 277.5 \\
   \hline
\multicolumn{7}{c}{\textbf{$\rho$ Oph F}}\\   
	5 & ISO-Oph 137 & 16:27:24.57& -24:41:03.86& $191$ &  $1.21$ & CRBR85, CRBR 2422.8-3423.8 & 222.0/455.1\tablefootmark{a}\\
\hline
\end{tabular} 
\begin{flushleft}
\tablefoot{Source coordinates are based on the peak of the 6.9 $\mu$m continuum emission from MIRI (see Fig. \ref{fig:cont_maps}), and source properties ($T_\mathrm{bol}$, $L_\mathrm{bol}$) are taken from \textit{Herschel} \citep[GSS30-IRS 1; ][]{karska18} and \textit{Spitzer} surveys of low-mass protostars: ISO-Oph 137 from \cite{enoch09} and the remaining sources from \cite{evans09}.  $L_\mathrm{bol}$ is re-scaled by a factor of 1.2 to account for the new estimate of the distance to Ophiuhus of 137 pc \citep{Ortiz17}. \tablefootmark{a} Integration time of 455.1 s was used only for grating B.} 
\end{flushleft}
\end{table*}

In this work, we aim to study the impact of external UV irradiation on the outflows and jets of low-mass protostars and on the broader process of star formation. To that end, we investigate the morphology, physical conditions, and origin of the pure rotational emission of H$_2$ in the outflows of 5 protostars in Ophiuchus. All targets are classified as Class I objects (Table \ref{table:cat}, see also \citealt{vK09}), which allows us to also search for possible evolutionary trends via comparisons with Class 0 objects with available JWST/MIRI data \citep{Francis25}.

The Ophiuchus molecular complex is a nearby (distance of 137 pc; \citealt{Ortiz17}), low-mass star-forming region \citep{evans09}.
In addition, Ophiuchus  contains a large number of prestellar cores, protostars and more evolved young stellar objects \citep{Wilking89, Leous1991,Motte98, Wilking08,lad20}. 
CO observations have shown that the complex primarily consists of two molecular clouds: -- L1688, which is associated with intense star formation in several dense clumps \citep{Loren1990,Motte98,Bontemps01}, and L1689, which appears much more quiescent \citep[e.g.,][]{Nutter06}. Both core properties and outflow activity were extensively studied using single-dish antennas by \cite{joh17} and \cite{vdM13}, at angular resolutions of 29\arcsec and 15\arcsec respectively.  The protostars in Ophiuchus are under the influence of the nearby Sco OB2 association and additional B-type stars \citep[e.g.,][]{Motte98,Nutter06,Wilking08}, resulting in UV fields of $\sim$5-10 times the interstellar radiation field toward diffuse clouds \citep{vD89}.
Thus, Ophiuchus is a unique laboratory to study the impact of external UV irradiation onto protostellar outflows and the star formation process.

The paper is structured as follows. In Sec. \ref{sec:observations}, we present the details of the observations analyzed in this work. In Sec. \ref{sec:results}, we discuss the morphology of the detected H$_2$ outflows and ionic jets, as well as spectra extracted along the outflows. In Sec. \ref{sec:analysis}, we show the results of the excitation analysis of the outflows emission as well as the results of comparisons with shock models. Then, in Sec. \ref{sec:discussion}, we investigate the possible origin of the H$_2$ emission and compare our results with those in earlier studies. Finally, Sec. \ref{sec:summary} contains a summary of our results and our conclusions.   

\begin{figure*}
\centering 
\includegraphics[width=0.49\linewidth]
{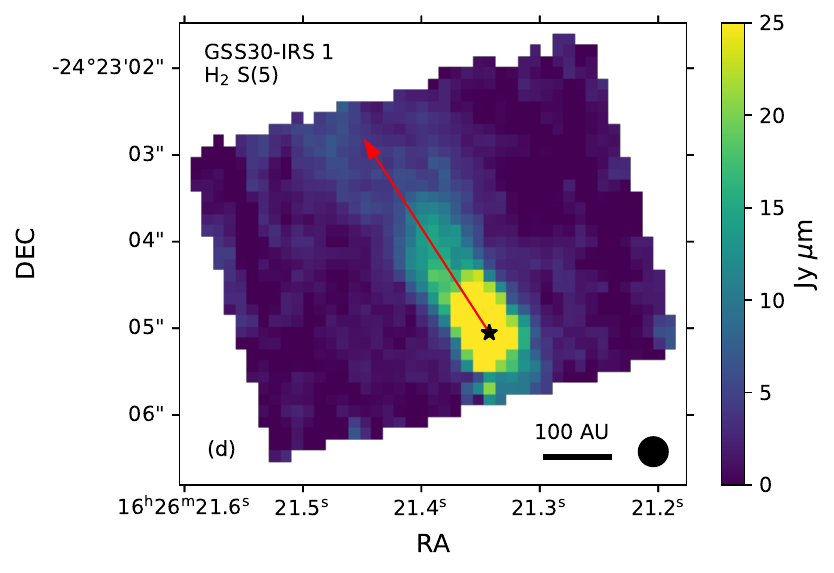} 
\includegraphics[width=0.49\linewidth]
{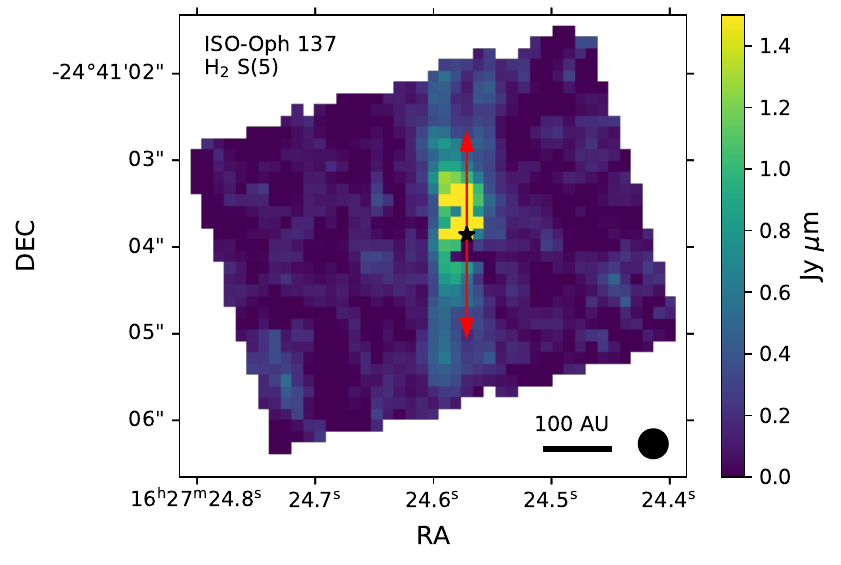} 
\includegraphics[width=0.49\linewidth]
{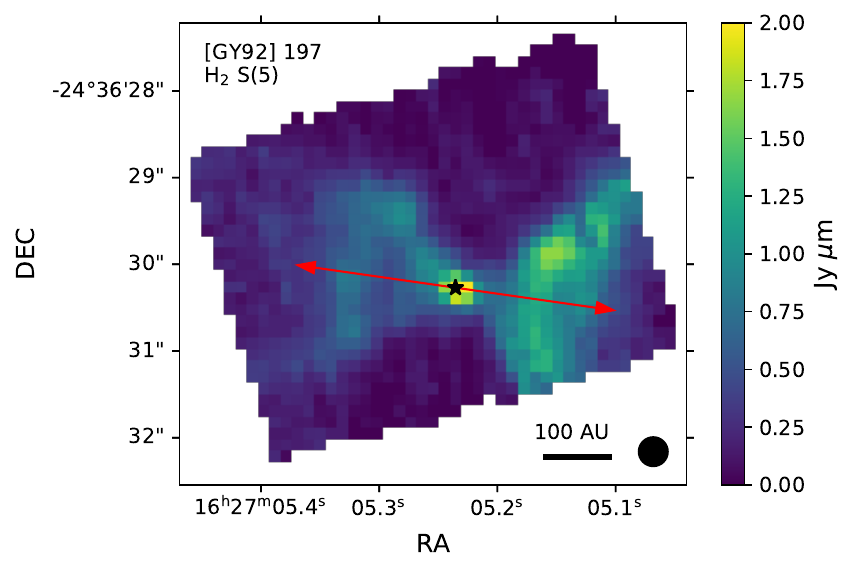} 
\includegraphics[width=0.49\linewidth]
{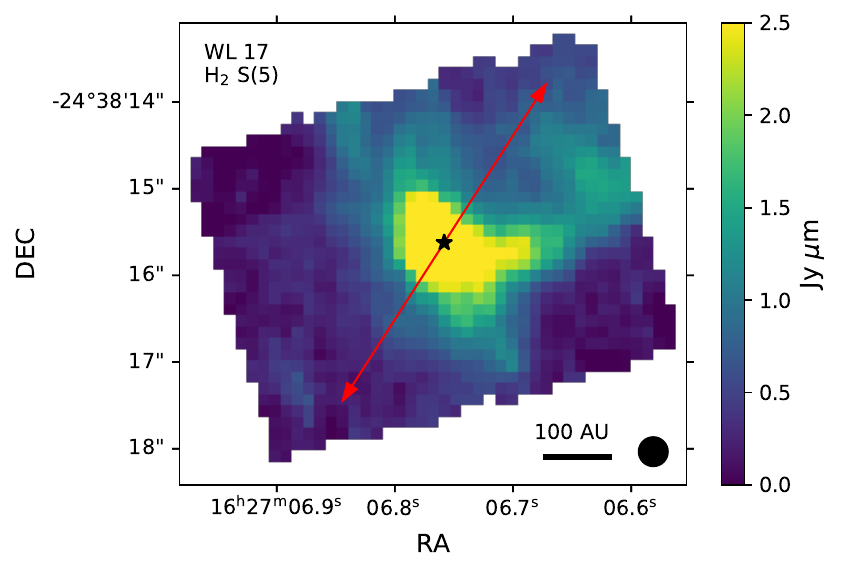}  
\caption{Integrated emission of H$_2$ S(5) line at 6.91 $\mu$m toward GSS30-IRS 1, ISO-Oph 137, [GY92] 197, and WL 17. The positions of the protostars, measured from the line-free region at 6.9 $\mu$m, are shown with black stars. The red arrows shows the apparent outflow direction based on the H$_2$ emission. The black filled circle marks the spatial resolution of MIRI at $\lambda = 6.9$~$\mu$m ($\sim0.2\arcsec$).}
\label{fig:h2_maps}
\end{figure*}
\begin{figure*}
\centering 
\includegraphics[width=0.49\linewidth]{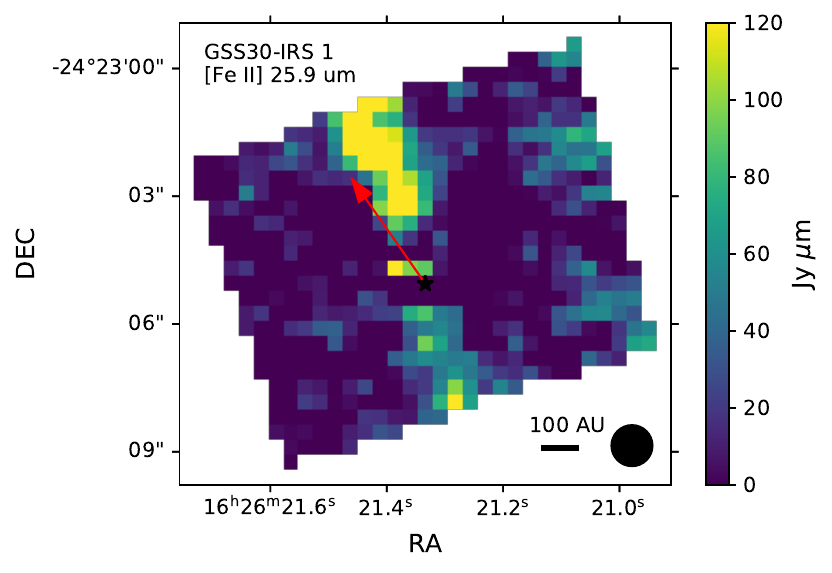}
\includegraphics[width=0.49\linewidth]{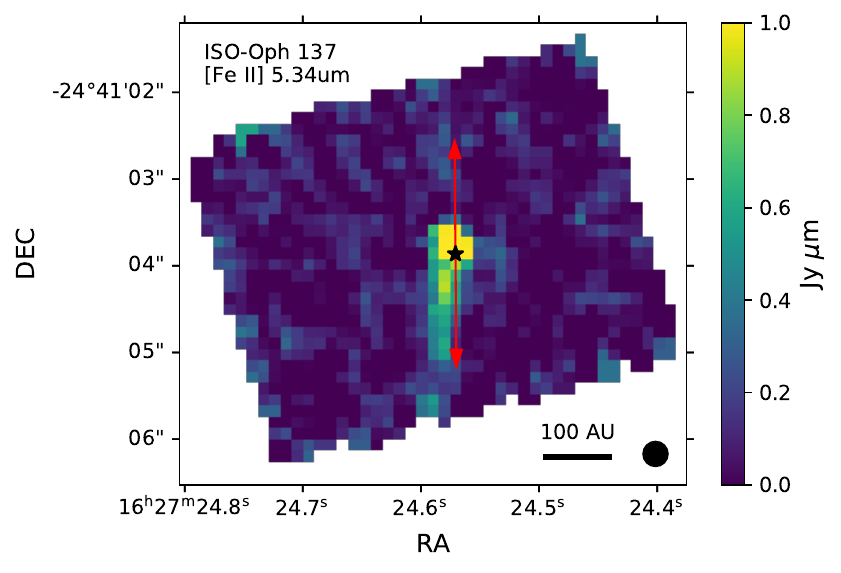}
\includegraphics[width=0.49\linewidth]{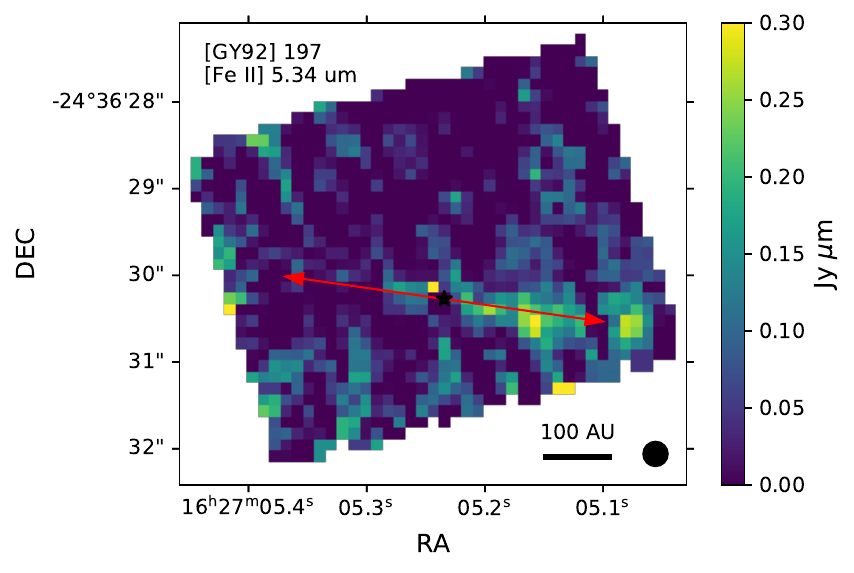}
\includegraphics[width=0.49\linewidth]{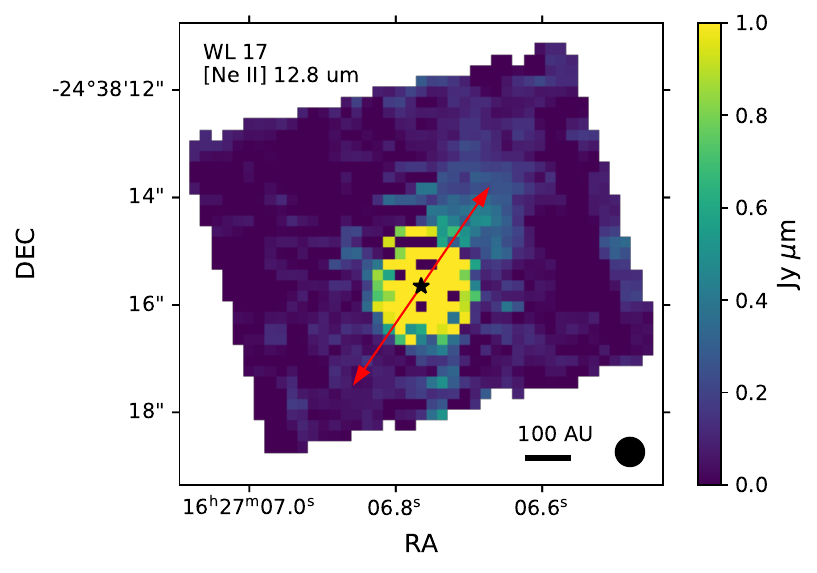}

\caption{Integrated emission of selected emission lines of ionized species: 
the [\ion{Fe}{II}] line at 25.9 $\mu$m toward GSS30-IRS 1, the [\ion{Fe}{II}] line at 5.34 $\mu$m toward ISO-Oph 137, and [GY92] 197, and -- since no extended [\ion{Fe}{II}] emission was detected -- the [\ion{Ne}{II}] line at 12.8 $\mu$m toward WL 17. The labels are the same as in Fig. \ref{fig:h2_maps}.} 
\label{fig:jet maps}
\end{figure*}

\section{Observations}
\label{sec:observations}

This work makes use of data taken as part of the JWST project 1959  "Ice chemical complexity toward the Ophiuchus molecular cloud" (PI: Will Rocha). The project consists of observations using the Medium Resolution Spectrometer (MRS) \citep{Argyriou23} of the Mid-Infrared Instrument (MIRI) \citep{wright23}, targeting 5 Class I protostars in the Oph A, E, and F regions of the Ophiuchus molecular cloud. The sources were selected such that they have varying distances from the nearby massive stars, especially the B3 V type star S1 and the B2 V type star HD147899 (see Fig. \ref{fig:region}). As a result, the sources are exposed to different levels of UV irradiation, enabling the study of its impact on the star-formation process. Detailed information for the targeted sources is presented in Table \ref{table:cat}, while an overview of the region showing the observed sources as well as notable nearby massive stars is shown in Fig. \ref{fig:region}.

The data was taken using the FASTR1 readout mode and a four-point dither pattern in the negative direction, optimized for extended sources. All three gratings (A, B, and C) were used thus covering the entire wavelength range available with MIRI (4.9 - 27.9 $\mu$m) with spectral resolution ranging from $\sim 0.0015 \mu$m at 4.9 $\mu$m to $\sim$0.02 $\mu$m at 27.9 $\mu$m. A single pointing was used for each source targeting the corresponding source coordinates (see Table \ref{table:cat}).The resulting field of view (FoV) depends on the MIRI Channel, and goes from 3.2\arcsec $\times$ 3.7\arcsec in Channel 1 (4.9–7.65 $\mu$m) up to 6.6\arcsec $\times$ 7.7\arcsec in Channel 4 (17.7–27.9 $\mu$m). The integration time per channel is also shown in Table \ref{table:cat}. Two separate background observations were taken for sources in Oph A and Oph E/F respectively.  

The data were processed using the JWST pipeline \citep{Bushouse24} with the reference context \texttt{jwst\_1188.pmap}  of the JWST Calibration Reference Data System \citep[CRDS;][]{Greenfield16}. The raw data were initially processed using the \texttt{Detector1Pipeline} set at default and next, the \texttt{Spec2Pipeline} was used to create the calibrated detector images. The corresponding background was then subtracted and the fringe flat as well as the residual fringe correction (Kavanagh et al. in prep.) were applied. Next, using the Vortex Imaging Processing package \citep{Christiaens23} an additional bad pixel map was created. The final datacubes were then created from the calibrated detector files by using the \texttt{Spec3Pipeline}, with the drizzle algorithm \citep{Law23}, separately for each sub-band and MIRI/MRS Channel. Both the master background and the outlier rejection steps were switched off at this stage, while the wavelength calibration from \cite{Pontoppidan24} was included in the data reduction.  

\section{Results}
\label{sec:results}
\subsection{Spatial distribution of mid-IR line emission}
\label{sec:maps}

The MIRI range covers a number of H$_2$ lines from the ground-state vibrational level, which are excellent probes of outflows from low-mass protostars \citep[e.g.,][]{Neufeld2009,nisini10spitzer,gia11}. The morphology of H$_2$ emission in various lines is a powerful tool to trace the interaction of the molecular jet with the surrounding envelope \citep[e.g.,][]{nisini10herschel,tappe12}. In this section, we compare the H$_2$ emission with those of atomic/ionic species, and discuss the results in the context of previous studies.

Figure \ref{fig:h2_maps} shows continuum-subtracted H$_2$ S(5) maps for four out of five sources in our sample, revealing their outflow characteristics. 
We detect two distinct outflow morphologies: (i) a narrow, collimated outflow structure (GSS30-IRS 1 and ISO-Oph 137), and (ii) a wide opening angle, \lq\lq butterfly''-shaped outflow with extended outflow cavities ([GY92] 197 and WL 17). The remaining source, ISO-Oph 21, shows weak H$_2$ emission in the S(1) and S(2) lines, has no clear outflow morphology, and no other transition is detected. 

Apart from H$_2$, several atoms and ions also show significantly extended line emission. Figure \ref{fig:jet maps} shows the morphology of the most extended ionic emission in each source, namely that of the [\ion{Fe}{II}] line at 5.34 $\mu$m for [GY92] 197 and ISO-Oph 137, the [\ion{Fe}{II}] line at 25.9 $\mu$m for GSS30-IRS 1, and the [\ion{Ne}{II}] line at 12.8 $\mu$m for WL 17. In the latter source, [\ion{Fe}{II}] emission is not detected. The full list of detections, for every source and location, is given in Table \ref{table:detections}.

The ionic emission reveals well-collimated jets in GSS30-IRS 1, ISO Oph 137, and [GY92] 197. The jet of GSS30-IRS 1 is detected in multiple transitions of [\ion{Fe}{II}], and shows a bipolar morphology extending in the North-South direction (Fig. \ref{fig:jet maps}).
Because of the offset of Channel 4 relative to Channels 1-3 of MIRI, the southern lobe that is not visible in H$_2$ can still be detected in the longer wavelength transitions of [\ion{Fe}{II}]. In any case, the limited FoV of MIRI does not allow to capture the full extent of the outflow.

In ISO-Oph 137 and [GY92] 197, the [\ion{Fe}{II}] emission tracing the jet is detected only toward one of the outflow lobes in each source (Fig. \ref{fig:jet maps}). Thus, the morphology differs from the emission seen in H$_2$, which shows a bipolar symmetry (Fig. \ref{fig:h2_maps}). Emission in additional ionic tracers, in particular [\ion{Ne}{II}] at 12.8 $\mu$m, shows a compact morphology close to the protostar position of ISO-Oph 137, but an extended pattern in [GY92] 197. In WL 17, which shows a particularly wide opening angle in the H$_2$ maps, no collimated jet is detected in atomic/ionic lines. 

In the following subsections, we discuss observations from MIRI/MRS in the context of previous studies of our sources from the literature.

\subsubsection{GSS30-IRS 1}

The outflow activity in GSS30-IRS 1 was identified by the detection of extended line wings of CO \citep[][]{Tamura90} and its association with a bright bipolar reflection nebula \citep{Weintraub93}. The outflow morphology was subsequently studied with single dish CO $J=3-2$ observations with a resolution of 14$\arcsec$ from the James Clerk Maxwell Telescope (JCMT) \citep{white15}. Subsequent, high-resolution ($\sim$0.6\arcsec) CO $J=2-1$ observations from ALMA revealed a significant overlap of the blue- and red-shifted outflow components of GSS30-IRS 1, suggesting a large inclination angle \citep{Friesen18}. While MIRI observations of the H$_2$ outflow are restricted to a single lobe, the spatial extent of [\ion{Fe}{II}] confirms the N-S direction of the outflow (see Fig. \ref{fig:jet maps}). The disk inclination angle of $i \sim  60\degree$ from \cite{Michel23} and \cite{dlV19} agrees with a scenario where the narrow collimated jet displays a well-separated bipolar morphology, and the CO emission shows a significant overlap between the outflow lobes.

\subsubsection{ISO-Oph 137}

ISO-Oph 137 is host to a well-studied protoplanetary disk \citep{Brandner00,Pontoppidan05,vK09}; however, a corresponding molecular outflow has not been reported. The N-S orientation of the H$_2$ outflow in the MIRI/MRS maps (see Fig. \ref{fig:h2_maps}), along with a clear separation of the two outflow lobes, agrees with the proposed edge-on disk orientation \citep{vK09}. The collimated morphology of the outflow traces highly-excited gas over a small volume, which could explain the lack of significant low$-J$ CO emission.

\subsubsection{[GY92] 197}

[GY92] 197 possesses a CO outflow which extends close to the plane of the sky in the E-W direction, revealed by low-$J$ CO single-dish observations (at a resolution of 22 and 40\arcsec, respectively) by \cite{Bussmann07} and \cite{Nakamura11}. The atomic/ionic emission from MIRI confirms the proposed outflow orientation (see Fig. \ref{fig:jet maps}) and is consistent with the almost vertical and edge-on disk \citep{Michel23}. The H$_2$ emission shows broad outflow cavities (Fig. \ref{fig:h2_maps}), which are likely associated with the hourglass-shaped nebulosity detected in near-IR images \citep{Duchene04}.  

\subsubsection{WL 17}

WL 17 is a confirmed embedded YSO \citep{vK09}, which drives a bipolar outflow extending in a NW-SE direction as seen in single dish CO $J = 3-2$ observations (at a resolution of 15\arcsec, \citealt{vdM13}). High-resolution ALMA observations ($\sim$0.03\arcsec) revealed prominent cavities in both outflow lobes \citep{shoshi24}. Both the direction of the outflow and the presence of cavities are also seen in the MIRI/MRS H$_2$ maps (Fig. \ref{fig:h2_maps}). 
In this extended source, [\ion{Ne}{II}] emission is seen, filling up the H$_2$ cavity (bottom right panel in Fig. \ref{fig:jet maps}). Given the broad shape of the emission, the \ion{Ne}{II} does not appear to be associated with any kind of jet launched from the protostar. Scattering of [\ion{Ne}{II}] emission from the source onto the cavity walls offers a likely explanation for the wide morphology of the emission.

\subsubsection{ISO-Oph 21}

The outflow of ISO-Oph 21 was identified via the blue- and red-shifted line wings of CO, but the resolution was insufficient to study its morphology \citep{white15}. The source was classified as a disk source, based on the distribution of dust continuum and dense gas emission, questioning its embedded nature \citep{vK09}. MIRI/MRS observations of ISO-Oph 21 show only very weak H$_2$ emission in the lowest rotational levels and no atomic/ionic emission. Thus, we cannot confirm the presence of an outflow from this source. The apparent lack of an outflow in ISO-Oph 21 is potentially due to its more evolved stage (Table \ref{table:cat}). The initial detection suffered from poor angular resolution, which could lead to the confusion with prominent outflows driven by nearby sources \citep[see Fig. 9 in][]{white15}. 

\subsection{Spectra}
\label{sec:spectra}
 
We used the information about the spatial distribution of the H$_2$ emission to identify regions for the subsequent analysis. Namely, depending on the H$_2$ emission patterns, we selected 3 to 8 circular apertures for each of the sources, making sure they lie within the FoV covered by the MRS spectrometer across the full MIRI wavelength range. The exact location of the apertures for each source are shown in Fig. \ref{fig:S3}. The radius of the circular apertures is 0.5$\arcsec$, ensuring they cover a significant number of pixels in all four MIRI Channels. For sources with collimated outflows, we selected four (ISO-Oph 137) and three (GSS30-IRS 1) regions along the outflow direction, omitting the central region, which is strongly affected by extinction from the protostellar envelope. In case of WL 17 and [GY92] 197, characterized by the butterfly-shaped emission (Section \ref{sec:maps}), we selected 4 apertures in each of the outflow lobes, 8 for each object in total. In the case of ISO-Oph 21, where no clear outflow is detected, the aperture locations where selected to cover the area of the most prominent H$_2$ emission.

\subsubsection{Line detections across the maps}

Several transitions of H$_2$ $\varv=0-0$ are detected toward all targeted low-mass protostars in Ophiuchus (Table \ref{table:cat}), except ISO-Oph 21, where only the S(1) and S(2) transitions are detected. None of the sources shows lines of H$_2$ $\varv=1-1$, which were detected toward other low-mass protostars \citep[e.g.,][]{tych24}. 
In addition, several atomic and ionic emission lines are present in the MIRI/MRS wavelength range. Among these, the most commonly detected lines are the [\ion{Fe}{II}] lines at 5.34 $\mu$m (4 sources), the [\ion{S}{I}] line at 25.2 $\mu$m (4 sources), and [\ion{Ne}{II}] line at 12.8 $\mu$m (3 sources). A summary of the line emission detected in each aperture and each source is presented in Table \ref{table:detections}. 

In GSS30-IRS 1, S(3) to S(8) H$_2$ lines are detected along the full outflow extent, whereas the  S(1) and S(2) lines are only detected in its outermost parts (Fig. \ref{fig:h2_maps}). The reason for the non-detection of lowest-excited lines is due to an increase in the noise level in the spectra, likely resulting from the imperfect fringe correction due to the bright continuum emission toward the protostar's position. Noteworthy, the [\ion{Ne}{II}] line at 12.8 $\mu$m line is not detected toward GSS30-IRS 1, and the [\ion{S}{I}] line at 25.2 $\mu$m is detected only in the outflow positions (B and C). 

In ISO-Oph 137, S(1) to S(7) H$_2$ lines are detected in all apertures, and the most highly-excited S(8) line is also detected toward the B and D positions -- at the tip of the outflow imaged with MIRI/MRS (Fig. \ref{fig:h2_maps}). The non-detection of the S(8) line close to the protostar is most likely due to blending with the H$_2$O and CO lines present in absorption toward the protostar position (Rocha et al., in preparation). Additionally, the [\ion{Fe}{II}] lines are detected toward all positions except position A. The [\ion{Ne}{II}] line at 12.8 $\mu$m and [\ion{S}{I}] line at 25.2 $\mu$m are detected along the entire outflow.

In [GY92] 197, H$_2$ lines up to S(8) are detected in the western outflow lobe, which shows a brighter H$_2$ emission in general (Fig. \ref{fig:h2_maps}). The lack of highly excited H$_2$ lines in the bulk of the eastern outflow lobe, except the position closest to the protostar, is likely due to sensitivity limits of the observations. While [GY92] 197 is  characterized by the largest number of detections of atomic and ionic species among all sources (Table \ref{table:detections}), such emission is absent in the eastern outflow lobe, except the [\ion{Ne}{II}] line at 12.8 $\mu$m. 

In WL 17, H$_2$ lines up to S(8) are detected across all apertures. From the ionic/atomic emission lines, the [\ion{Ne}{II}] line at 12.8 microns is the most prominent, being detected in all positions except position D. In contrast, [\ion{Fe}{II}] is only seen in aperture E, close to the location of the protostar. 

In ISO-Oph 21, the lack of emission from H$_2$ transitions above S(2) is consistent with the lack of a clear outflow morphology (Section \ref{sec:maps}). It also lacks any emission from atomic and ionic lines due to the absence of a jet.

\begin{figure}
\centering 
\includegraphics[width=0.49\linewidth]{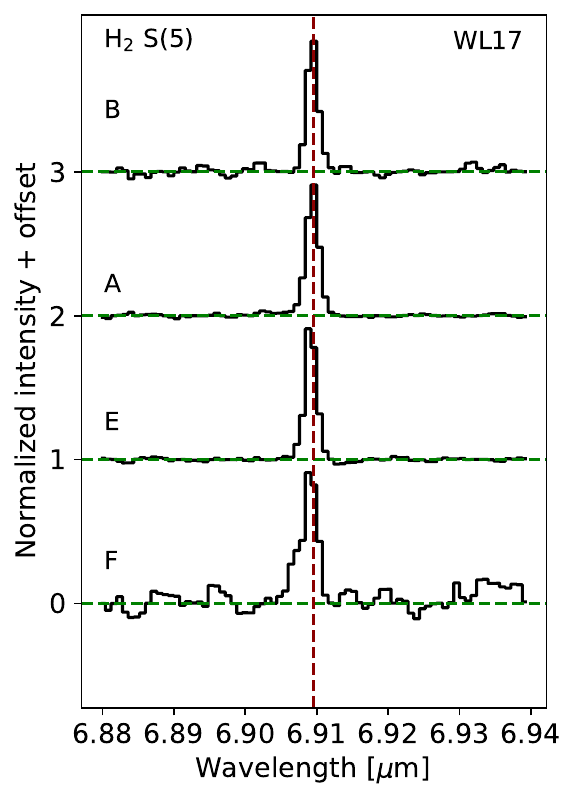}
\includegraphics[width=0.49\linewidth]{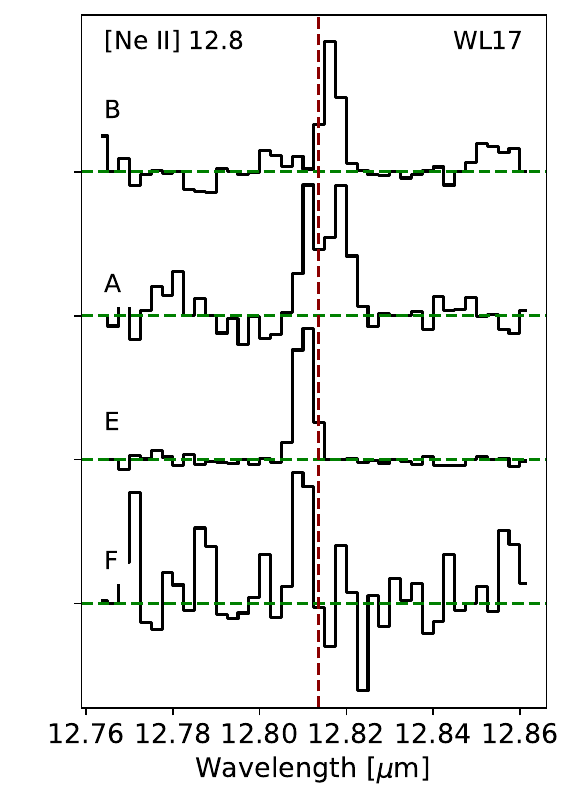}
\caption{Continuum-subtracted spectra of H$_2$ S(5) (left) and [\ion{Ne}{II}] 12.8 $\mu$m (right) toward WL 17. An offset is added to the spectra at various positions on the maps for better visualization. The dashed lines show the laboratory wavelengths of the respective transitions.}
\label{fig:velshift_spectra}
\end{figure}

\subsubsection{Gas kinematics}
\label{sec:kinematics}
Jets from protostars are often characterized by gas velocities exceeding 100 km s$^{-1}$, primarily seen in atomic/ionic lines \citep{Frank14,Bal16}. Observations with MIRI, with a velocity resolution from $\sim$85 (in MIRI/MRS Channel 1) to $\sim$200 km s$^{-1}$ (in Channel 4), may provide some information about the gas kinematics. For instance, a fast-moving jet resulting in velocity shifts of molecular and atomic mid-IR lines was reported in HH46 \citep[][]{nisini24}. 

We investigated the line profiles of molecular and ionic lines to search for possible velocity shifts across the maps. We focused on apertures corresponding to the location of H$_2$ jets (Fig. \ref{fig:h2_maps}) to maximize the chance to detect any velocity gradients. 

Figure \ref{fig:velshift_spectra} shows the results obtained for WL 17, the only source in the sample where velocity shifts can be noticed in the [\ion{Ne}{II}] line at 12.8 $\mu$m. 
However, similar shifts are not as clear in the H$_2$ lines toward this source. Noteworthy, WL 17 does not possess a collimated H$_2$ jet, and is characterized by a wide opening angle in molecular emission, and a rather compact pattern of ionic emission (Figs. \ref{fig:h2_maps}, \ref{fig:jet maps}). We note here that the one sided detection of ionic emission in ISO-Oph 137 and [GY92] 197 restricts our ability to compare the velocities between the two lobes of the jet, where differences in velocity would be most prominent.
Additionally, the incomplete coverage of the GSS30-IRS 1 outflow, where only one lobe is covered in all wavelengths makes such a comparison impossible in its case.   

We also estimated gas velocities of the different H$_2$ transitions, using Gaussian fits to the spectra of each source and from each aperture. This approach, when applied to sufficiently strong lines, enables us to estimate velocity shifts at a resolution better than the instrumental MIRI/MRS resolution. Table \ref{table:velocities} shows the resulting velocities for a warm gas component, estimated as the average velocity of the S(2) and S(3) transitions, and the velocity of the only the S(7) transition for a hot component. The presence of two separate gas components, warm and hot respectively, is established through the rotational diagram analysis in Sec. \ref{sec:analysis}. Here, we selected these transitions as they represent the best detected lines associated with the warm and hot component respectively. In most cases, despite the similar spatial distribution of the two components, the hot component displays somewhat higher velocities than the warm component, potentially indicating that the hot component has a stronger connection to the fast moving jet. Similar to the direct spectral comparison, the hot component of WL 17 is the only case in which a noticeable velocity shift between the two lobes can be seen. We detect redshifted emission in the northern lobe and blueshifted towards the south, matching the orientation of the entrained CO outflow \citep{shoshi24}.

Still, velocities for all H$_2$ lines lie within a single spectral channel of MIRI/MRS indicating that the molecular gas component of the protostellar ejections is associated with material moving at lower velocities than the dominant atomic/ionized gas component \citep{harsono23,tych24}.

\begin{table}[htb!]
\caption{Gaussian fitting velocity estimates for the warm and hot H$_2$ gas component.}
\label{table:velocities} 
\small
\centering 
\begin{tabular}{l c c} 
\hline\hline 
Pos. &\begin{tabular}[c]{@{}c@{}}$v_\text{warm}$\\  $\left( \text{km s}^{-1} \right)$ \end{tabular}  &\begin{tabular}[c]{@{}c@{}}$v_\text{hot}$\\  $\left( \text{km s}^{-1} \right)$ \end{tabular}\\
\hline
\multicolumn{3}{c}{GSS30-IRS 1}\\
\hline
A & -12.4 $\pm$ 5.9\phantom{-1} & -9.8 $\pm$ 2.0\phantom{-} \\
B & -9.9 $\pm$ 2.7\phantom{-} & -14.7 $\pm$ 2.3\phantom{-1}\\
C & -16.7 $\pm$ 1.3\phantom{-1} & -28.8 $\pm$ 6.9\phantom{-1}\\
\hline
\multicolumn{3}{c}{[GY92] 197}\\
\hline
A & -9.0 $\pm$ 1.3\phantom{-} & -17.4 $\pm$ 1.7\phantom{-1} \\
B & -5.4 $\pm$ 1.4\phantom{-} & -11.4 $\pm$ 4.3\phantom{-1} \\
C & -8.0 $\pm$ 0.5\phantom{-} & -15.8 $\pm$ 2.6\phantom{-1} \\
D & -8.9 $\pm$ 0.9\phantom{-} & -12.0 $\pm$ 1.4\phantom{-1} \\
E & -5.1 $\pm$ 1.4\phantom{-} & -12.5 $\pm$ 2.4\phantom{-1} \\
F & -8.5 $\pm$ 1.2\phantom{-} & -13.1 $\pm$ 5.4\phantom{-1} \\
G & -7.0 $\pm$ 0.8\phantom{-} &  -7.6 $\pm$ 3.5\phantom{-} \\
H & -5.1 $\pm$ 1.9\phantom{-} & -11.4 $\pm$ 10.7\phantom{-} \\
\hline
\multicolumn{3}{c}{WL 17}\\
\hline
A & -2.4 $\pm$ 0.5\phantom{-} & 1.6 $\pm$ 1.7 \\
B & -4.4 $\pm$ 1.6\phantom{-} & 3.8 $\pm$ 2.1 \\
C & -0.2 $\pm$ 0.2\phantom{-} & 3.8 $\pm$ 1.0\\
D & 1.7 $\pm$ 1.0 & 11.4 $\pm$ 1.5\phantom{1}\\
E & -10.0 $\pm$ 0.9\phantom{-1} & -20.1 $\pm$ 1.5\phantom{-1} \\
F & -4.5 $\pm$ 1.4\phantom{-} & -23.9 $\pm$ 9.9\phantom{-1} \\
G & -9.9 $\pm$ 1.1\phantom{-} & -19.0 $\pm$ 1.1\phantom{-1} \\
H & -7.0 $\pm$ 0.8\phantom{-}  & -15.8 $\pm$ 3.0\phantom{-1} \\
\hline
\multicolumn{3}{c}{ISO-Oph 137}\\
\hline
A & -10.8 $\pm$ 7.4\phantom{-1} & -1.6 $\pm$ 5.9\phantom{-}\\
B & -10.4 $\pm$ 1.0\phantom{-1} & 7.1 $\pm$ 3.3 \\
C & -1.8 $\pm$ 0.1\phantom{-}& \phantom{1}-8.7 $\pm$ 11.9\phantom{-} \\
D & -8.7 $\pm$ 1.3\phantom{-} & -38.1 $\pm$ 18.7\phantom{-}\\
\hline
\end{tabular}
\end{table}

\section{Analysis}
\label{sec:analysis}

MIRI/MRS imaging of H$_2$ emission on 100 AU scales allows us to associate molecular gas with the outflows from Class I protostars. Detection of multiple H$_2$ lines can constrain both the gas excitation as well as the shock properties responsible for the emission. However, external UV radiation in the Ophiuchus region likely also influences the shock structure and its observational signatures. In the following sections, the gas excitation toward four protostars in Ophiuchus and its spatial variation are discussed (Section \ref{sec:h2}), and the ratios of selected H$_2$ lines are compared with the predictions of shock models including the effects of external UV radiation (Section \ref{sec:shocks}).

\subsection{Rotational diagrams of H$_2$}
\label{sec:h2}

\begin{figure*}
\centering 
\includegraphics[width=0.49\linewidth]{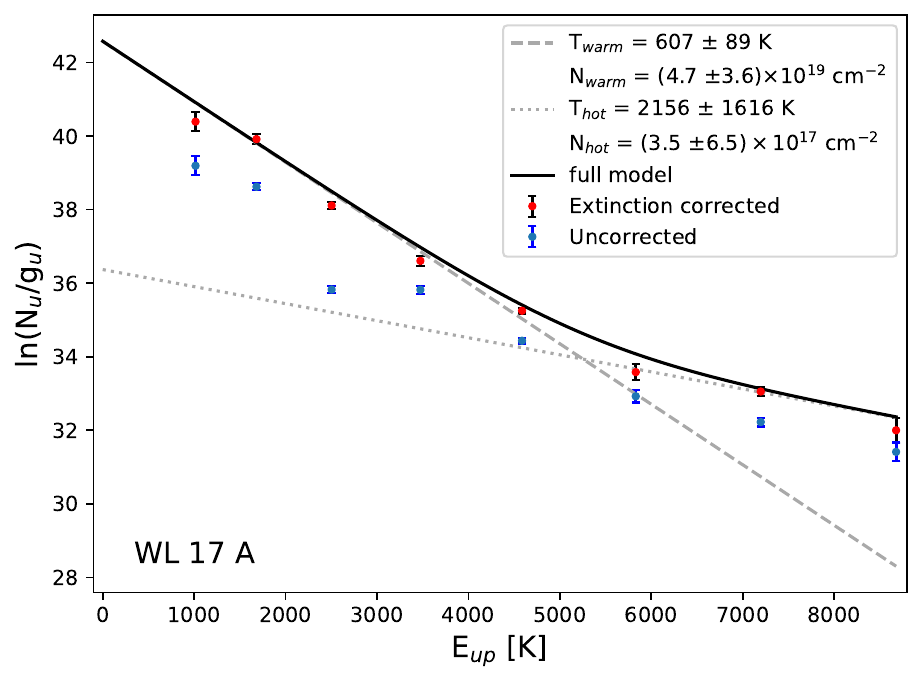} 
\includegraphics[width=0.49\linewidth]{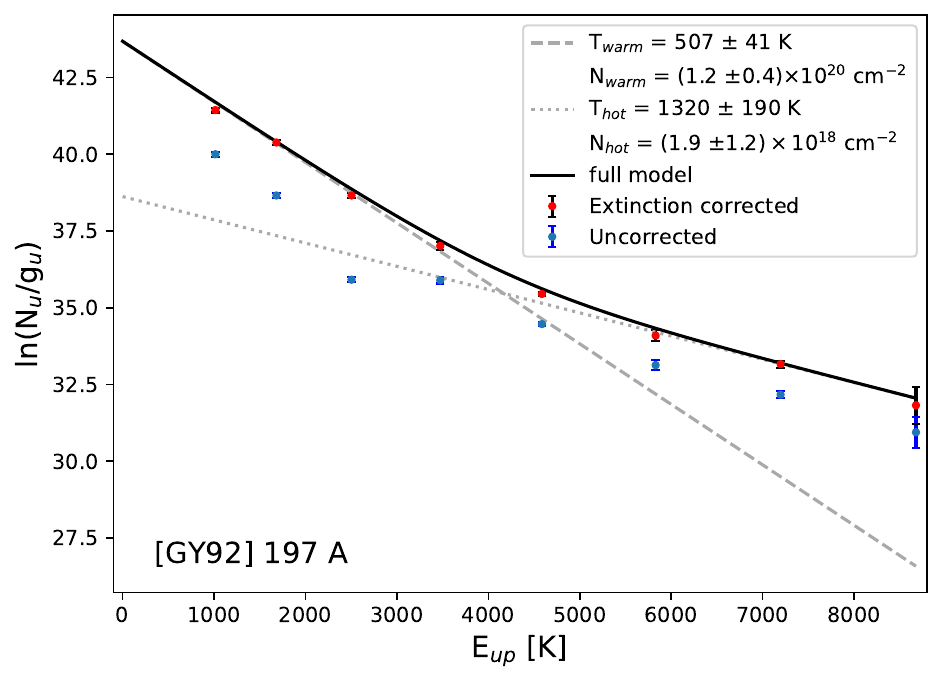} 
\includegraphics[width=0.49\linewidth]{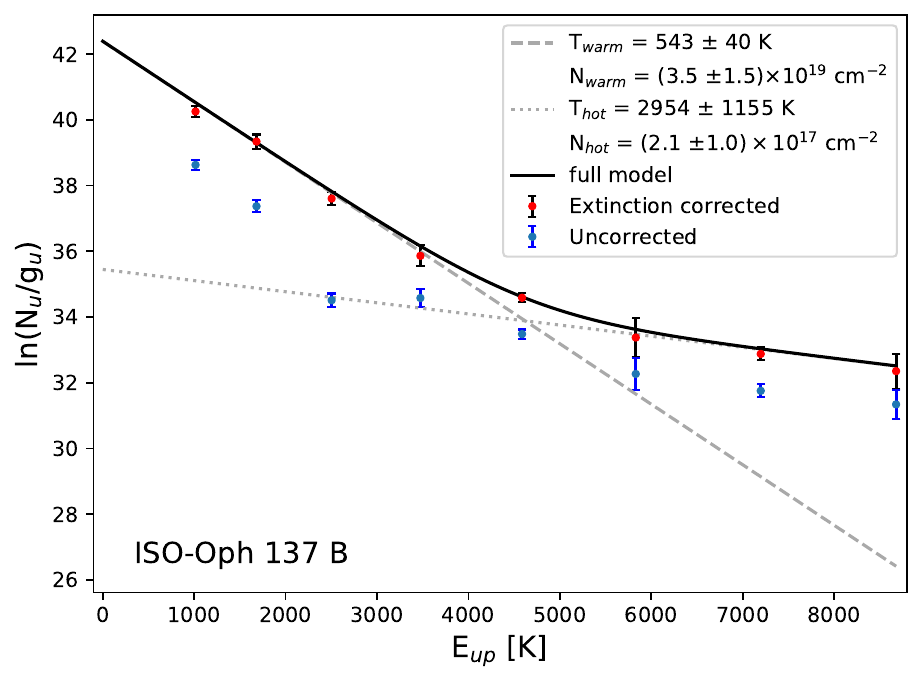} 
\includegraphics[width=0.49\linewidth]{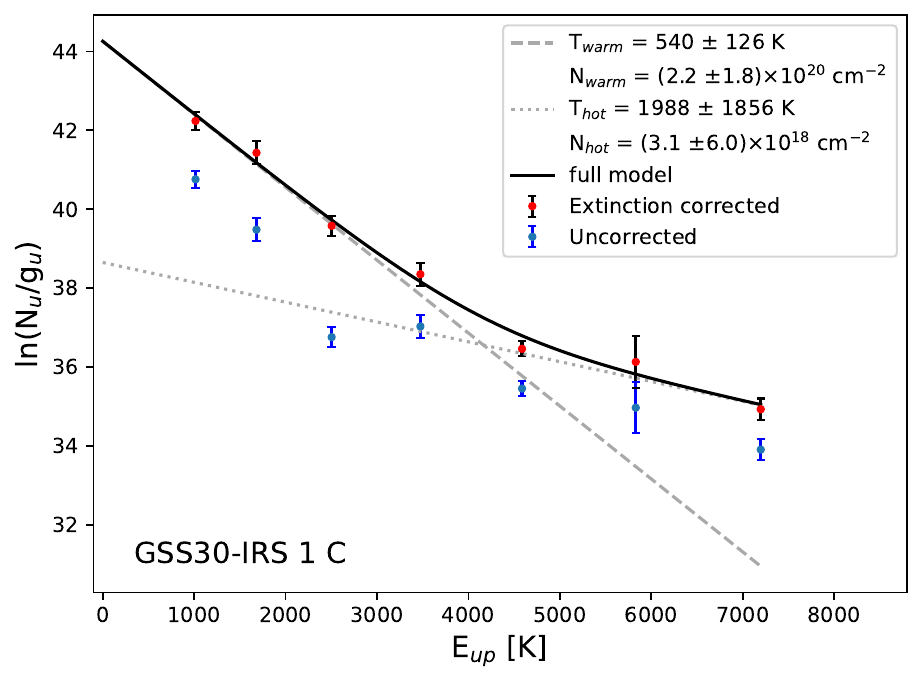} 
\caption{H$_2$ rotational diagrams for the outflow positions in WL 17, [GY92] 197, ISO-Oph 137, and GSS30-IRS 1 with the largest number of line detections. The natural logarithm of
the column density from a level $u$, $N_\mathrm{u}$, divided by the
degeneracy of the level, $g_\mathrm{u}$, is written on the Y-axis. The extinction-corrected values are shown in red, and the ones before the correction in blue. The two-component fits cover the transitions below and above $E_\mathrm{u}$$\sim$4000 K for the \lq warm' (dashed line) and \lq hot' (dotted line) component (see text). The combined fit is shown as a black line. Finally, letters next to source name mark the aperture where the H$_2$ were measured.}
\label{fig:rotation_diagrams}
\end{figure*}

Rotational diagrams are commonly used to estimate gas physical conditions and column densities of molecules along outflows from low-mass protostars \citep[e.g., ][]{Neufeld2009,herczeg12,karska13,green16,manoj2013,yang18}. 
The detection of multiple H$_2$ lines with MIRI/MRS (Section \ref{sec:spectra}) allows the determination of the gas rotational temperatures and the corresponding H$_2$ column densities across the maps of protostars in Ophiuchus.

For optically thin, thermalized lines, the natural logarithm of the column density of the upper level $N_\mathrm{u}$ of a given transition over its degeneracy $g_\mathrm{u}$ is related linearly to the energy $E_\mathrm{u}$ of that level \citep{Mangum15}:

\begin{equation}
   \mathrm{ln} \left( \frac{N_\mathrm{u}}{g_\mathrm{u}} \right) = \mathrm{ln} \left( \frac{N_\mathrm{tot}}{Q(T_\mathrm{rot})} \right) - \frac{E_\mathrm{u}}{T_\mathrm{rot}} ,
\end{equation}

\noindent where $Q(T_\mathrm{rot})$ is the rotational partition function at a temperature $T_\mathrm{rot}$ for a given molecule and  $N_\mathrm{tot}$ is the total column density.

The level-specific column density $N_\mathrm{u}$ is calculated from the measured line intensities,  $I_\mathrm{ul}$:
\begin{equation}
\label{eq2}
  N_\mathrm{u}  = \left( \frac{4 \pi I_\mathrm{ul}}{A_\mathrm{ul} g_\mathrm{u} h\nu} \right),
\end{equation}

\noindent where $A_\mathrm{ul}$ the Einstein emission coefficient, $h$ the Planck constant, and $\nu$ the frequency of the line. The line intensities, following the approach presented in \cite{Francis25}, are calculated using the integrated fluxes ($F_\text{ul}$; shown in Table \ref{table:H2_fluxes}) as $I_\text{ul} = F_\text{ul}/\Omega$, where $\Omega$ is the opening angle for an aperture with a diameter of 1\arcsec at the distance of Ophiuchus \citep[137 pc;][]{Ortiz17}.

Due to significant dust content in the envelopes of low-mass protostars, mid-IR lines of H$_2$ have to be corrected for extinction. We follow the procedure described in detail in \cite{Francis25}, in which the wavelength-dependent correction factor is determined from the fit to the rotational diagram using uncorrected intensities. Due to the positive curvature of our H$_2$ rotational diagrams, we fit the natural logarithm of the column densities obtained from observations (Eq. \ref{eq2}) over the level degeneracy, $g_\mathrm{u}$, with a two-temperature component fit and an additional term for the ortho-to-para ratio (OPR) correction:
\begin{equation}
    y = \mathrm{ln} \left[ \left(N_{warm}\frac{\mathrm{exp}(-E_\mathrm{u}/T_\mathrm{warm})}{Q(T_\mathrm{warm})} + N_\mathrm{hot}\frac{\mathrm{exp}(-E_\mathrm{u}/T_\mathrm{hot})}{Q(T_\mathrm{hot})}\right) 10^{\frac{-A_\lambda}{2.5}}\right] + z(J)
\end{equation}

\noindent where $A_{\lambda}$ is taken from the KP5 extinctions curve \citep{Pontoppidan24}, and with
$$
    z(J)= 
\begin{cases}
    \ln\left(\frac{\mathrm{OPR}}{3}\right)& \text{odd } J\\
    0,              & \text{even } J
\end{cases},
$$
where OPR is the ortho-to-para ratio.
This way, we obtain the H$_2$ column densities and temperatures of two gas components, as well as the value of the local extinction. In the two-component fit, the low and high temperature components are primarily constrained by the S(1)-S(4) and S(5)-S(8) transitions respectively.

\begin{figure*}
\centering 
\includegraphics[width=0.49\linewidth]{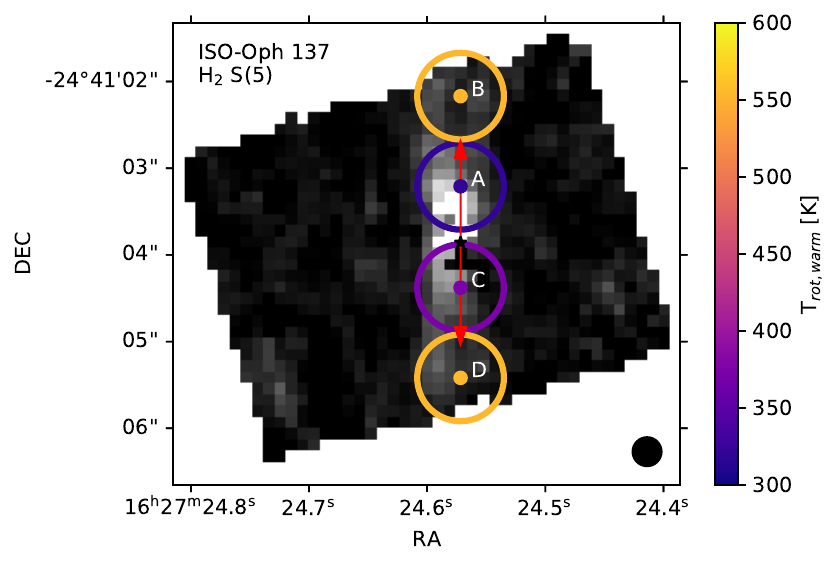} 
\includegraphics[width=0.49\linewidth]{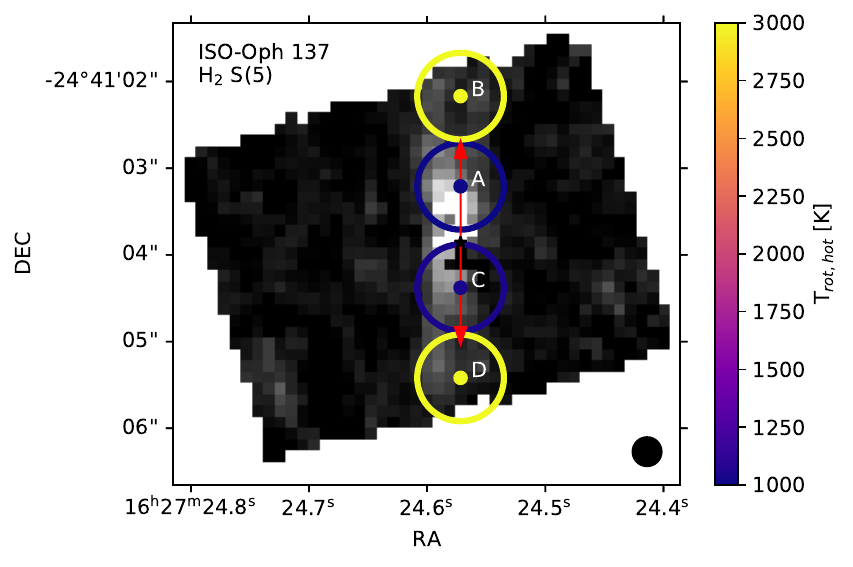} 
\includegraphics[width=0.49\linewidth]{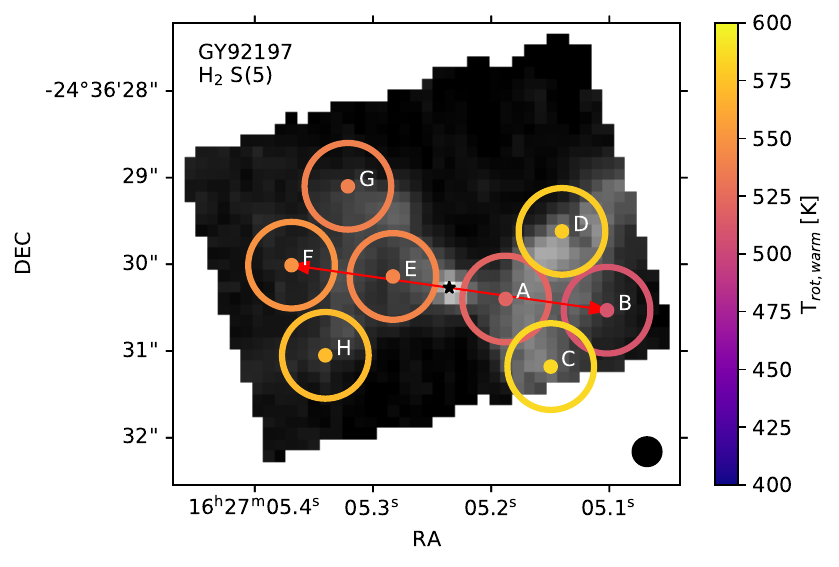} 
\includegraphics[width=0.49\linewidth]{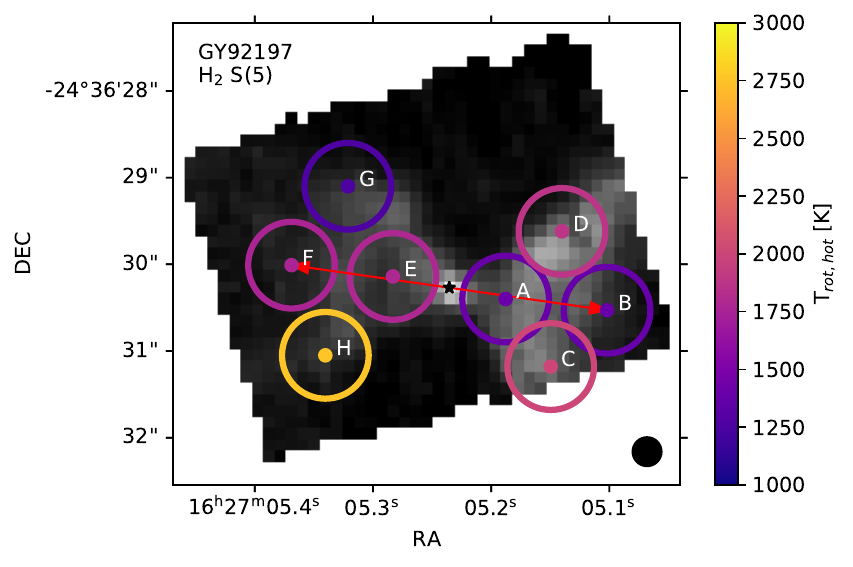} 
\caption{Integrated emission of H$_2$ S(5) line toward ISO-Oph 137 (top) and [GY92] 197 (bottom) with circles color-coded with the rotational temperatures of warm (left) and hot (right) gas components, respectively. Red arrows show the outflow directions, and capital letters refer to apertures used for the spectra extraction.}
\label{fig:CRBR2422_results}
\end{figure*}

\begin{table*}[htb!]
\caption{Rotational temperatures, column densities, extinction, and ortho-to-para ratios for all sources and apertures}
\label{table:Ntot Tkin Ak results} 
\small
\centering 
\begin{tabular}{l c c c c c c} 
\hline\hline 
Pos. &  \begin{tabular}[c]{@{}c@{}}$T_\text{warm}$\\  $\left(  \text{K} \right)$ \end{tabular}& \begin{tabular}[c]{@{}c@{}}$T_\text{hot}$\\  $\left(  \text{K} \right)$ \end{tabular} & \begin{tabular}[c]{@{}c@{}}$N_\text{warm}$\\ $\left(\times10^{19}   \text{cm}^{-2} \right)$ \end{tabular}&\begin{tabular}[c]{@{}c@{}}$N_\text{hot}$\\ $\left( \times10^{17}  \text{cm}^{-2} \right)$ \end{tabular} & A$_K$ & OPR\\
\hline
\multicolumn{7}{c}{GSS30-IRS 1}\\
\hline
A\tablefootmark{a} & 679$\pm$0.5\phantom{.} & -- & 65.5$\pm0.4$\phantom{1} & -- & 4.7$\pm$2.0 & 3.0\tablefootmark{b} \\
B\tablefootmark{a} & 1495$\pm$206\phantom{1} & -- & 1.4$\pm$0.8 & -- & 2.0$\pm$6.5 & 3.0\tablefootmark{b} \\
C & 540$\pm$126 & 1988$\pm$1856 & 22.0$\pm$18.0 & 31.0$\pm$60.0& 3.0$\pm$1.2 & 3.0\tablefootmark{b} \\
\hline
\multicolumn{7}{c}{[GY92] 197}\\
\hline
A & 507$\pm$41\phantom{1} & 1320$\pm$190\phantom{1} & 12.0$\pm$4.0\phantom{1} & 19.0$\pm$12.0 & 2.9$\pm$0.5 & 2.5$\pm$0.3 \\
B &444$\pm$56\phantom{1} & 1269$\pm$72\phantom{10} & 3.8$\pm$0.7 & 17.0$\pm$4.0\phantom{1} & 2.3$\pm$0.5 & 2.1$\pm$0.2 \\
C &560$\pm$10\phantom{1} &1698$\pm$84\phantom{10} &6.1$\pm$0.5 &9.2$\pm$1.5 &2.8$\pm$0.1 & 2.2$\pm$0.1 \\
D &545$\pm$32\phantom{1} &1607$\pm$166\phantom{1} &8.0$\pm$1.6 &19.0$\pm$7.0\phantom{1} & 3.0$\pm$0.3 & 2.1$\pm$0.2 \\
E &543$\pm$55\phantom{1} &2198$\pm$2082 &9.9$\pm$1.9 &4.0$\pm$7.1 & 3.3$\pm$0.7 & 2.2$\pm$0.5 \\
F &532$\pm$164 & 1894$\pm$2198 &3.8$\pm$4.2 &\phantom{1}5.2$\pm$13.4 & 3.3$\pm$1.6 & 2.0$\pm$0.9\\
G &507$\pm$116 & 1140$\pm$325\phantom{1} &6.1$\pm$4.4 & 19.0$\pm$29.0& 3.4$\pm$1.1 & 2.1$\pm$0.6\\
H &566$\pm$38\phantom{1} & 2520$\pm$1194 &4.0$\pm$1.8 & 1.2$\pm$1.1 & 2.9$\pm$0.6 & 2.5$\pm$0.4 \\
\hline
\multicolumn{7}{c}{WL 17}\\
\hline
A &607$\pm$89\phantom{1} &2156$\pm$1616 &4.7$\pm$3.6 &3.5$\pm$6.5 & 2.4$\pm$1.0 & 2.3$\pm$0.7 \\
B &600$\pm$44 \phantom{1} &1964$\pm$444\phantom{1} &2.9$\pm$0.9 & 4.8$\pm$3.0 & 2.4$\pm$0.4 & 2.3$\pm$0.3 \\
C &649$\pm$73\phantom{1} &1863$\pm$746\phantom{1} &3.5$\pm$1.5 &6.7$\pm$8.5 & 2.4$\pm$0.6& 2.6$\pm$0.5 \\
D &532$\pm$206 &1185$\pm$347\phantom{1} &3.8$\pm$2.8 &28.0$\pm$40.0 & 2.5$\pm$1.3 & 2.9$\pm$0.9 \\
E &544$\pm$65\phantom{1} & 1813$\pm$608\phantom{1} &7.0$\pm$3.9 &7.9$\pm$7.9 & 2.2$\pm$0.8 & 2.0$\pm$0.4 \\
F &460$\pm$42\phantom{1} & 2456$\pm$550\phantom{1}&4.2$\pm$1.6 & 3.2$\pm$1.1 & 2.7$\pm$0.6 & 2.2$\pm$0.3 \\
G &502$\pm$69\phantom{1} & 1283$\pm$120\phantom{1} &4.0$\pm$1.0 &23.0$\pm$9.0\phantom{1} & 1.8$\pm$0.5 & 2.2$\pm$0.2\\
H &541$\pm$50\phantom{1} & 1993$\pm$695\phantom{1} &3.9$\pm$2.0 &2.6$\pm$2.4 & 1.8$\pm$0.7 & 2.3$\pm$0.5 \\
\hline
\multicolumn{7}{c}{ISO-Oph 137}\\
\hline
A &317$\pm$15 & 1000$\pm$70\phantom{11} &560.0$\pm$100.0 & 230.0$\pm$30.0\phantom{1} & 5.5$\pm$0.3 & 2.7$\pm$0.1 \\
B &543$\pm$40 & 2954$\pm$1155 &3.5$\pm$1.5 &2.1$\pm$1.0 & 3.3$\pm$0.6 & 2.5$\pm$0.4 \\
C &380$\pm$40 &1061$\pm$100\phantom{1} &84.0$\pm$8.4\phantom{1} &98.0$\pm$9.9\phantom{1} & 4.3$\pm$0.4 & 2.4$\pm$0.2 \\
D &541$\pm$33 &2787$\pm$762\phantom{1} &2.5$\pm$0.8 &1.9$\pm$0.7 & 3.1$\pm$0.4 & 2.5$\pm$0.3\\
\hline
\end{tabular}
\tablefoot{\tablefoottext{a}{Values derived from single temperature component fit due to the non-detection of the S(1), S(2) and S(8) lines in these apertures.}
\tablefoottext{b}{OPR equilibrium value of 3.0 assumed due to the non-detection of the S(1) and S(2) transitions in apertures A and B. For aperture C, an OPR ratio of 3 led to a better fit, and is therefore used.}
}
\end{table*}

Figure \ref{fig:rotation_diagrams} shows example rotational diagrams for each source in our sample, except ISO-Oph 21 (see Section \ref{sec:results}), and for the aperture in which the number of H$_2$ line detections was the largest. Table \ref{table:Ntot Tkin Ak results} shows the temperatures and column densities obtained for all sources and apertures.
 
\begin{figure}[ht!]
\centering 
\includegraphics[width=\linewidth]{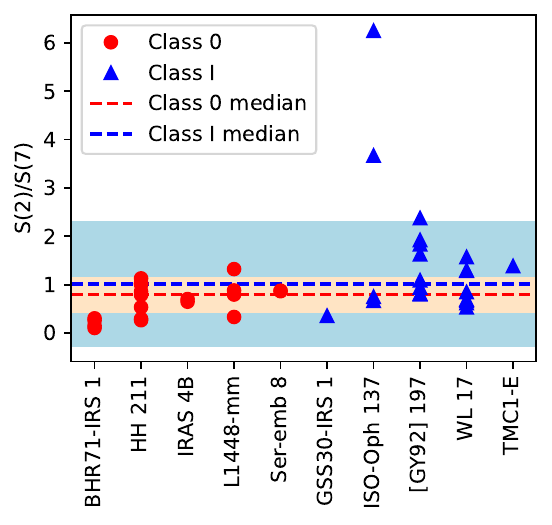} 
\caption{\label{fig:ratio} Ratio of the S(2)/S(7) H$_2$ emission lines in different low-mass protostellar sources. Red points mark Class 0 sources from \cite{Francis25} and blue points are Class I sources from this work and TMC1-E from \cite{tych24}. The dashed lines mark the median values, 0.3 for Class 0 (in red) and 1.0 (in blue). Colored areas mark the median $\pm$ standard deviation range.}
\end{figure}

\begin{figure*}[!]
\centering 
\includegraphics[width=0.99\linewidth]{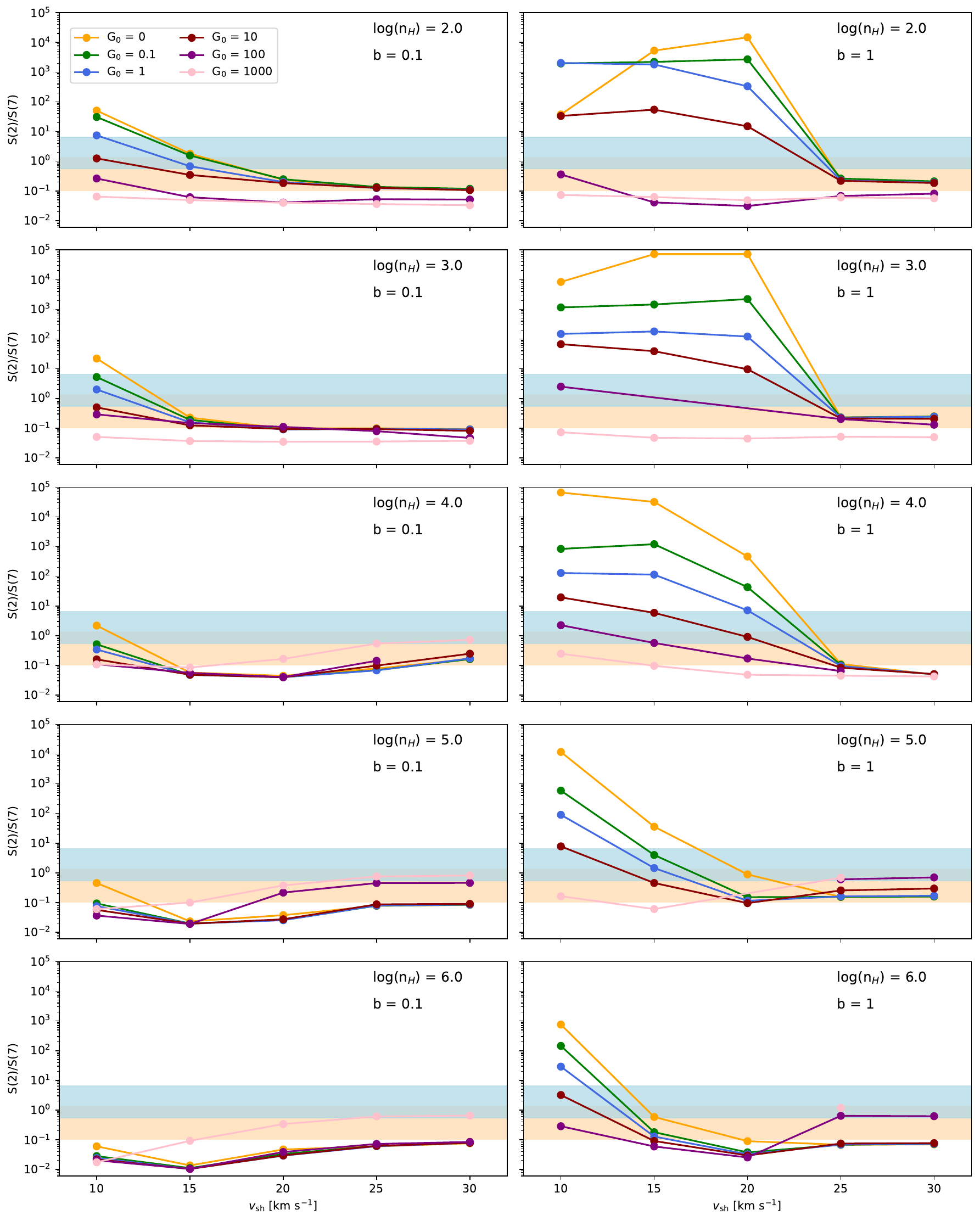} 
\caption{\label{fig:G0_comparison} Ratio of the S(2) over the S(7) transition of H$_2$ against the shock velocity for different $\mathrm{UV}$ field strengths. The log(n$_H$) ranges from 2 to 6 from top to bottom, while magnetic field strength is b = 0.1 for the left column and 1 for the right. For all panels $\zeta$ = 10$^{-17}$ s$^{-1}$, and X(PAH)= $10^{-7}$ are used. The blue shaded region marks the observed range of S(2)/S(7) ratios for the sources in Ophiuchus, while the beige region marks the range for Class 0 sources from \cite{Francis25}. Note that line intensities from \cite{Kristensen23} are not available for all combinations of model parameters covered in the figure.}
\end{figure*}

The H$_2$ rotational diagrams are well-described by the two component model with distinct temperatures, corresponding to $T_\mathrm{warm}$ of $\sim$500-600 K, and $T_\mathrm{hot}$ of $\sim$1000-3000 K (Fig. \ref{fig:rotation_diagrams}). The warm component temperatures show a relatively narrow range considering also other apertures, whereas the range of the hot component values spans thousands of K (Table \ref{table:Ntot Tkin Ak results}). Noteworthy, a small number of line detections corresponding to the hot component results in significant uncertainties of the fit. Similarly, the warm component of GSS30-IRS 1 in apertures A and B is also less constrained, and likely overestimated, due to the lack of detections of S(1)-S(2) lines caused by the strong continuum emission in the region (see Section \ref{sec:spectra}). We find OPR = 2-3 for most cases, consistent the results of previous studies on the OPR within protostellar outflows \citep[e.g.][]{neu06}. For GSS30-IRS 1 we adopt the equilibrium OPR ratio of 3 due to the non-detection of S(1) and S(2) lines, where the effects of deviation from the OPR = 3  are most prominent.

The spatial variations of the rotational temperatures toward two sources with distinct morphologies are presented in Fig. \ref{fig:CRBR2422_results}. ISO-Oph 137 is a source with a bipolar collimated outflow, which shows a clear increase of the gas temperatures toward the map edges -- a trend which is confirmed by the detections of S(8) lines only toward the outermost apertures (Table \ref{table:detections}). Such a morphology is consistent with the presence of \lq\lq shock spots'' tracing the interaction of the jet/outflow with the surroundings, seen e.g., in L1157 \citep{nisini10herschel}. [GY92] 197, on the other hand, drives an outflow with a wide opening angle, and shows significantly less variations in the gas temperatures (Fig. \ref{fig:CRBR2422_results}). The western outflow lobe, which shows brighter H$_2$ emission, seems to be associated with warmer gas than the eastern outflow lobe. Generally, the highest temperatures correspond well with the positions along the outflow cavity rather than the ionic jet. This is seen also in the eastern outflow lobe, where one of the positions is associated with especially high hot component (aperture H). The higher rotational temperatures found along the edges of the outflow cavity suggest that the H$_2$ is heated by shocks produced at the cavity walls instead of along the central jet. Such shocks likely arise from the interaction between the outflowing gas and the surrounding cold envelope.
 
The variation of gas excitation in different positions along the outflows are further explored in Fig. \ref{fig:ratio}. Here, we  calculate the ratio of two H$_2$ transitions which are detected toward most of our sources: the S(2) line from the warm gas component and the S(7) line from the hot gas component (see also Section \ref{sec:shocks}). 
The median ratio for our sources is $\sim$1.0. The highest values of the ratio correspond to the central positions of the ISO-Oph 137, which show significantly lower excitation that the outermost parts of the outflow (see Fig. \ref{fig:CRBR2422_results}). Overall, the S(2)/S(7) ratio for the Ophiuchus protostars, constituting only Class I protostars, is about a factor of 3 higher than the same ratio calculated for other low-mass protostars using the literature data \citep{tych24,Francis25}. We will discuss a possible impact of evolutionary trends on the H$_2$ excitation in Section \ref{sec:discussion}.

To summarize, the analysis of H$_2$ lines confirms the presence of highly-excited molecular gas with properties showing some changes depending on the exact origin of the emission, e.g. along the outflow cavities or jet. The differences between sources in our sample are not significant, suggesting that neither source properties or environment conditions play a strong role in the H$_2$ excitation.

\subsection{Comparisons to shock models}
\label{sec:shocks}

In this section, we compare the measured H$_2$ intensities from our observations to the results of models of UV-irradiated shocks propagating in a broad range of physical conditions of the ISM \citep{Kristensen23}. We aim to constrain the shock properties for the Ophiuchus protostars and identify the impact of external UV radiation from nearby B-type stars on the H$_2$ excitation in their outflows.

The model grid from \cite{Kristensen23} was obtained using the Paris-Durham\footnote{\url{http://ism.obspm.fr/shock.html}} shock code  that simulates the changes in the chemical, thermal and dynamical conditions of interstellar matter due to steady state, plane parallel shocks \citep{fp03}. In its latest version, the code accounts for the impact of external UV irradiation on the shock structure and chemistry \citep{lesaffre13,Godard19}.

\cite{Kristensen23} considers the following input parameters of the environment and shock properties: (i) the pre-shock gas density $n_\text{H}$, (ii) the traverse magnetic field strength, $B = b\sqrt{n_\textrm{H}}$ in units of $\mu$Gauss, where $b$ is a scaling factor for magnetic field, (iii) the shock velocity $v_\text{sh}$ (iv) the external UV radiation field in units of G$_0$, (v) the cosmic ray ionization rate $\zeta_{H_2}$, and (vi) the abundance of polycyclic aromatic hydrocarbons, $X$(PAH). The code calculates a chemical steady-state for a given gas density $n_\text{H}$ and UV field (step 1), the chemical equilibrium results are used as input for a PDR calculation (step 2), and the resulting PDR conditions are used as input on the shock model (step 3), for details see Section 2 of  \cite{Kristensen23}. The excitation of H$_2$ is subsequently calculated taking into considerations collisions with H \citep{flower97,fr98,Martin95}, H$_2$ \citep{fr98}, and He \citep{Flower98}. The resulting H$_2$ line intensities, along with a number of atomic line intensities, are provided in a machine readable format on CDS\footnote{\url{https://cdsarc.cds.unistra.fr/viz-bin/cat/J/A+A/675/A86\#/article}}. 

The shock model predictions from \cite{Kristensen23} are provided for the following ranges of parameters: $n_\text{H}$ from 10$^{2}$ to 10$^{8}$ cm$^{-3}$, $b$ from 0.1 to 10, $v_\text{sh}$ from 2 to 90 km s$^{-1}$, G$_0$ from 0 to 10$^3$, $\zeta_{H_2}$ from 10$^{-17}$ to 10$^{-15}$ s$^{-1}$, and $X$(PAH) from 10$^{-8}$ to 10$^{-6}$. Here, we restrict the considered parameters to those expected in outflows from low-mass protostars. In particular, typical (post-shock) gas densities in outflows are in the range from 10$^{5}$-10$^{8}$ cm$^{-3}$ \citep{mot14}, given the high compression factors predicted in \cite{Kristensen23} all available pre-shock densities, from 10$^{2}$-10$^{6}$ cm$^{-3}$ are considered. 
For such densities, the $b$ scaling factor of 0.1--1 provide magnetic field strengths of the order of 0.001--1 mG, consistent with the recent measurements for the Ophiuchus molecular cloud \citep{ngan24}. The shock velocities above 10 km s$^{-1}$ are expected based on the resolved profiles of H$_2$O and high$-J$ CO lines with \textit{Herschel} \citep{kri12,kri17co}. Since the line profiles are not resolved with MIRI, the maximum shock velocities are $\sim$60 km s$^{-1}$, but more likely they are not exceeding $\sim$30 km s$^{-1}$ if they trace a similar gas component as H$_2$O. The ionization by cosmic rays and PAH abundances are not known for the studied region in Ophiuchus; however, neither of these parameters have a significant impact on our analysis (Appendix \ref{app:shocks}). Here, we adopt the values from the low-end of the grid, $\zeta$ = 10$^{-17}$ s$^{-1}$ and X(PAH) = 10$^{-7}$.

Figure \ref{fig:G0_comparison} shows the comparison of the shock model results with the S(2)/S(7) ratios measured in the outflows from Ophiuchus protostars (see Section \ref{sec:results}). Depending on the assumed magnetic field strength, the S(2)/S(7) ratio shows significant differences for a given value of pre-shock density but varying UV fields. Here, we consider magnetic fields with $b$ = 0.1, representative of predominantly dissociative $J$-type shocks where ionic and neutral fluids are well coupled \citep{neu89,fp10}, and $b$ = 1, typically adopted for $C$-type shocks, where ions and neutrals are decoupled \citep{kn96,fp10}.

The behavior of the S(2)/S(7) line ratio is notably different between the two types of shocks. For the lower density scenarios ($n_\text{H} = 10^{2}-10^3$ cm$^{-3}$) $J$-type shocks appear capable to reproduce the observed line ratios for a range of relatively low shock velocities ($v_\text{sh}$ = $10-20$ km s$^{-1}$), which becomes smaller for higher pre-shock densities and is already below 15 km s$^{-1}$ at $n_\text{H} = 10^3$ cm$^{-3}$. The models also require an external UV field strength in the range of 1-100 G$_0$ to match the observed line ratios. We note here that, the external UV radiation field is contributed either by the nearby stars or produced in accretion shocks and escaping through the outflow cavity; the self-irradiation is negligible in such low-velocity shocks \citep{Kristensen23}. 
For the same range of densities, $C$-type shocks only agree with the observed line ratios for a limited range of higher shock velocities ($v_\text{sh}$ = 20-25 km s$^{-1}$). An exception to this limited velocity range is the case with $n_\text{H} = 10^3$ cm$^{-3}$ and UV field of 100 G$_0$, which reproduces the observed ratios for $v_\text{sh}$ = 10-25 km s$^{-1}$. The restricted velocity range in which $C$ shocks are capable of reproducing the observed line ratios suggests that for pre-shock densities of $n_\text{H} = 10^{2}-10^3$ cm$^{-3}$ $J$ shocks are more likely responsible for the excitation of H$_2$.

At pre-shock densities above $10^{4}$ cm$^{-3}$, the model S(2)/S(7) ratios for the $J$-type shocks decrease and span less than 2 orders of magnitude, and are generally not consistent with the ratios observed in the Ophiuchus sources. 
The only exceptions are the models assuming $n_\text{H} = 10^4$ cm$^{-3}$ and a complete absence of UV irradiation, or models with UV fields of $10^3$ G$_0$, which are too high for most low-mass star-forming regions \citep{karska18,kri17co,vanDish11}.
In contrast, the model ratios for the $C$-type shocks span several orders of magnitude and reproduce the observations for an extended range of shock velocities ($v_\text{sh}$ = 10-25 km s$^{-1}$). At the pre-shock density of 10$^{4}$ cm$^{-3}$, the S(2)/S(7) ratio decreases roughly by an order of magnitude with increasing values of UV field strengths. At the same time, the ratio does not depend strongly on the assumed shock velocity, facilitating meaningful comparisons with observations. For shock velocities in a broad range of values, from 10 to 20 km s$^{-1}$, the best-fit value of $G_0$ is 10-100. Slightly higher velocities would be also consistent with lower UV fields, but already at 25 km s$^{-1}$ none of the models would reproduce the observations; thus, lower UV fields are rather unlikely.

At the pre-shock densities of 10$^{5}$-10$^{6}$ cm$^{-3}$, the model ratios for C-type shocks decrease strongly with the increasing UV field strength and shock velocity (right column, middle and bottom panels of Fig.~\ref{fig:G0_comparison}). At 15 km s$^{-1}$, the best fit models suggest $G_\mathrm{0}$ of 1-10 for the pre-shock density of 10$^{5}$ cm$^{-3}$, and $G_\mathrm{0}$ of 0 (a fully-shielded scenario) for 10$^{6}$ cm$^{-3}$. Noteworthy, $G_\mathrm{0}$ of 100 is also consistent with observations for faster shocks ($\varv_\mathrm{sh}$ of 25-30 km s$^{-1}$) at those higher pre-shock densities.

The sample of Class 0 sources from \cite{Francis25} displays lower values for the observed S(2)/S(7) line ratios, suggesting higher excitation conditions within their outflows. The lower line ratios mean that the low-density, low-velocity $J$-type shocks require even higher values of UV irradiation (10-100 G$_0$) in order to reproduce the observations. In addition, both higher pre-shock density (up to 10$^{4}$ cm$^{-3}$) $J-$ shocks with low velocities, as well as, high-velocity ($v_\text{sh} \ge$  20 km s$^{-1}$), high-density $n_\text{H} \ge 10^4$ cm$^{-3}$ solutions with UV fields of 100-1000 G$_0$ are compatible with Class 0 sources.

In Appendix \ref{app:shocks}, we show that the choice of other H$_2$ lines representing the warm and hot gas components does not significantly change the shock conditions required to reproduce the observations. We have also investigated various pairs of lines located either in the warm or in the hot component, but these ratios do not constrain well the shock models. A similar trend was noted with CO, where a broad range of shock conditions fitted the observations of the warm component alone \citep{Dio13,karska13}, whereas the inclusion of the external UV-irradiation of C-shocks allowed to explain the curvature seen in rotational diagrams \citep{karska18}. 

\begin{figure}[ht!]
\centering 
\includegraphics[width=\linewidth]{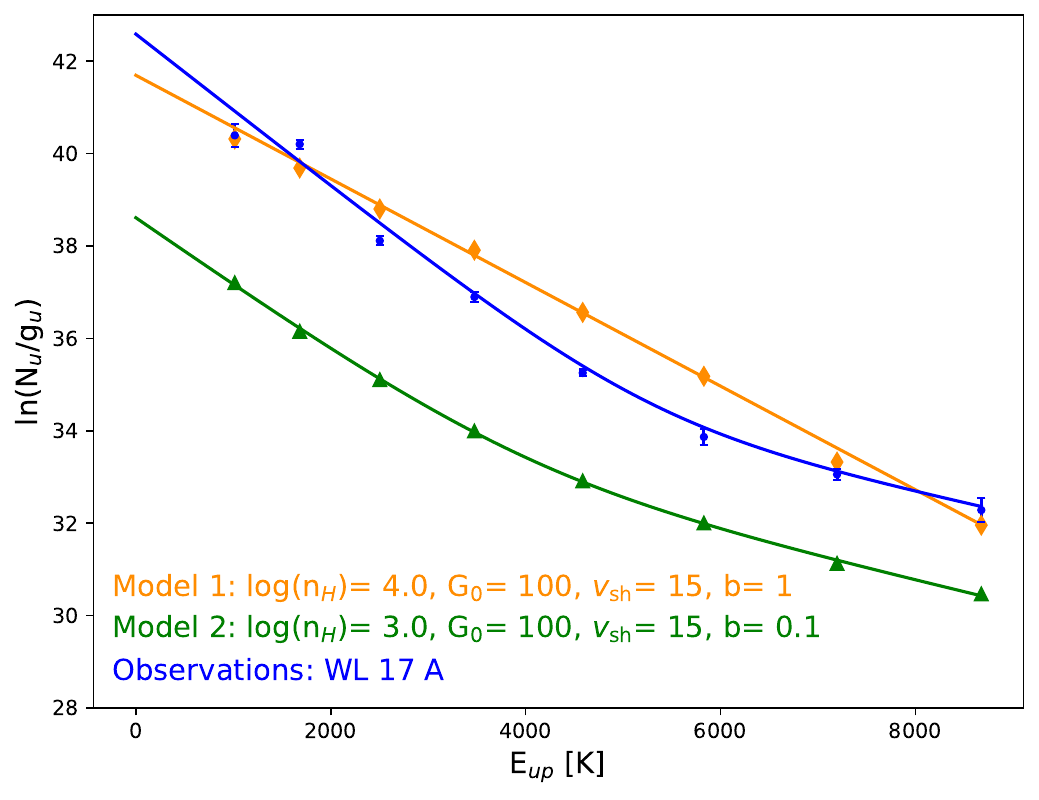} 
\caption{\label{fig:model rotational diagrams} Synthetic rotational diagrams based on the H$_2$ intensities predicted for different combinations of parameters in the \cite{Kristensen23} shock models. Green triangles show model results for n$_\text{H}$ = 10$^{3}$ cm$^{-3}$, G$_0$ = 100, $v_\text{sh}$ = 15 km s$^{-1}$, b = 0.1, $\zeta$ = 10$^{-17}$ s$^{-1}$, and X(PAH) = 10$^{-7}$. Orange diamonds show the model with n$_\text{H}$ = 10$^{4}$ cm$^{-3}$, G$_0$ = 100, $v_\text{sh}$ = 15 km s$^{-1}$, b = 1, $\zeta$ = 10$^{-17}$ s$^{-1}$, and X(PAH) = 10$^{-7}$. Blue points correspond to the observed values for aperture A of WL 17. Solid lines represent the best fit in each case.}
\end{figure}

The shock model results presented in \cite{Kristensen23} allow also for the reconstruction of full rotational diagrams, offering an alternate approach of comparing them to observations.
Figure \ref{fig:model rotational diagrams} shows a comparison between the rotational diagrams for two of the best models along with the extinction corrected rotational diagram from aperture A of WL 17 (see also Fig. \ref{fig:rotation_diagrams}), as an example among our observed rotational diagrams. The first model corresponds to a $C-$type shock with pre-shock density of $n_\text{H}$ = 10$^{4}$ cm$^{-3}$, that best reproduces the S(2)/S(7) ratio. The second model corresponds to a $J-$type shock with pre-shock density $n_\text{H}$ = 10$^{3}$ cm$^{-3}$, which in turn best reproduces the two component rotational temperatures fit of the observations.  
Despite the $C-$shock model matching well the observed S(2)/S(7) line ratio, in the range of velocities studied (10 km s$^{-1}$ < $v_\text{sh}$ < 30 km s$^{-1}$), it fails to reproduce the curvature seen in the rotational diagrams of our observations. Instead, it is best described by a single temperature component with $T_\text{rot} \sim$ 900 K, a value that lies between that of the two components found in our observations. The $J-$shock model on the other hand, is best described by a two components fit, with T$_\text{warm} \sim 680$ K and T$_\text{hot} \sim$ 2000 K, values similar to those measured in our sample, and also displays the shift between the two components at a similar $E_\text{up}$. Despite better predicting the shape of the rotational diagram, model 2 predicts column densities of $N_\text{hot} = 6.4 \times 10^{16}$ cm$^{-2}$ and $N_\text{warm} = 9.7 \times 10^{17}$ cm$^{-2}$, which are consistently one to two orders of magnitude lower than the values estimated for our Ophiuchus sources (see Table \ref{table:Ntot Tkin Ak results}). In contrast, model 1 predicts $N_\text{tot} = 2.8 \times 10^{19}$ cm$^{-2}$, in line with the observed values. We note here though that column densities predicted in shock models depend heavily on the assumed geometry of the shocks. Since models in \cite{Kristensen23} assume a simple planar shock morphology whereas real shocks have complex morphologies, the reported differences in column densities are not sufficient to dismiss $J$-shocks.
Overall, while comparing the full rotational diagram produced by different shock models to the observations has the benefit of considering all the information from the available transitions, it comes with some significant caveats in terms of the column densities and total fluxes which rely on the assumed shock sizes and area.

To summarize, H$_2$ observations of outflows from four Class I protostars in Ophiuchus and five Class 0 sources studied by \cite{Francis25} are consistent with the line ratios in both $C$- and $J$-type shocks. For the examined range of shock velocities, $J$-type shocks are also better reproducing the overall shape of rotational diagrams albeit they seem to under-predict the total column densities. Regardless, both kinds of models show that external irradiation by UV fields of the order of 10-100 in $G_\mathrm{0}$ units is required to best match the observations. As a next step, we determine UV radiation using other indirect methods to verify the results from shock models.

\begin{table*}[ht!]
\caption{External UV field strengths in units of G$_0$ \label{tab:UV fluxes}} 
\centering 
\begin{tabular}{l c c c c c | c c c}
\hline \hline 
Protostar & HD147899 & S1 (GSS35)& $\sigma$-Sco & $\alpha$-Sco & $\rho$-Oph & UV fields (stars) & UV fields (dust) & Difference\\
\hline    
GSS30-IRS 1 & 149 & 967 & 51 & 2 & 21 & 1190 & 3923 & 0.70\\
ISO-Oph 21  & 176 & 552 & 52 & 2 & 20 & 803 & 528 & 0.34\\
{[}GY92{]} 197 & 48 & 43 & 51 & 3 & 13 & 157 & 417 & 0.62\\
WL 17 & 44 & 35 & 51 & 3 & 12 & 146 & 314 & 0.53 \\
ISO-Oph 137 & 31 & 21 & 49 & 3 & 11 & 115 & 344 & 0.67\\ 
\hline
\end{tabular} 
\begin{flushleft}
\tablefoot{Difference estimated as 1$-(\text{UV}_\text{(stars)}$/$\text{UV}_\text{(dust)})$, except for ISO-Oph 21 where 1$-(\text{UV}_\text{(dust)}$/$\text{UV}_\text{(stars)})$ was used since the UV field estimate from stars is higher. }
\end{flushleft}
\end{table*}

\subsection{The impact of irradiation environment on the protostars in Ophiuchus}
\label{sec:UV fields}
To assess the strength of the external UV irradiation on the protostars in Ophiuchus, we consider two additional methods: (i) UV radiation from nearby stars, (ii) emission from dust assuming that all UV radiation was absorbed and re-emitted in the far-IR. 

We follow the procedure from \cite{Sch16,Sch23} to calculate the UV flux at the position of each of our sources and originating from young, massive stars in the region. We consider several nearby stars and/or binaries: $\sigma$-Sco, a B1 III and B1 V type binary \citep{Apellaniz21}; $\alpha$-Sco, a M1.5 I$_\mathrm{ab}$ and B2 V type binary \citep{Apellaniz21}; $\rho$-Oph, a B2 IV and B2 V type binary \citep{Helmut11}; HD 147889, a B2 V type star \citep{Chini81,Liseau99}, and S1 (GSS35), a B3 V type star \citep{wilking05}.
The total UV luminosity emitted by each of those stars is calculated from:
\begin{equation}
   L_\mathrm{UV} = 4\pi^2 r^2 \int^{\lambda_{2066}}_{\lambda_{910}}B(\lambda,T)d\lambda
\end{equation}
\noindent where $r$ is the stellar radius and $T$ is the effective temperature of the star. Values for the radius and effective temperature of each source are taken from \cite{Drilling00}, based on their spectral type. The integration between 910 and 2066 $\AA$ covers far-UV photons relevant for the star-forming regions, with energies between 6 and 13.6 eV. 

To obtain the UV flux at the position of protostars, the total UV luminosity from nearby stars  is scaled by a factor of $(4\pi R^2)^{-1}$, where $R$ is the distance between the source of UV radiation and the respective target in Ophiuchus. Due to the high uncertainties of line-of-sight distances, we assume all considered objects to be at the same distance from Earth, and thus use the projected distances for the calculation \citep{Sch16,Sch23}. The distances between the B-type stars and the protostars range between $\sim4.2$-4.3 pc for $\sigma$-Sco, $\sim$4.3-5.2 pc for $\alpha$-Sco, $\sim$2.3-3.1 pc for $\rho$-Oph, $\sim$0.6-1.3 pc for HD147899, and $\sim$0.1-0.9 pc for S1. The resulting UV fluxes for each of our sources are presented in Table \ref{tab:UV fluxes}. Noteworthy, this approach does not correct for dust attenuation, which is significant in dusty molecular clouds, and thus represents only an upper limit to the radiation reaching our protostars.
\begin{figure}[ht!]
\centering 
\includegraphics[width=\linewidth]{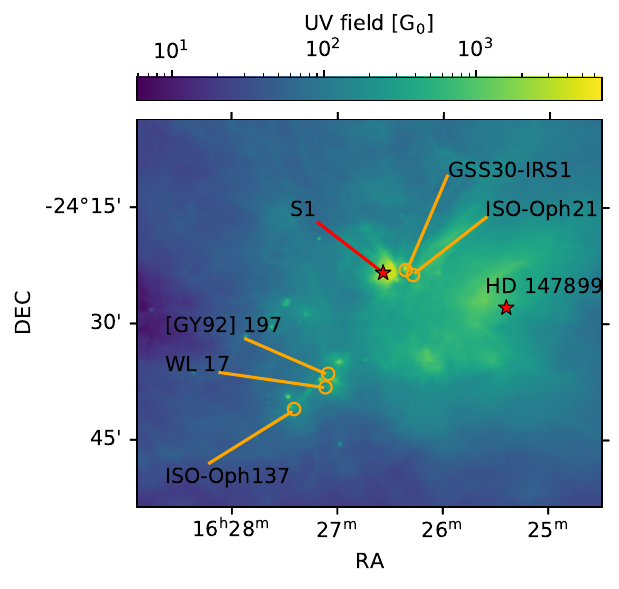} 
\caption{\label{fig:G0map} Distribution of the interstellar UV radiation field estimated from the dust continuum emission from \textit{Herschel} under the assumption that all UV radiation from nearby stars was absorbed and re-emitted in the far-IR. The G$_0$ values presented here are likely overestimated due to contribution from thermal dust emission to the FIR continuum. Orange circles mark the location of the protostars, while red stars mark the position of the massive stars S1 and HD147899. $\alpha$-Sco, $\sigma$-Sco, and $\rho$-Oph are located outside the FoV of this figure (similar to Fig. \ref{fig:region}).}
\end{figure}

Perhaps a more accurate way to estimate the UV radiation field in the region is by using the far-IR dust emission. Under the assumption that the UV emission of nearby massive stars is fully absorbed and re-emitted in the far-IR, the local UV radiation field can be estimated from the total FIR intensity between 60 and 200 $\mu$m \citep[][]{Kramer08,Rocca13,Sch16}.  Here, we use the \textit{Herschel}/PACS continuum intensities at 70 and 160 $\mu$m, collected as part of the \textit{Herschel} Gould Belt Survey \citep{lad20}, to obtain an estimate of the far-IR intensity ($I_\mathrm{FIR}$). Then, the total UV field, in units of G$_0$, is given by \citep{Sch16}:
\begin{equation}
    F_\mathrm{UV} = \frac{4\pi I_\mathrm{FIR} 1000}{1.6}.
\end{equation}

Figure \ref{fig:G0map} shows the resulting map of the estimated $F_\mathrm{UV}$ in the Ophiuchus where our targets are located. To obtain an estimate for each of our sources, we calculate the average $F_\mathrm{UV}$ in a $12\arcsec$ aperture around each protostar, corresponding to the resolution of the \textit{Herschel}/PACS continuum (see Table \ref{tab:UV fluxes}).

The two methods presented above yield somewhat different results, with the general trend of the dust-based method predicting values higher by approximately a factor of 3, with the exception of ISO-Oph 21. 
Since the estimate from the nearby stars is expected to represent an upper limit of the $F_\mathrm{UV}$, the even higher values determined from the dust could be explained by the dust-based method accounting also for locally produced UV radiation, or from a significant contribution of thermal dust emission from the cloud \citep[see also][]{Sch16}.

Both methods predict, however, that GSS30-IRS 1 stands apart from the other sources with a significantly higher $F_\mathrm{UV}$ flux of the order of $10^3$. ISO-Oph 21 has the second strongest $F_\mathrm{UV}$, with $F_\mathrm{UV}$ $\sim$ 500 G$_0$ based on the $I_\mathrm{FIR}$ and $\sim$ 800 G$_0$ based on the stellar emission. The remaining three sources show consistently similar UV fields of $\sim$300 - 400 G$_0$ based on $I_\mathrm{FIR}$, and $\sim$ 100 G$_0$ based on the stellar emission. 

These results confirm the presence of significant variance in the UV conditions among protostars in Ophiuchus, allowing us to examine its potential impact on the star formation process. They are consistent with the estimates from UV-irradiated shock models implying UV fields of 10-100 times the average interstellar radiation fields.

\section{Discussion}
\label{sec:discussion}

\subsection{Origin of H$_2$ emission in outflows from low-mass protostars}

Mapping of pure rotational H$_2$ lines with ISO and the InfraRed Spectrograph (IRS) onboard \textit{Spitzer} Space Telescope was a stepping stone to understand the molecular excitation in low-mass protostellar outflows \citep[e.g.,][]{vD04,neu06,Neufeld2009,tappe12}. The low-$J$ ($J<4$) H$_2$ emission in the prototypical outflow L1157 was associated with the wider cold molecular outflow traced by CO, while that of higher-$J$ ($J>7$) lines was connected to the precessing jet and more localized shocked regions \citep{nisini10spitzer}. On the contrary, the morphology of outflows seen in various H$_2$ transitions was rather similar \citep{gia11}. Such a pattern resembles also our observations of low-mass protostars in Ophiuchus, where only slight variations of extent and opening angles are seen over different transitions even at the much smaller scales probed with JWST (see Section \ref{sec:maps} and also Fig. \ref{fig:opening angle}).

Detection of several H$_2$ transitions allowed the determination of excitation temperatures of the emitting gas, accounting also for the non-equilibrium H$_2$ ortho-to-para ratios \citep{neu08,Neufeld2009}. For an assumed power-law distribution of gas temperatures, with the column densities of material at temperature T to $T+dT$ proportional to $T^{-\beta}dT$, the best-fit power-law indices were found in the range 3.8-4.2 toward BHR71, L1448, and NGC 2071 \citep{gia11}. The range of indices of 3.9-5.4 in L1157 indicated a large variation in the physical parameters along the outflows \citep{nisini10spitzer}, in line with MIRI maps covering  smaller spatial scales (Fig. \ref{fig:CRBR2422_results}).

\cite{nisini10spitzer} and \cite{gia11} also provide temperature estimates for selected locations along the outflows based on linear fits to the S(0)-S(2) and S(5)-S(7) lines, corresponding to a warm and hot gas component, respectively. The resulting values of $T_\text{warm} \sim$240 - 400 K and $T_\text{hot} \sim$1000 - 1500 K are lower than the ones estimated in this work: 540$\pm$70 K and 2000$\pm$700 K for warm and hot components, respectively (Section \ref{sec:h2}). However, the difference in temperature could be due to the inclusion of the S(0) line and the lack of detections of higher$-J$ lines due to limited sensitivity of \textit{Spitzer}.

Important insights on the the origin of hot molecular gas were enabled by \textit{Herschel}, and the detection of high$-J$ CO lines up to $J=48$ and at a resolution of 9.4\arcsec, even toward outflows from low-mass protostars \citep{herczeg12}. Similar to H$_2$, CO rotational diagrams revealed two dominant gas physical components with surprisingly similar temperatures of the warm component of $\sim$ 300 K toward all low-mass protostars \citep{karska13,green2013,manoj2013}. This gas temperature is consistent with the \textit{Spitzer} results, but about 200 K lower than the respective warm component seen in H$_2$ with MIRI (Section \ref{sec:h2}). 

\begin{figure}[ht!]
\centering 
\includegraphics[width=\linewidth]{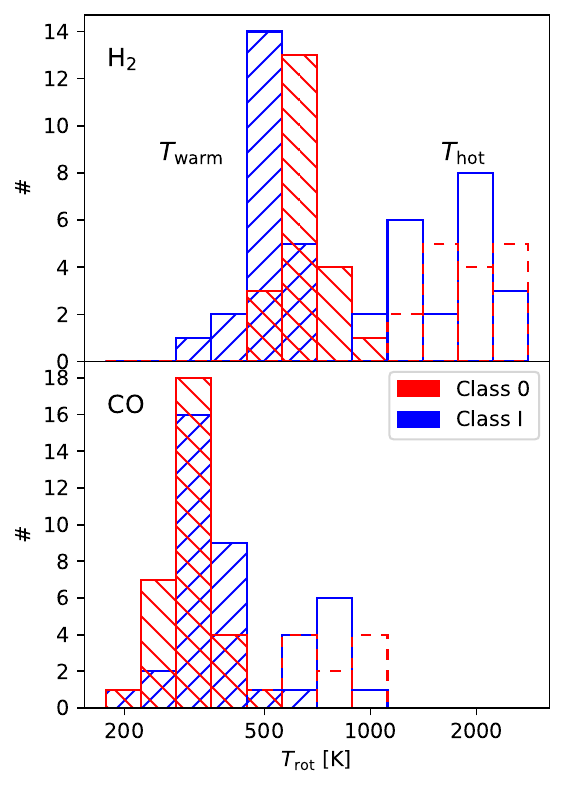} 
\caption{\label{fig:Trot_comparison} Distribution of the rotational temperatures of H$_2$ (top) and high-$J$ CO (bottom) in low-mass protostars. Temperatures in Class 0 protostars are shown in red, and in Class I sources in blue. The warm gas components is shown in stripped bins, and the hot one in clear bins. 
H$_2$ results for Class 0 protostars are adopted from \cite{Francis25}, and CO results for both Class 0 and I are from \cite{karska18}. }
\end{figure}

To investigate possible changes of gas temperature as the protostar evolves, Figure \ref{fig:Trot_comparison} shows a comparison of rotational temperatures of H$_2$ and CO toward Class 0 and I objects. The H$_2$ results for Class 0 objects are adopted from a recent study of on-source positions of five objects by \cite{Francis25}, whereas Class I temperatures refer to outflows from Ophiuchus protostars (Section \ref{sec:analysis}), and a single source in Taurus -- TMC1 \citep{tych24}. The distribution of rotational temperatures based on CO and for Class 0 and I objects are adopted from a consistent analysis of $\sim$90 protostars with \textit{Herschel}/PACS \citep{karska18}. 

 We find that the distribution of $T_\mathrm{warm}$ and $T_\mathrm{hot}$ using H$_2$ is tentatively shifted toward higher values for less evolved protostars (top panel of Fig. \ref{fig:Trot_comparison}). However, the range of the hot component temperature is relatively broad for both Class 0 and I objects, reaching values of up to $3000$ K. The difference is reflected by the average H$_2$ rotational temperatures for the Class 0 objects of $660\pm100$ K (warm) and $2200\pm600$ K (hot) versus for the Class I objects of $520\pm70$ K (warm) and $1800\pm550$ K (hot).

Similar trend of decreasing rotational temperature with the evolutionary stage was not present using CO observations (bottom panel of Fig. \ref{fig:Trot_comparison}). Noteworthy, the maximum temperatures obtained with CO did not exceed $\sim1100$ K, and the distribution of the hot component temperatures with \textit{Herschel} of $760\pm170$ K (\citealt{karska18}) suggests that high$-J$ CO lines might trace a less energetic part of the flow than H$_2$.

Combined analysis of H$_2$O and CO from \textit{Herschel} provided a solid evidence that both gas temperature components discussed above originate in non-dissociative shocks, with the hot component additionally tracing regions irradiated by UV radiation \citep{karska14a,karska18,kri17co}. The impact of far-UV photons was further confirmed by high abundances of OH and other hydrides in the immediate surrounding of both low- and high-mas protostars \citep{wamp13,benz16}. The H$_2$ line ratios measured toward Ophiuchus protostars, favor the origin of H$_2$ emission in the UV-irradiated shocks (Section \ref{sec:shocks}). The ratio of the S(2)/S(7) lines for Class 0 objects is at the lower end of values for Class I objects (Fig. \ref{fig:ratio}), which would imply higher UV-fields in those sources assuming the same gas densities (Fig. \ref{fig:G0_comparison}). Consequently, gas temperatures would be higher for Class 0 objects, consistent with the results of our rotational analysis (Fig. \ref{fig:Trot_comparison}). However, the densities of the envelopes of Class 0 objects are typically higher than for Class I objects \citep[e.g., ][]{kri12}, which should reduce the impact of external UV radiation, in particular for the outflow positions closer to the central source. The small variation of excitation conditions among the Oph sources despite the significant difference of the UV field strength from the surrounding environment, as estimated in Sec. \ref{sec:UV fields}, suggests that the the environmental UV radiation has little to no effect on the excitation within protostellar outflows. We note though that the positions examined in this work are located very close to the protostars, and therefore likely to be well shielded from the external UV radiation.
In addition, some UV radiation can be also produced in-situ by the shock itself, but velocities above 30 km s$^{-1}$, higher than measured using velocity-resolved H$_2$O lines for low-mass protostars, would be required \citep{lehmann20,lehmann22,Kristensen23}. 
Hence, accretion shocks and high velocity shocks along the protostellar jets are more likely sources for the UV radiation required to reproduce the excitation conditions observed in the Oph sources. Such an origin for the UV radiation also aligns with the slightly higher excitation conditions seen in Class 0 sources, since they are in general believed to have higher accretion rates and drive more powerful jets. Prompt OH emission, shown to probe well the UV field within protostellar outflows \citep[e.g.,][]{neu24}, could serve as an independent approach to confirm such evolutionary trends. Regardless, observations of larger number of sources with additional line tracers are needed to confirm the trend of higher gas temperatures in Class 0 objects and its interpretation.

\begin{figure}[ht]
\centering 
\includegraphics[width=\linewidth]{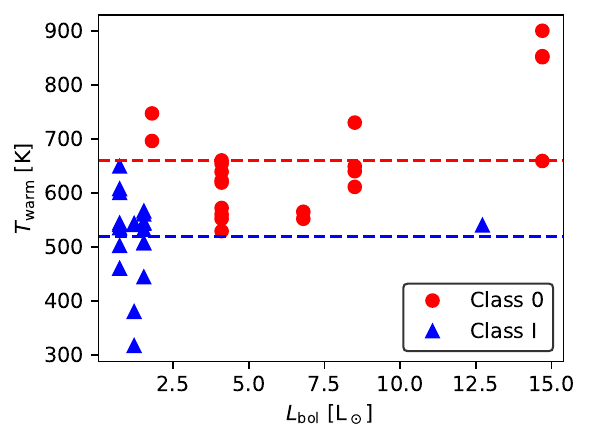} 
\caption{\label{fig:Twarm vs Lbol} Temperature of the warm gas component along the protostellar outflows over the bolometric luminosity of the corresponding driving source. Blue triangles show the Class I sources from this work and \cite{tych24}, while red circles show the Class 0 sample from \cite{Francis25}. Blue and red dashed lines show the mean values for the Class I and 0 sources, respectively.}
\end{figure}

In addition to the evolutionary stage, source properties such as the bolometric luminosity or envelope mass have also been closely associated with outflow properties \citep[e.g.][]{Cabrit92,yi15,mot17,Skretas22}. To explore whether the observed shifts in temperatures and excitation conditions are associated with source properties, instead of the evolutionary stage of the sources, we look into the comparison of the warm component temperature over the bolometric luminosity of the source, for the same sample that was used in Figs. \ref{fig:ratio} and \ref{fig:Trot_comparison}. Even though the sample is small, Fig. \ref{fig:Twarm vs Lbol} shows no clear correlation between the source properties and the resulting H$_2$ temperatures. In contrast, the slight shift between the Class I and Class 0 sources, initially noted in Fig. \ref{fig:Trot_comparison}, is apparent. The lack of impact of the source properties on the excitation conditions within the outflows are further supported by the remarkable similarity of temperatures between low- and high-mass sources reported in \cite{Francis25}.

To summarize, H$_2$ emission from protostellar outflows in Ophiuchus likely originates in shocks irradiated by UV fields produced in shocks along the protostellar jets or by accretion onto the protostar. 
By comparisons with other protostars observed with JWST \citep{Francis25}, we find a trend of decreasing H$_2$ rotational temperatures from Class 0 to Class I objects. In addition, we find no clear connection between the driving source properties and the excitation conditions within their outflows. The upcoming analysis of larger samples of protostars with JWST will certainly allow us to draw more solid conclusions on the origin of H$_2$ emission.

\subsection{Gas energetics in outflow shocks}
 
Theoretical studies of line cooling from dense cores predict that most of the released energy is emitted in mid-infrared emission from H$_2$, followed by far-infrared lines fines structure lines of [\ion{O}{i}], high$-J$ CO, and H$_2$O rotational transitions 
\citep{GL78,Ta83,NK93,CC96,DN97}. Similarly, H$_2$ emission is the most significant cooling pathway for shock excited gas in protostellar outflows \citep{kn96,fp10,Kristensen23}. Thus, JWST observations of H$_2$ lines allow us to study the energetics of protostars, at spatial scales significantly smaller than previously achieved with \textit{Spitzer} \citep[e.g., ][]{nisini10spitzer,gia11}.

For a non-dissociative shock propagating in a medium with a pre-shock density of $n_\text{H} \sim 10^4$ cm$^{-3}$ and $v_\text{sh} \sim 10-50$ km s$^{-1}$, H$_2$ luminosity corresponds to $\sim$ 20-70\% of the total outflow cooling rate \cite{kn96}. 
This allows for the calculation of the total kinetic luminosity ($L_\text{kin}$) of the outflow from the observed rotational lines of $H_2$ \citep{Maret09}:  

\begin{equation}
\label{equ4}
L_\mathrm{kin}=\frac{1}{2} \dot M v_\text{sh} ^2 = \frac{(1 - f_m)}{f_c} L_\mathrm{H_2},
\end{equation}
where $(1-f_m)$ is the fraction of shock mechanical energy that is converted into internal excitation and $f_c$ is the fraction of cooling attributed to the H$_2$ emission. In turn, the outflow mass rate can be calculated as 
\begin{equation}
\label{equ5}
\dot M  = 1.24\times10^{-2} \frac{(1 - f_m)}{f_c} \left(\frac{L_{\mathrm{H}_2}}{\mathrm{L}_{\odot}}\right) \left(\frac{v}{\mathrm{km}/\mathrm{s}}\right)^{-2}  \mathrm{M}_{\odot} \mathrm{yr}^{-1}
\end{equation}
Adopting $f_c = 0.25$ - 0.5, $(1-f_m)$=0.75, and $v_\text{sh} \sim$ 15-35 km s$^{-1}$, \cite{Maret09} report mass loss rates of the order of 10$^{-6}$ M$_\odot$ yr$^{-1}$ for five Class 0 sources in NGC 1333 in Perseus. 

We calculate the mass loss rate for the four outflow sources in Ophiuchus, by summing up the H$_2$ line intensity measured in all the apertures (see Table \ref{table:line_luminosities}). In addition, we adopt $f_c =$ 0.5 and $(1-f_m)$ = 0.75, similar to \cite{Maret09} and assume $v_\text{sh}$ = 15 km s$^{-1}$ which best matches our observations (see Section  \ref{sec:shocks}). We find values of $\dot M \sim$ 4-7 $\times 10^{-9}$ M$_\odot$ yr$^{-1}$ except for GSS30-IRS 1, where $\dot M$=2.8 $\times 10^{-8}$ M$_\odot$ yr$^{-1}$. The significantly higher value measured for GSS30-IRS 1 reflects the much stronger H$_2$ emission compared to the rest of the sources, and could be partly due to inclusion of the extended and bright protostellar envelope in some of the apertures. 

\begin{table}[ht!]
 \caption{Mass outflow rates and kinetic luminosities} 
 \centering 
\begin{tabular}{l c c }
 \hline \hline 
 & $\dot M$ ($\times$ 10$^{-9}$ M$_\odot$ yr$^{-1}$)& $L_\text{kin}$ ($\times 10^{-5}$ L$_\odot$)\\
 \hline
 GSS30-IRS 1 & 27.8 & 50.4 \\
 \text{[GY92]} 197 & 5.0 & 9.1 \\ 
 WL 17 & 4.2 & 7.5 \\ 
 ISO-Oph 137 & 6.7 & 12.2 \\
 \hline    
 \hline
 \end{tabular} 
 \end{table}
 
Noteworthy, the values estimated here represent lower limits for the true $\dot M$ values for our outflows due to the limited field of view of MIRI. The mass loss rates of $\sim$ 7.8 $\times 10^{-7}$ M$_\odot$ yr$^{-1}$ were obtained using low$-J$ CO lines for the entire outflow from [GY92] 197 \citep{vdM13}, so almost 2 orders of magnitude higher than the JWST estimates. However, those CO observations trace the entrained outflow gas, not the shocked gas inside the cavity seen in H$_2$. Nevertheless, \cite{nisini10spitzer} found a good agreement between the outflow parameters estimated from H$_2$ and CO observations in L1157 \citep{Bachiller01}, where the full extent of the outflow was considered. This study supported the scenario that the CO outflow is accelerated by the H$_2$ flow; however, the angular resolution of existing CO observations for protostars in Ophiuchus is too low for similar comparisons. 

\begin{table}[ht!]
 \caption{\label{tab:mass loss rates}Mass outflow rates using Eq. \ref{eq:massloss}} 
 \centering 
\begin{tabular}{l c c }
 \hline \hline 
 &\begin{tabular}[c]{@{}c@{}}$\dot M_\text{warm}$\\  $\left( \times 10^{-8} \text{M}_\odot \text{yr}^{-1} \right)$ \end{tabular}& \begin{tabular}[c]{@{}c@{}}$\dot M_\text{hot}$\\  $\left( \times 10^{-10} \text{M}_\odot \text{yr}^{-1} \right)$ \end{tabular}\\
 \hline
 GSS30-IRS 1 & 11.4 & 14.0 \\
 \text{[GY92]} 197 & 11.4 & 29.4 \\ 
 WL 17 & 1.8 & 10.0 \\ 
 ISO-Oph 137 & 0.6 & 0.7 \\
 \hline    
 \hline
 \end{tabular} 
 \tablefoot{The values for GSS30-IRS 1 are derived only for the single outflow lobe covered by our observations. For all other sources the values represent the sum over the two outflow lobes.}
 \end{table}

A alternative calculation method, proposed by \cite{delabrosse24} offers a more direct approach to estimate mass loss rates using the H$_2$ lines within the MIRI range. For this method, the mass loss rate is given by 
\begin{equation}
    \dot{M} = 2m_\text{H}N_{\text{H}_2}l_\text{perp}\frac{v_\text{rad}}{cos(i)}
    \label{eq:massloss}
\end{equation}
where $m_\text{H}$ is the hydrogen atom mass, $N_{\text{H}_2}$ the column density of H$_2$, $l_\text{perp}$ the width of the molecular outflow, $v_\text{rad}$ the radial velocity of the flow and $i$ the inclination angle of the outflow. Table \ref{tab:mass loss rates} shows the resulting mass loss rates for all sources, calculated separately for the warm and hot components of the outflows. The mass loss rates are calculated independently for each lobe and then summed to give the total rate for each source. For GSS30-IRS 1, where only one lobe is covered by our FoV, the given value corresponds only to the observed outflow lobe. 
We use the column densities derived from the rotational diagram analysis (Table \ref{table:Ntot Tkin Ak results}). For the wide angle outflows of [GY92] 197 and WL 17, we use the average column density of the three apertures spanning the width of the outflow lobes (In both cases, apertures A, C, and D and E, G, and H respectively for each lobe). For GSS30-IRS 1, we use the values derived from aperture C due to the higher uncertainties in the cold component properties in apertures A and B, caused by the non-detection of the S(1) and S(2) lines due to the bright continuum (see Sec. \ref{sec:spectra}). For ISO-Oph 137, we use the values from the apertures further away from the source (B and D) as the column densities derived for apertures A and C are likely overestimated. The radial velocities for each source are estimated from the Gaussian fitting (Table \ref{table:velocities}) and using the same apertures (or averages) as for the $N_{\text{H}_2}$. The width of the outflows ($l_\text{perp}$) was measured from the directly from the integrated intensity maps, using the S(3) and S(7) transitions for the warm and hot component respectively. Finally, for the inclination angle, we adopt the literature values of 60$\degree$ for GSS30-IRS 1 \citep{dlV19}, and 34$\degree$ for WL 17 \citep{shoshi24}, and for the edge-on disks of [GY92] 197 \citep{Michel23} and ISO-Oph 137 \citep{vK09} we assume $i=80\degree$.

Overall, we find a broad range of mass loss rate among our sources, with values ranging from $\sim 10^{-7}$ - $10^{-9}$ M$_\odot$ yr$^{-1}$ for the warm and $\sim 10^{-9}$ - $10^{-11}$ M$_\odot$ yr$^{-1}$ for the hot component, respectively. Despite that fact that we only measure a single outflow lobe, the highest values are associated with the outflow of GSS30-IRS 1, the brightest source in our sample (see Table \ref{table:cat}). Surprisingly, the outflow of [GY92] 197 also shows directly comparable mass loss rates, despite having an order of magnitude lower bolometric luminosity. In addition to that, the lowest mass loss rates we find are associated with ISO-Oph 137, a source with comparable bolometric luminosity to that of [GY92] 197. This result may suggest that the morphology of the outflow, collimated or wide-angled, has an impact on the estimated outflow properties, with wide-angled outflows displaying higher outflow rates compared to collimated flows for sources with similar bolometric luminosities. Our sample though is extremely limited to draw any safe conclusions, and additional observations would be required to confirm or refute this suggestion.

In addition, we find that the warm component carries most of the mass in the outflow, with mass loss rates approximately two orders of magnitude higher that those of the hot component. This difference directly reflects  the two orders of magnitude difference seen in the column densities of the two components. Comparing our results with previous works employing the same method we find that our results lie between the values estimated for the hot gas component of the Class I source DG Tau \citep{delabrosse24} and those estimated for the Class 0 source HH211 \citep{caratti24}. Namely, the mass loss rates for the hot component of ISO-Oph 137 ($\sim 10^{-11}$ M$_\odot$ yr$^{-1}$) are directly comparable the results of \cite{delabrosse24}, while the mass loss rates for the warn component of GSS30-IRS 1 and [GY92] 197 ($\sim 10^{-7}$ M$_\odot$ yr$^{-1}$) appear comparable to those of HH211. Finally, the mass loss rate for [GY92] 197, estimated using the "moving slab" method is also directly comparable to the low$-J$ CO estimate of $\sim 7.8 \times 10^{-7}$ M$_\odot$ yr$^{-1}$ \citep{vdM13}. This results suggest that indeed, this method yields more accurate results, as it overcomes many of the limitations of the luminosity method discussed above.

To summarize, our observations highlight the potential of JWST in constraining the energetics of outflows via the H$_2$ emission. 

\section{Conclusions}
\label{sec:summary}

In this work, we present JWST MIRI/MRS observations of 5 Class I protostars in the Ophiuchus molecular cloud. We analyze the H$_2$ excitation associated with the protostellar outflows, and study the origin of the emission in shocks irradiated by external UV photons. The conclusions are as follows:
\begin{itemize}
    \item H$_2$ outflows are detected in four out of five sources. These outflows display two distinct morphologies, two of them appear narrow and well-collimated (GSS30-IRS 1 and ISO-Oph 137) and the other two show wide opening angles with well defined cavities (WL 17 and [GY92] 197).
    \item  Several atomic/ionic emission lines are detected, with most common those of [\ion{Ne}{II}], [\ion{Fe}{II}], and [\ion{S}{I}]. Their emission is associated with narrow, well-collimated jets, even in cases where the H$_2$ emission traces outflow cavities.
    \item The line emission along the outflows shows no significant velocity shifts between the outflow lobes, constraining the line-of-sight velocity of the H$_2$ outflows to $v<90$ km s$^{-1}$. The only case where a noticeable velocity shift is detected is the \ion{[Ne]}{II} line at 12.8 $\mu$m in WL 17.
    \item Rotational diagram analysis reveals that the H$_2$ emission within the protostellar outflows is well described by a two-temperature-components model. The two components are found to have rotational temperatures in the range of $T_\text{warm} \sim$500 - 600 K and $T_\text{hot} \sim$1000 - 3000 K.
    \item Based on shock model results from \cite{Kristensen23}, the external UV irradiation is found to have a significant impact on the S(2)/S(7) H$_2$ line ratios. In contrast, cosmic rays ionization rate and PAH abundance appear to have minimal to no impact. The magnetic field strength seems to shift the observed behavior of the ratio to higher velocities, by allowing higher velocity $C$-type shocks.
    \item The observed S(2)/S(7) ratios agree with predictions from $C$-type shocks at pre-shock densities n$_\text{H}\ge 10^{4}$ cm$^{-3}$, as well as with low-density (n$_\text{H}\le 10^{3}$ cm$^{-3}$), low-velocity ($v_\text{sh} \le 15$ km s$^{-1}$) $J$-type shocks. 
    The $J$-type shocks reproduce better the observed positive curvature of the rotational diagrams but at the same time underestimate the column densities.
    \item The observations require the presence of substantial ($\sim$10 - 100 G$_0$) UV irradiation in shock models. The strength of the UV fields is higher than that derived for the nearby diffuse cloud \citep{vD86}, suggesting that some UV photons originate from the accretion shocks in protostellar systems.
\end{itemize}

Overall, our results present a detailed view of four different protostellar outflows at close distances from their driving source highlighting their morphological differences and physical properties. The lack of substantial variation in outflow properties, despite the predicted variation in external UV field and model results suggests there is still more work required both on the aspect of observationally constraining the UV field strength but also on accurately modeling its impact.

\begin{acknowledgements}
We are grateful to Ewine F. van Dishoeck for useful insights and suggestions enhancing the overall quality of the manuscript. 
This work is based on observations made with the NASA/ESA/CSA James Webb Space Telescope. The data were obtained from the Mikulski Archive for Space Telescopes at the Space Telescope Science Institute, which is operated by the Association of Universities for Research in Astronomy, Inc., under NASA contract NAS 5-03127 for JWST. These observations are associated with program ID 1959. This publication is also based upon work from COST Action PLANETS CA22133, supported by COST (European Cooperation in Science and Technology). 
AK acknowledges support from the Polish National Science Center SONATA BIS grant No. 2024/54/E/ST9/00314.
Astrochemistry in Leiden is supported by funding from the European Research Council (ERC) under the European Union’s Horizon 2020 research and innovation programme (grant agreement No. 101019751 MOLDISK), by the Netherlands Research School for Astronomy (NOVA), and by grant TOP-1 614.001.751 from the Dutch Research Council (NWO). MF acknowledges support from the Polish National Agency for Academic Exchange grant No. BPN/BEK/2023/1/00036/DEC/01 and from the Polish National Science Centre SONATA grant No. 2022/47/D/ST9/00419. The material is based upon work supported by NASA under award number 80GSFC24M0006 (M.S.).
\end{acknowledgements}

\bibliographystyle{aa} 
\bibliography{main}

\appendix
\section{6.9 $\mu$m continuum maps}
\begin{figure*}
\centering 
\includegraphics[width=0.48\linewidth]{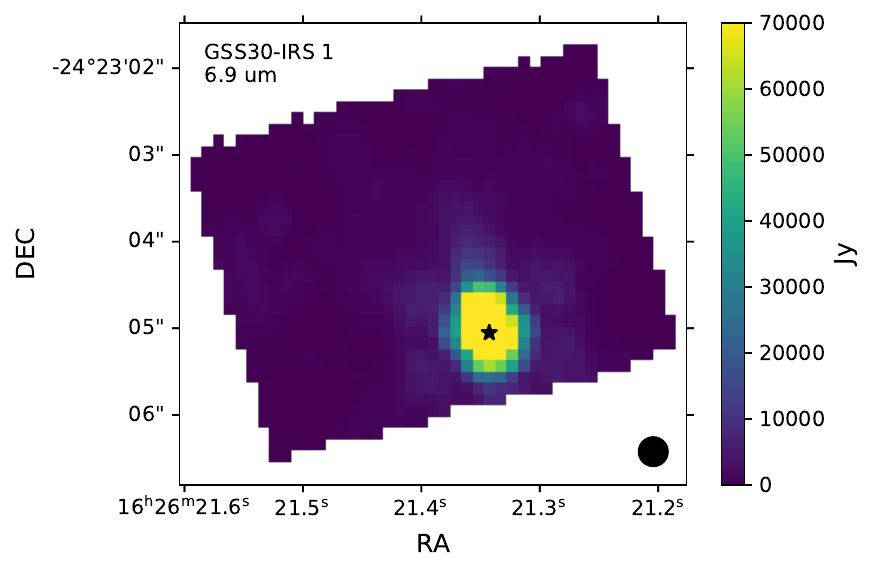} 
\includegraphics[width=0.48\linewidth]{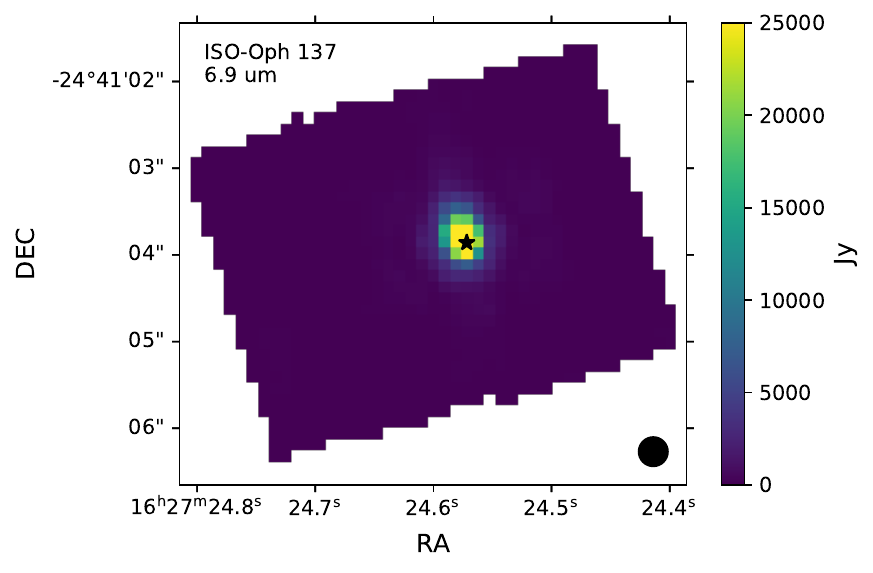} 
\includegraphics[width=0.48\linewidth]{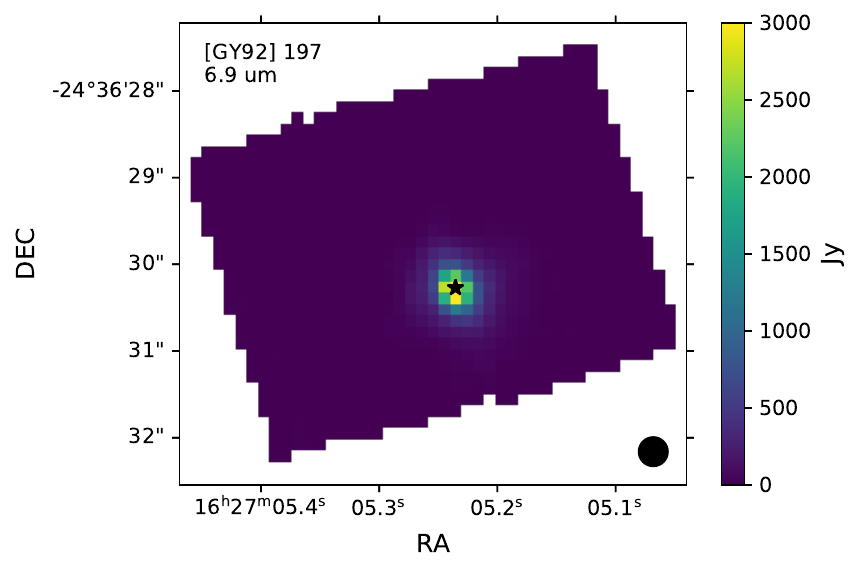} 
\includegraphics[width=0.48\linewidth]{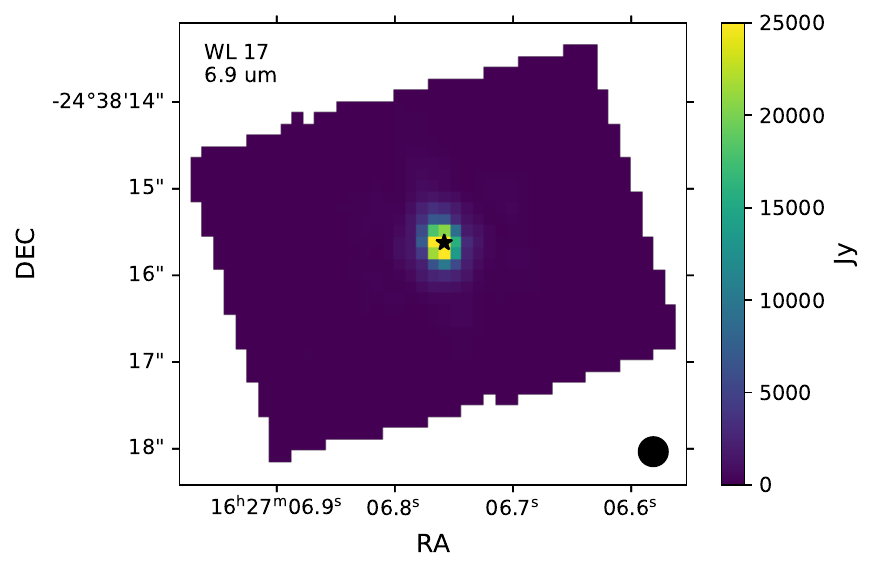} 
\includegraphics[width=0.48\linewidth]{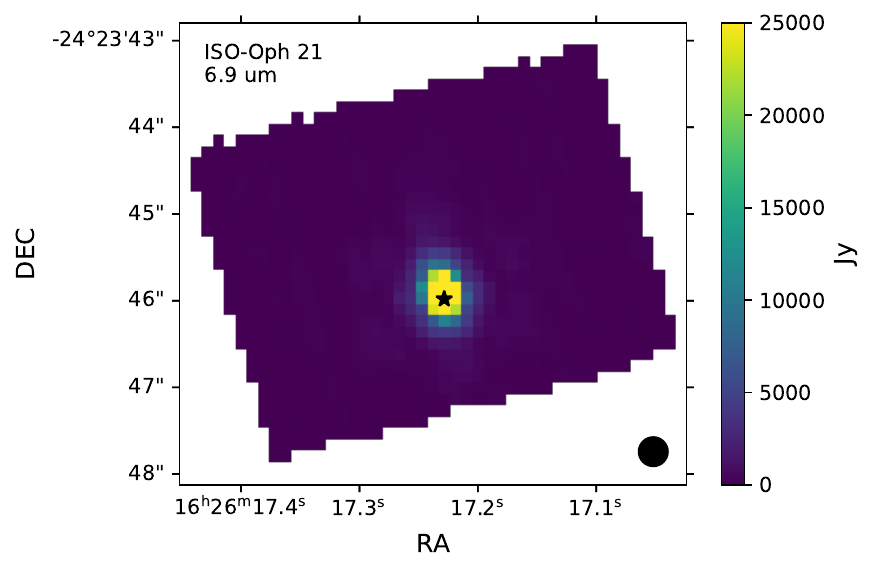} 
 
\caption{Continuum emission measured from line free channels at 6.9 microns. Black stars mark the peak of continuum emission.}
\label{fig:cont_maps}
\end{figure*}

Figure \ref{fig:cont_maps} shows the MIRI continuum emission, estimated at 6.9 $\mu$m avoiding any line emission. The peak of the continuum is in each case adopted as the location of the protostar. 

\section{Additional H$_2$ maps}
\begin{figure*}
\centering 
\includegraphics[width=0.45\linewidth]{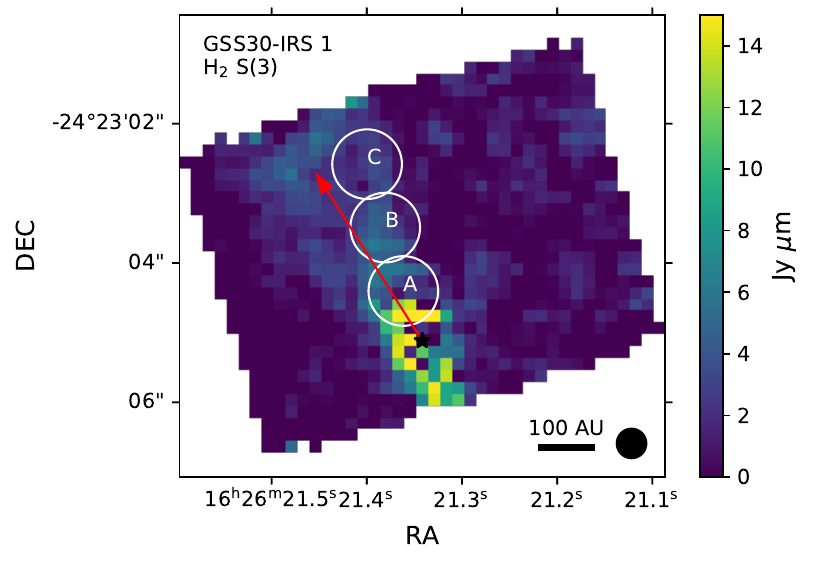} 
\includegraphics[width=0.45\linewidth]{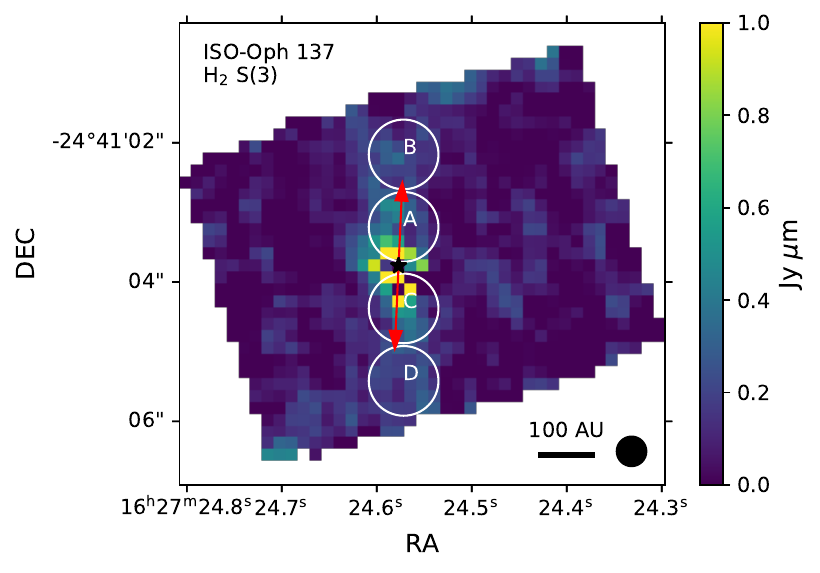} 
\includegraphics[width=0.45\linewidth]{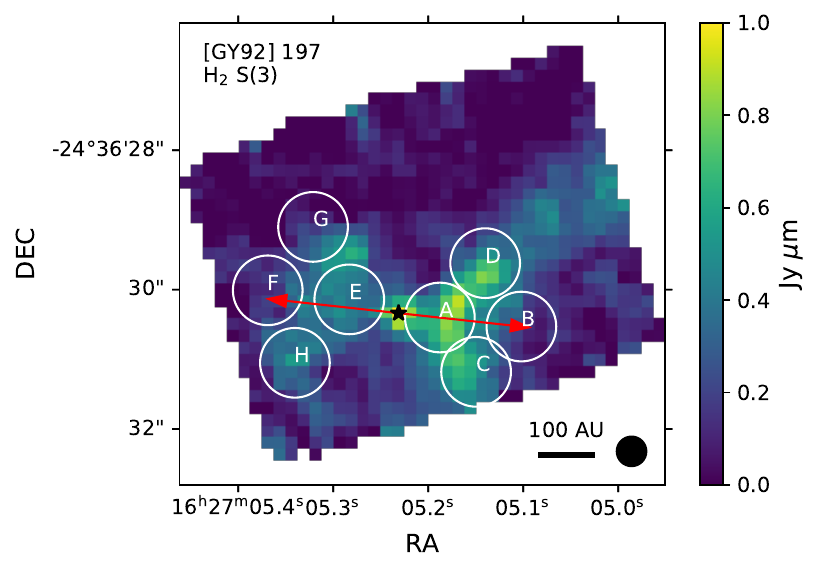} 
\includegraphics[width=0.45\linewidth]{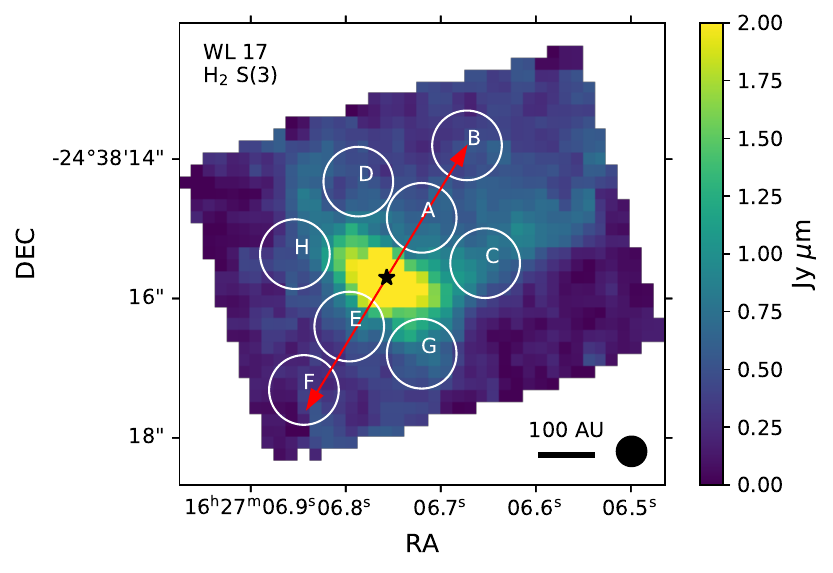} 
\includegraphics[width=0.45\linewidth]{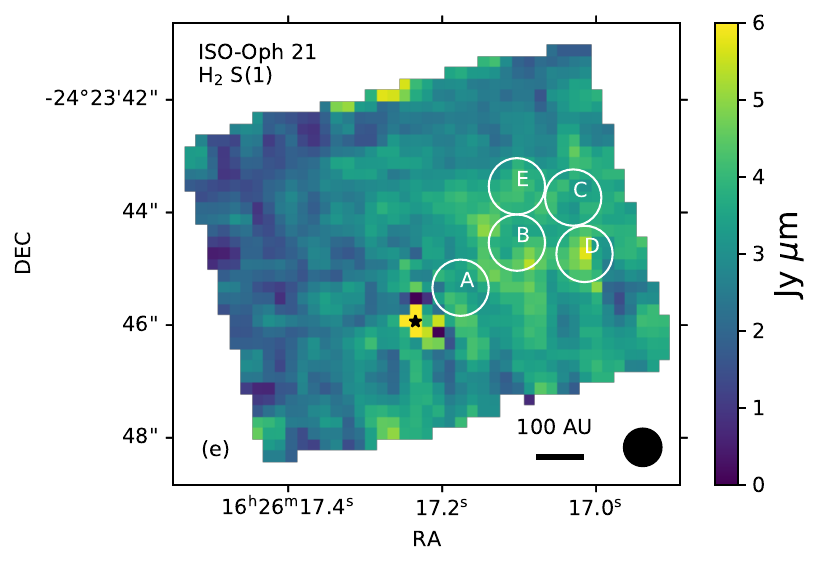} 
\caption{Same as Fig. \ref{fig:h2_maps} but for the S(3) transition. White circles show the apertures used for the spectra extraction (see Section \ref{sec:spectra}). For ISO-Oph 21, where the S(3) line is not detected, the S(1) transition is shown instead.}
\label{fig:S3}
\end{figure*}

\begin{figure*}
\centering 
\includegraphics[width=0.45\linewidth]{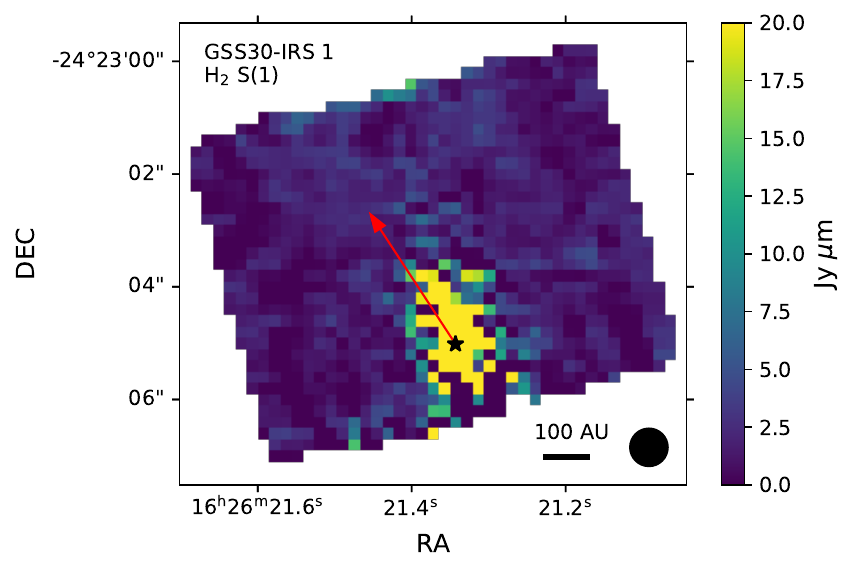} 
\includegraphics[width=0.45\linewidth]{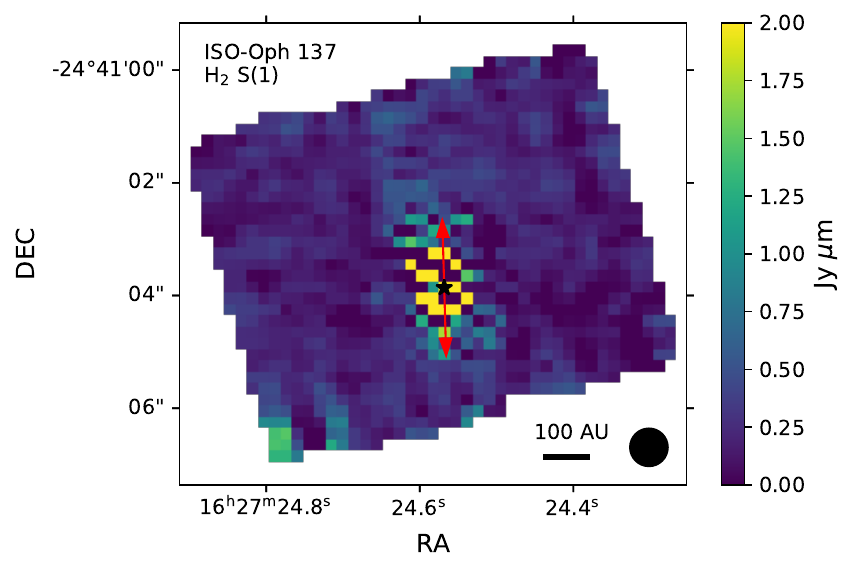} 
\includegraphics[width=0.45\linewidth]{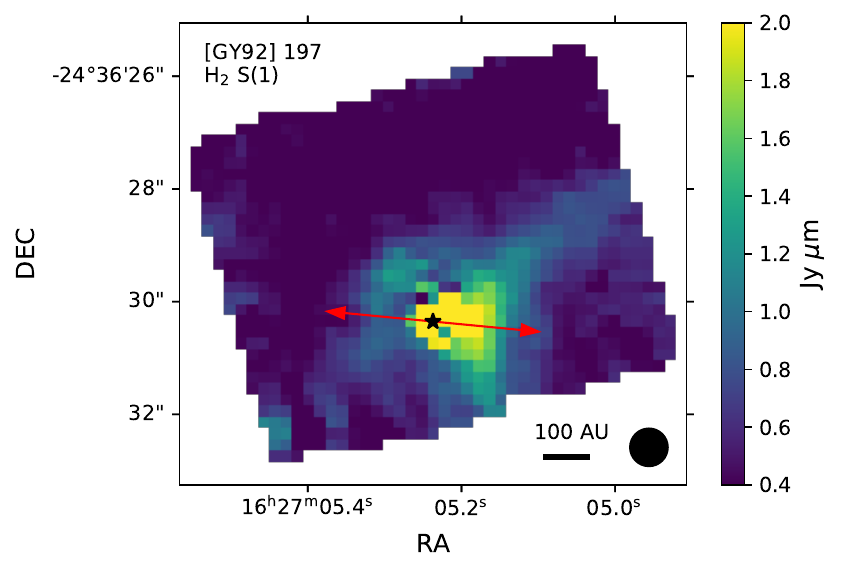} 
\includegraphics[width=0.45\linewidth]{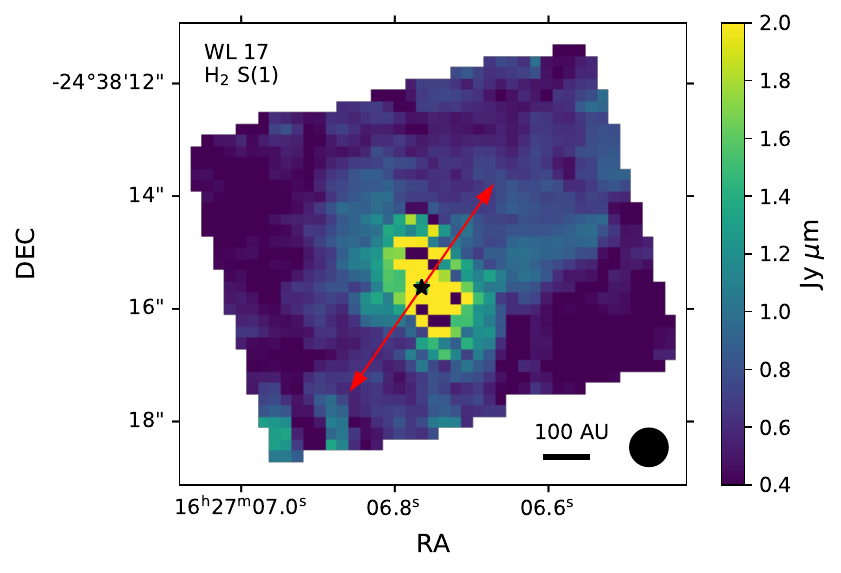} 
\includegraphics[width=0.45\linewidth]{Figures/appendix_figures/ISO-Oph_21_H2_S1_map_with_apertures.pdf} 
\caption{Same as Fig. \ref{fig:h2_maps} but for the S(1) transition. White circles show the apertures used for the spectra extraction (see Section \ref{sec:spectra}).}
\label{fig:S1}
\end{figure*}

\begin{figure*}
\centering 
\includegraphics[width=0.45\linewidth]{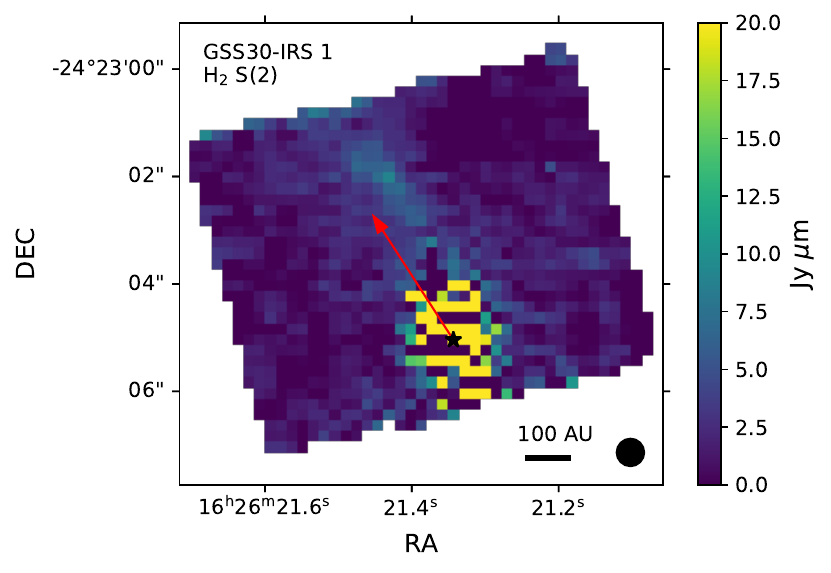} 
\includegraphics[width=0.45\linewidth]{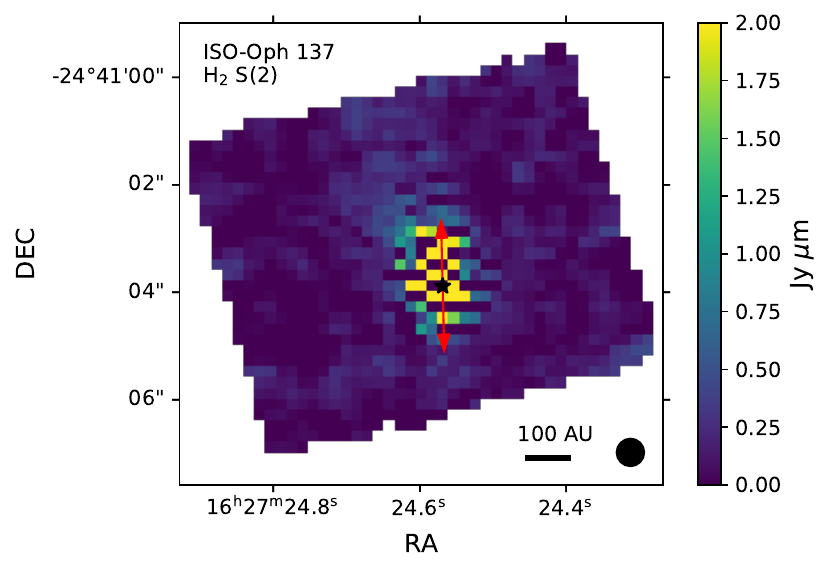} 
\includegraphics[width=0.45\linewidth]{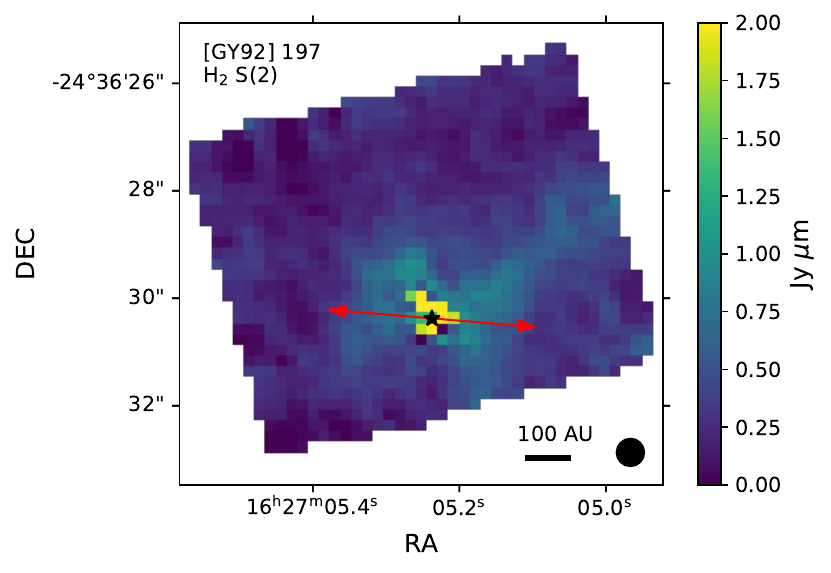} 
\includegraphics[width=0.45\linewidth]{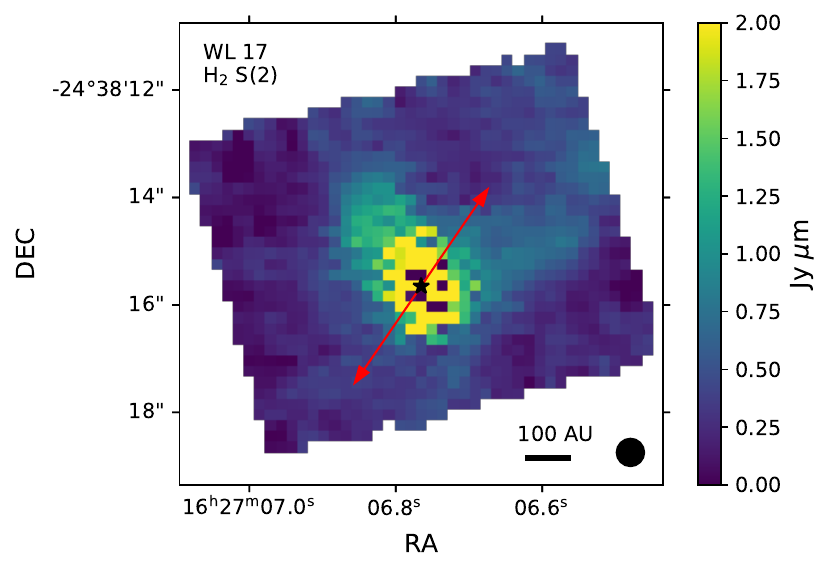} 
\includegraphics[width=0.45\linewidth]{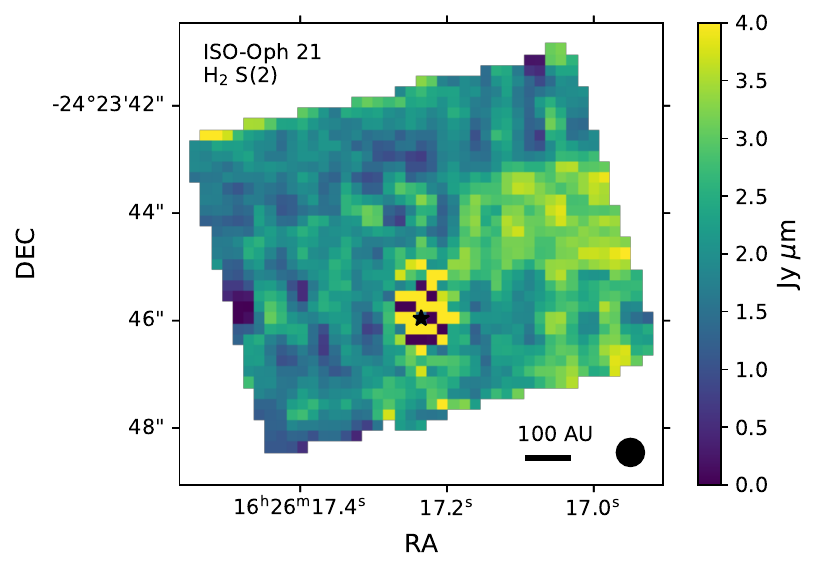} 
\caption{Same as Fig. \ref{fig:h2_maps} but for the S(2) transition.}
\label{fig:S2}
\end{figure*}

\begin{figure*}
\centering 
\includegraphics[width=0.45\linewidth]{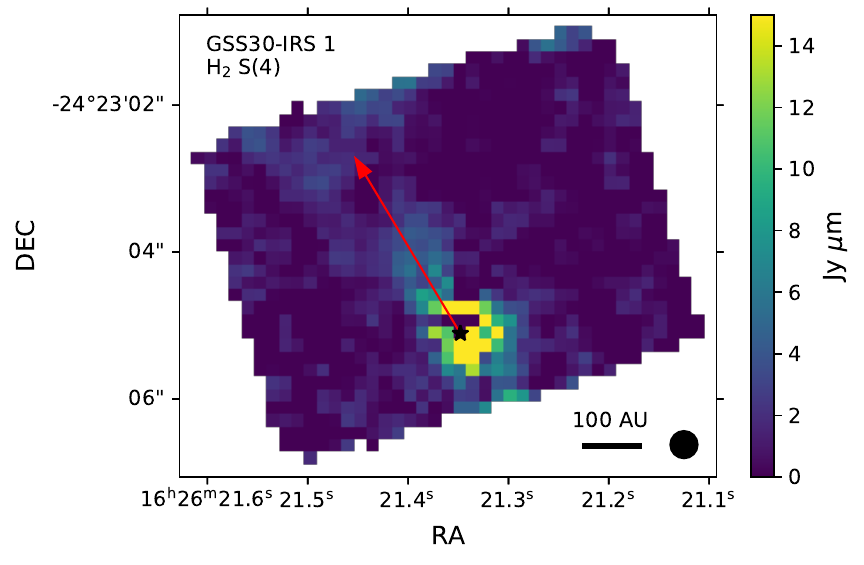} 
\includegraphics[width=0.45\linewidth]{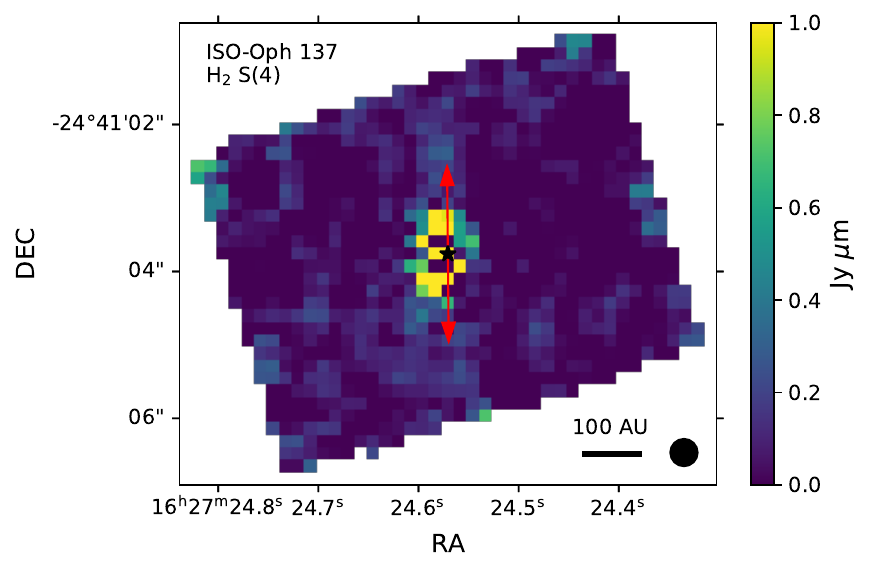} 
\includegraphics[width=0.45\linewidth]{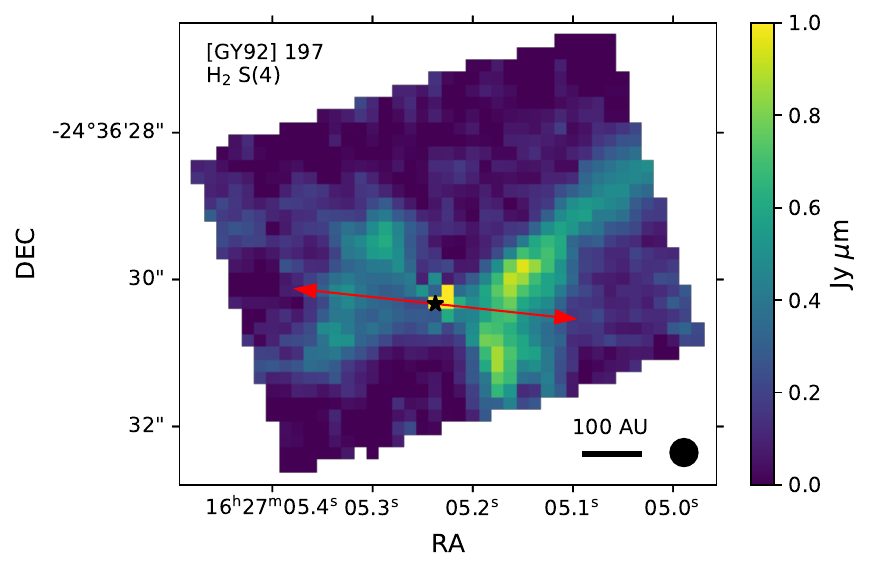} 
\includegraphics[width=0.45\linewidth]{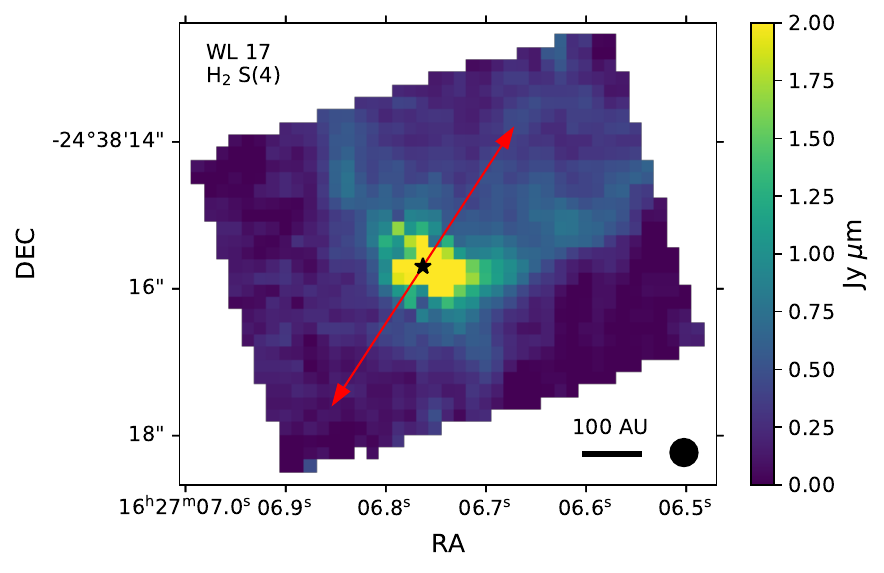} 
\caption{Same as Fig. \ref{fig:h2_maps} but for the S(4) transition.}
\label{fig:S4}
\end{figure*}

\begin{figure*}
\centering 
\includegraphics[width=0.45\linewidth]{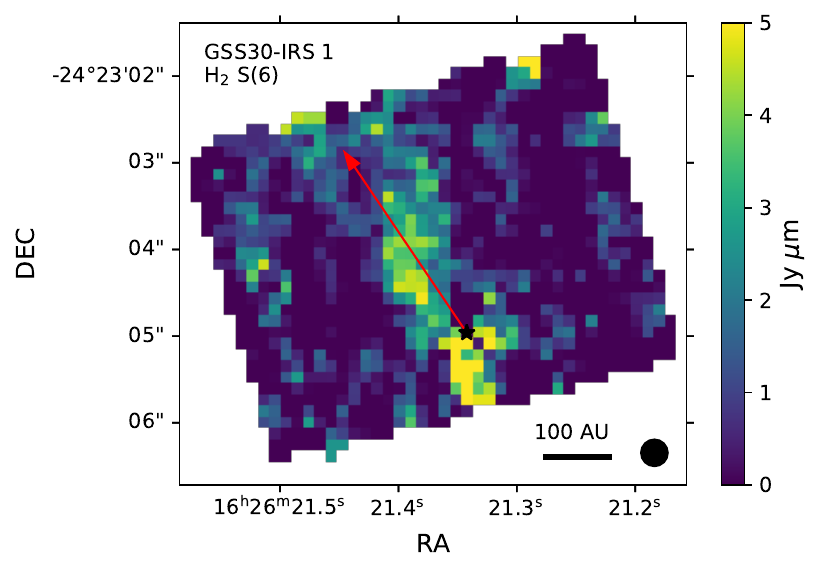} 
\includegraphics[width=0.45\linewidth]{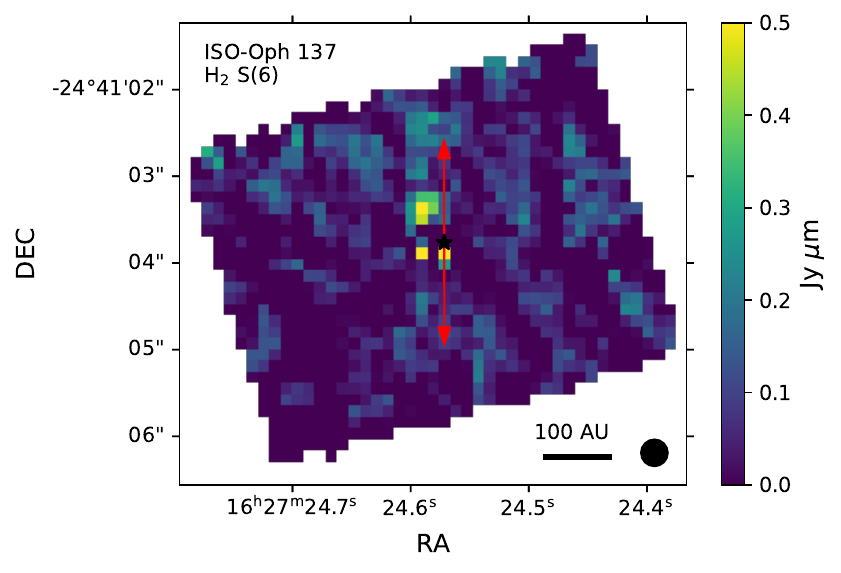} 
\includegraphics[width=0.45\linewidth]{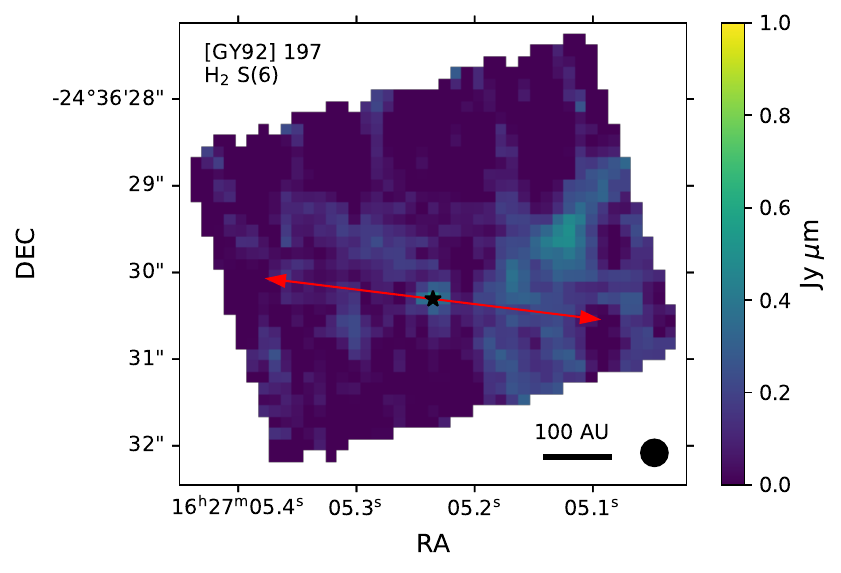} 
\includegraphics[width=0.45\linewidth]{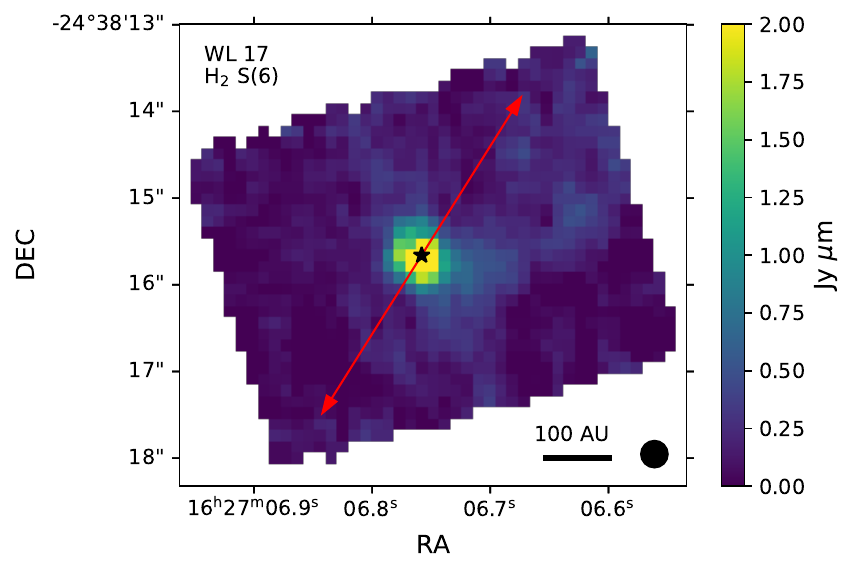} 
\caption{Same as Fig. \ref{fig:h2_maps} but for the S(6) transition.}
\label{fig:S6}
\end{figure*}

\begin{figure*}
\centering 
\includegraphics[width=0.45\linewidth]{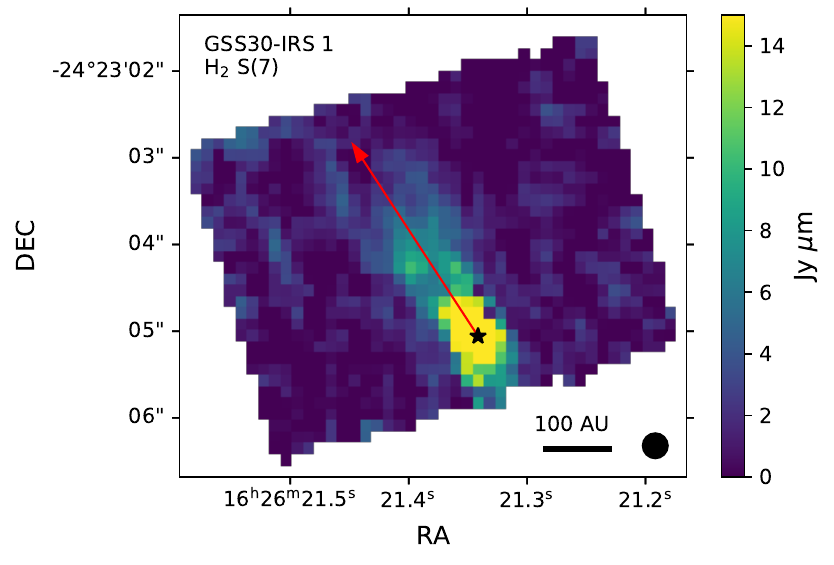} 
\includegraphics[width=0.45\linewidth]{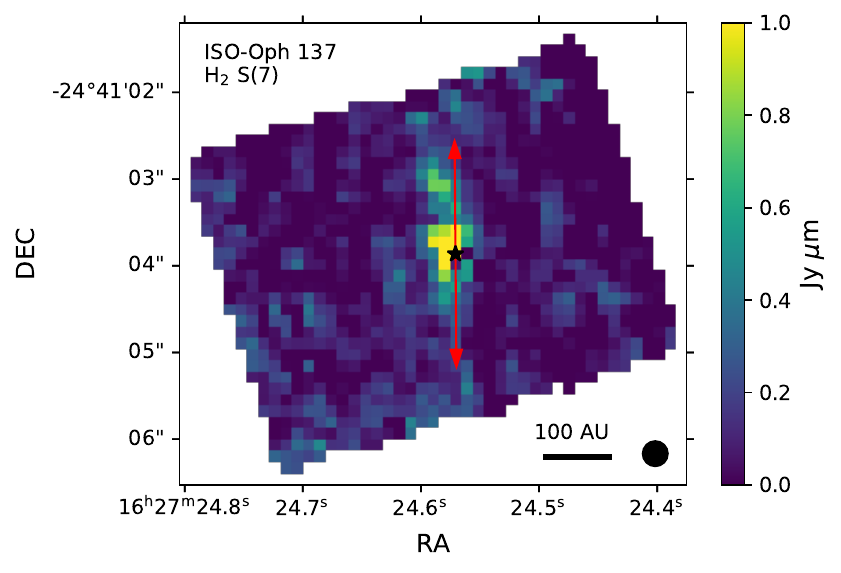} 
\includegraphics[width=0.45\linewidth]{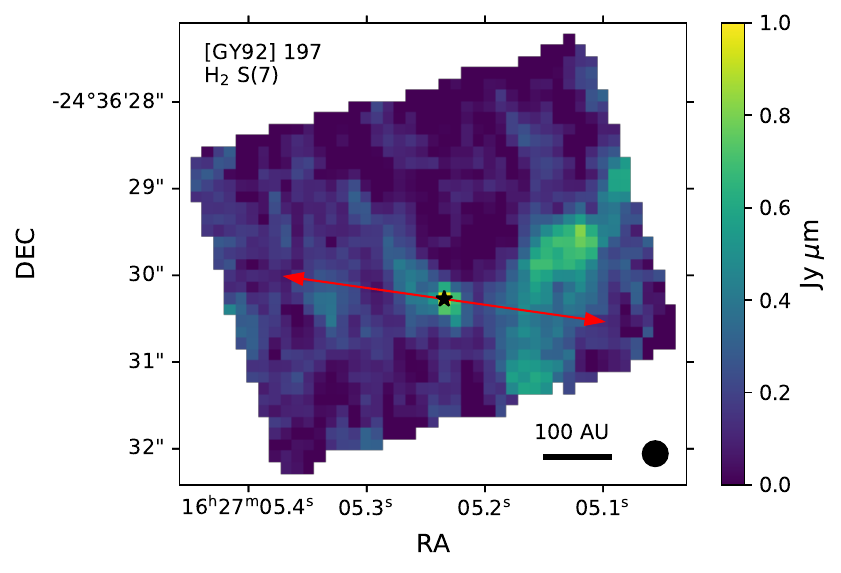} 
\includegraphics[width=0.45\linewidth]{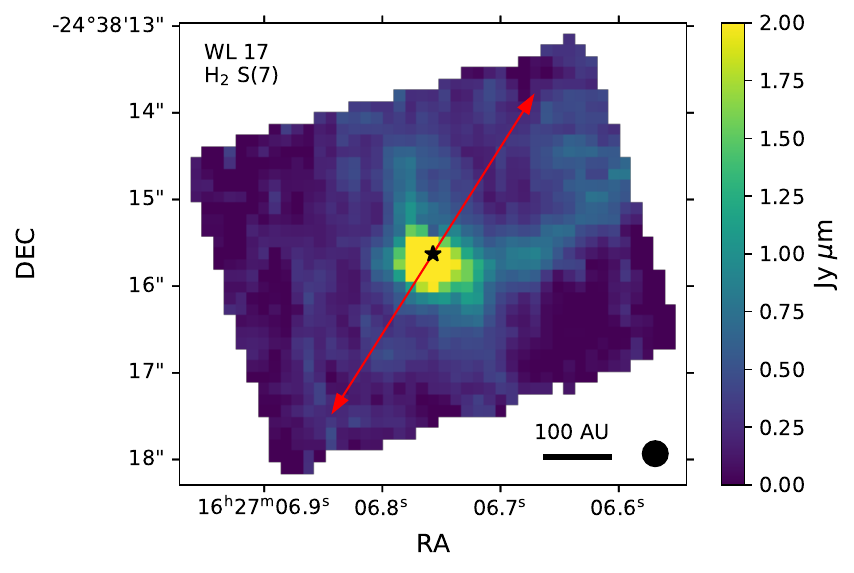} 
\caption{Same as Fig. \ref{fig:h2_maps} but for the S(7) transition.}
\label{fig:S7}
\end{figure*}

\begin{figure*}
\centering 
\includegraphics[width=0.45\linewidth]{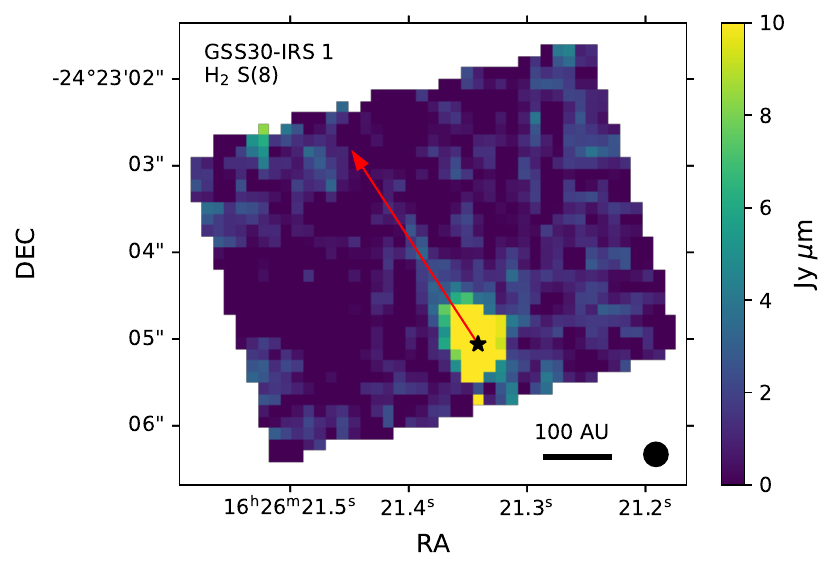} 
\includegraphics[width=0.45\linewidth]{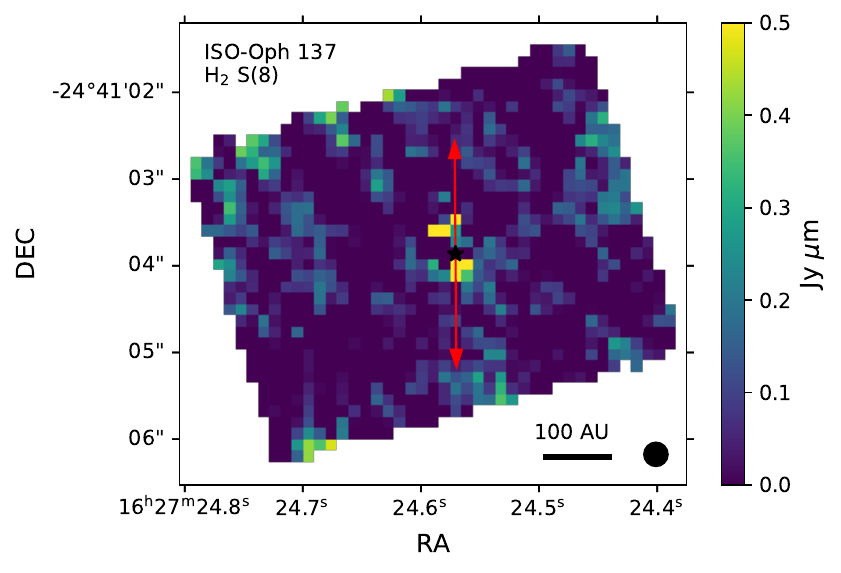} 
\includegraphics[width=0.45\linewidth]{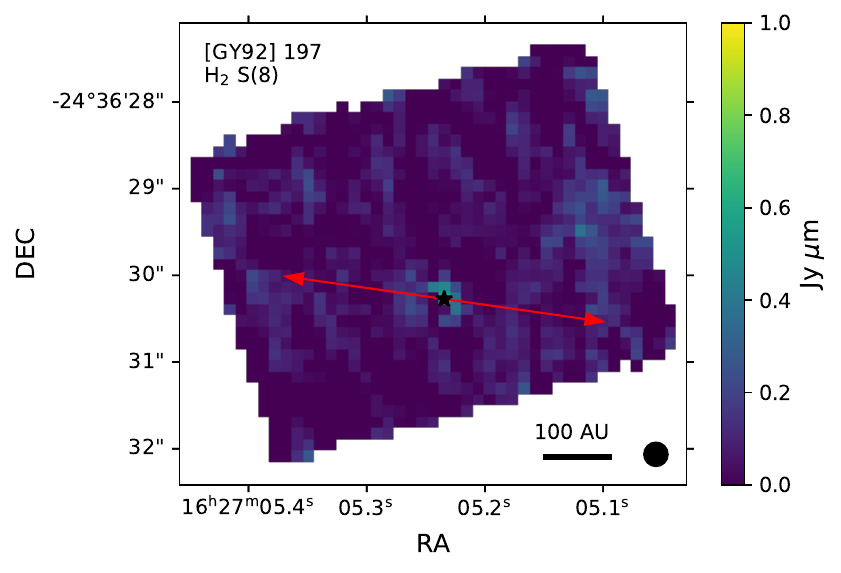} 
\includegraphics[width=0.45\linewidth]{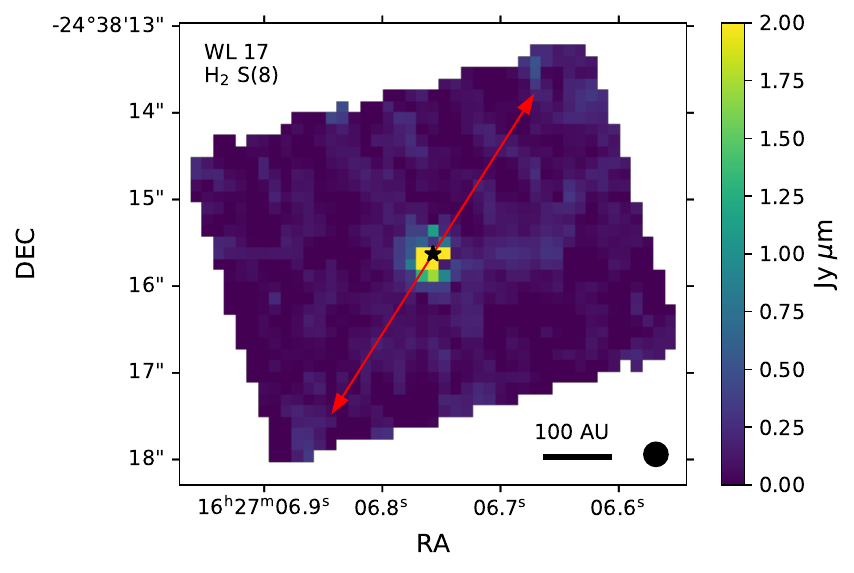} 
\caption{Same as Fig. \ref{fig:h2_maps} but for the S(8) transition.}
\label{fig:S8}
\end{figure*}

Figures \ref{fig:S3} - \ref{fig:S8} show continuum-subtracted H$_2$ maps for the different transitions. Noted with white circles in Fig. \ref{fig:S3} are the apertures used for the spectral extraction discussed in Section \ref{sec:spectra}.

\section{Additional ionic line maps}
\begin{figure*}
\centering 
\includegraphics[width=0.45\linewidth]{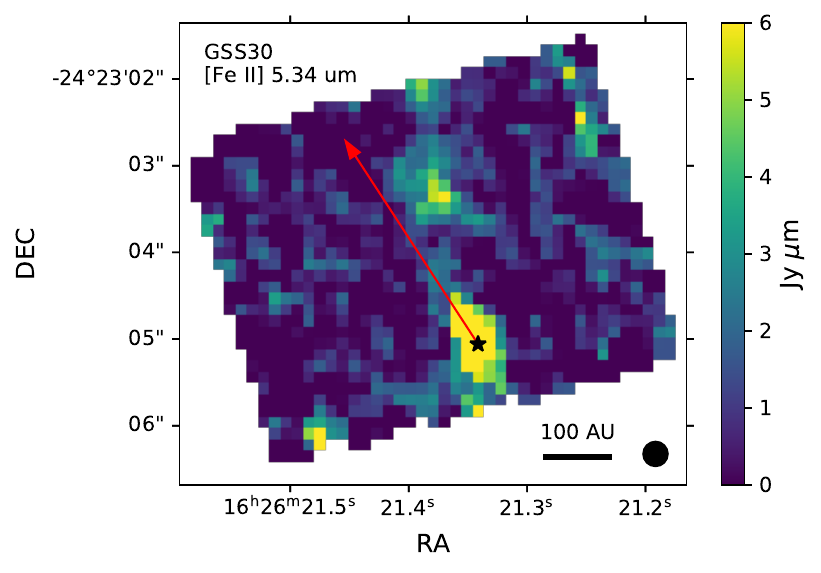} 
\includegraphics[width=0.45\linewidth]{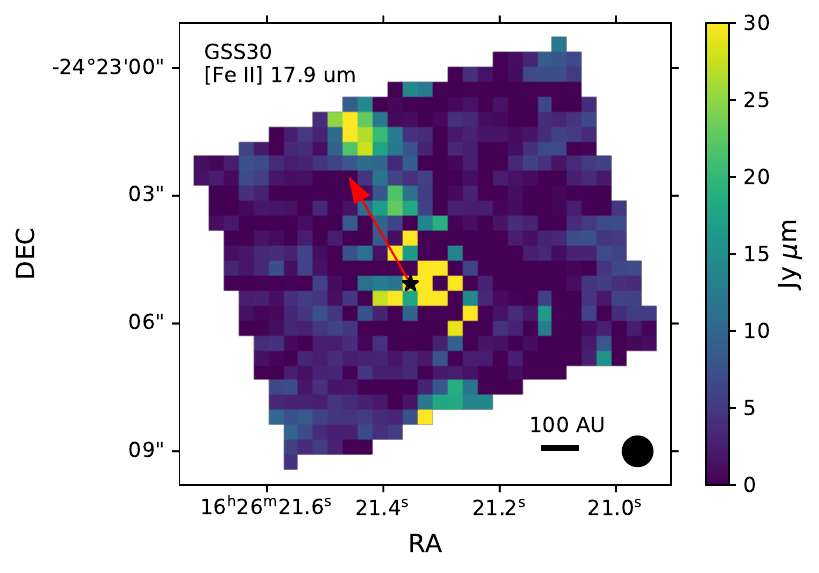} 
\caption{Continuum subtracted intensity maps of GSS30-IRS 1 in the 5.34 and 17.9 $\mu$m [\ion{Fe}{II}] transitions.}
\label{fig:GSS30 ion lines}
\end{figure*}

\begin{figure*}
\centering 
\includegraphics[width=0.45\linewidth]{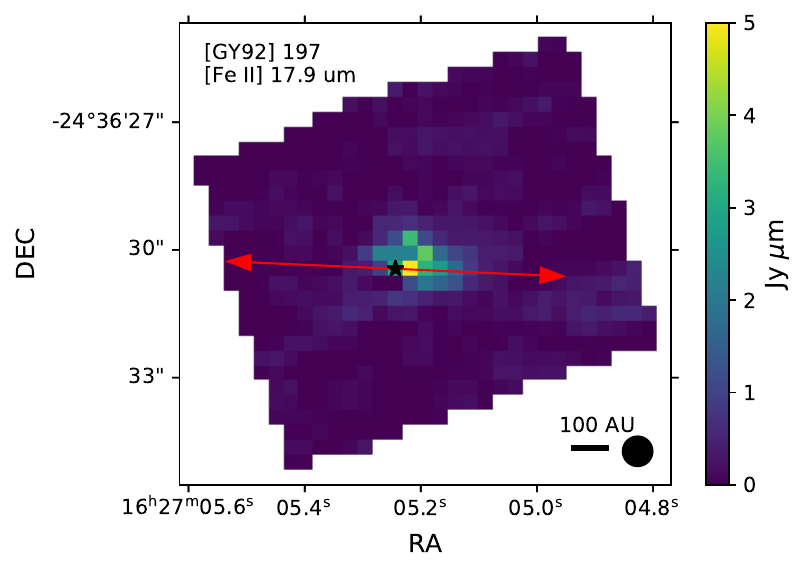} 
\includegraphics[width=0.45\linewidth]{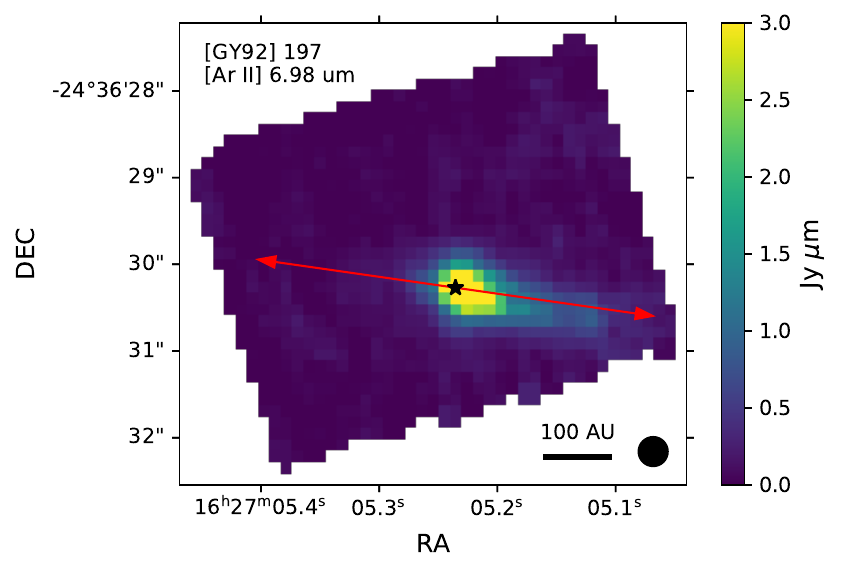}
\includegraphics[width=0.45\linewidth]{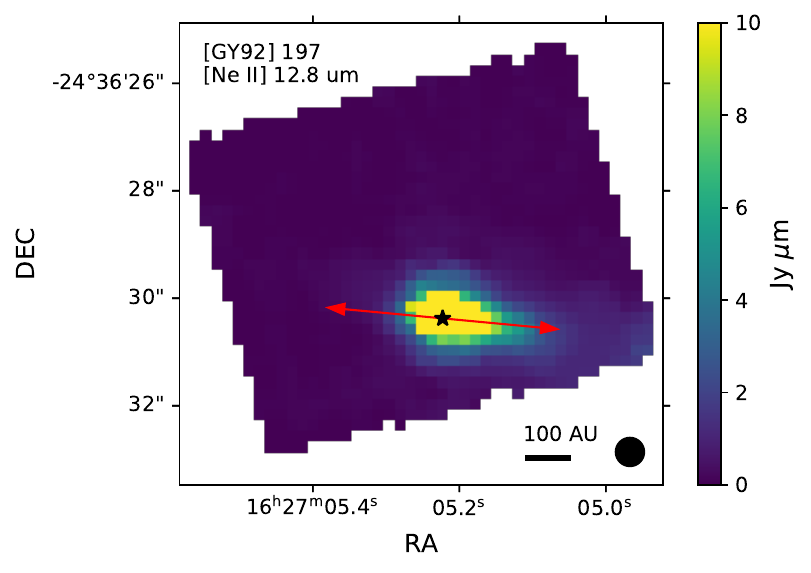} 
\includegraphics[width=0.45\linewidth]{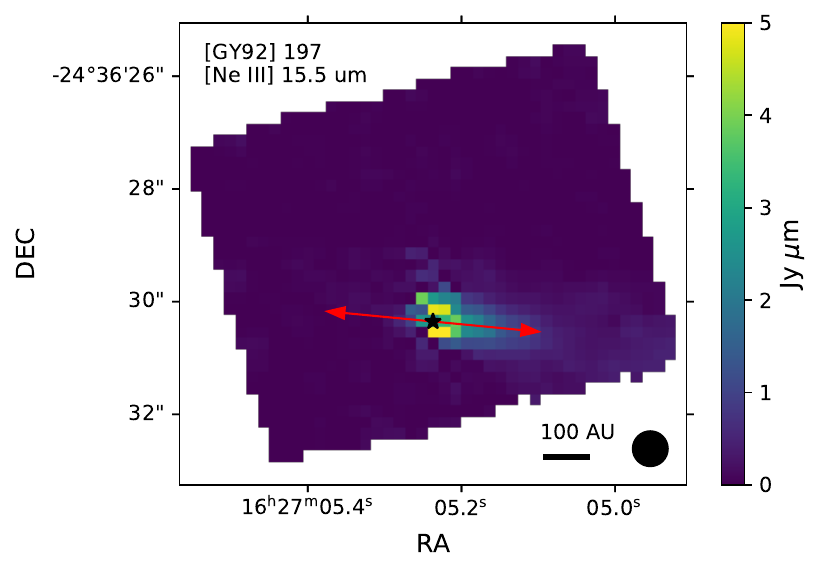}
\caption{Continuum subtracted intensity maps of [GY92] 197 in the 17.9 $\mu$m [\ion{Fe}{II}], the 6.98 $\mu$m [\ion{Ar}{II}], the 12.8 [\ion{Ne}{II}], and the 15.5 $\mu$m [\ion{Ne}{III}] transitions.}
\label{GY92197 ion lines}
\end{figure*}

\begin{figure*}
\centering 
\includegraphics[width=0.45\linewidth]{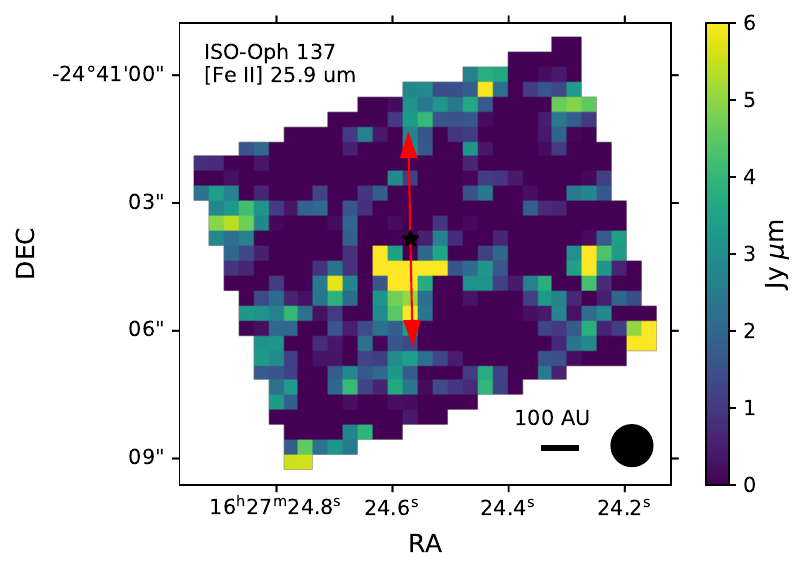}
\includegraphics[width=0.45\linewidth]{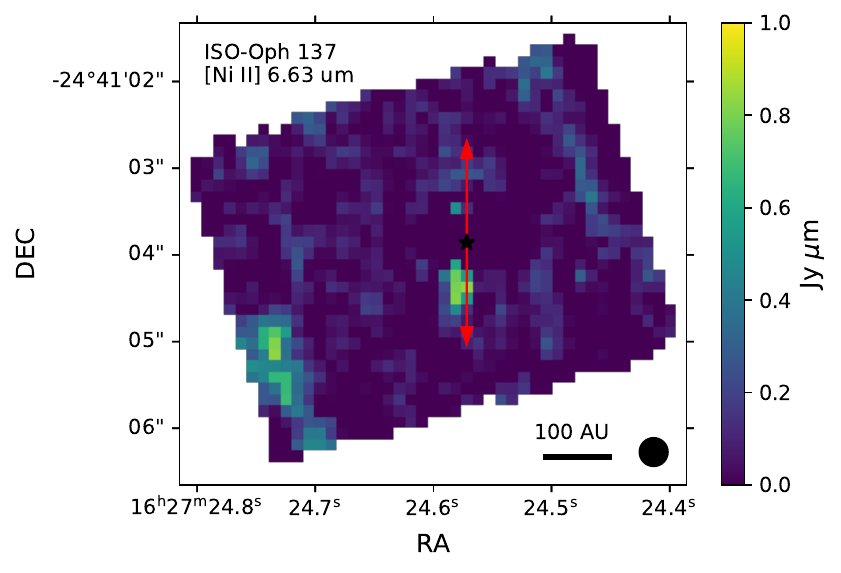} 
\caption{Continuum subtracted intensity maps of ISO-Oph 137 in the 25.9 $\mu$m [\ion{Fe}{II}] and the 6.63 $\mu$m [\ion{Ni}{II}] transitions.}
\label{fig:ISO-Oph 137 ion lines}
\end{figure*}

Figures \ref{fig:GSS30 ion lines} - \ref{fig:ISO-Oph 137 ion lines} show continuum-subtracted intensity maps for the remaining ionic transitions with strong extended emission for each source in our sample.

\section{Comparison of outflow morphology in different H$_2$ transitions}

\begin{figure*}[ht!]
\centering 
\includegraphics[width=\linewidth]{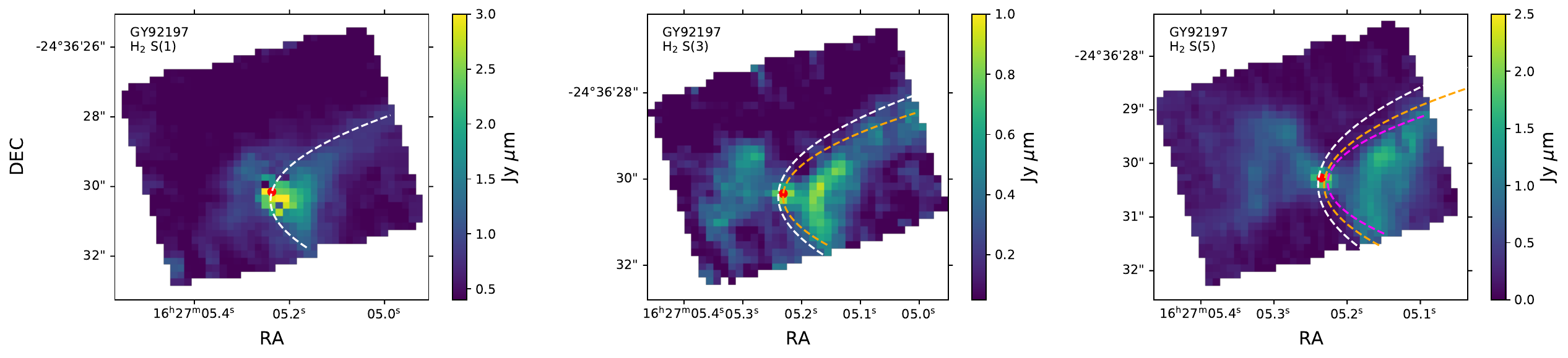}
\includegraphics[width=\linewidth]{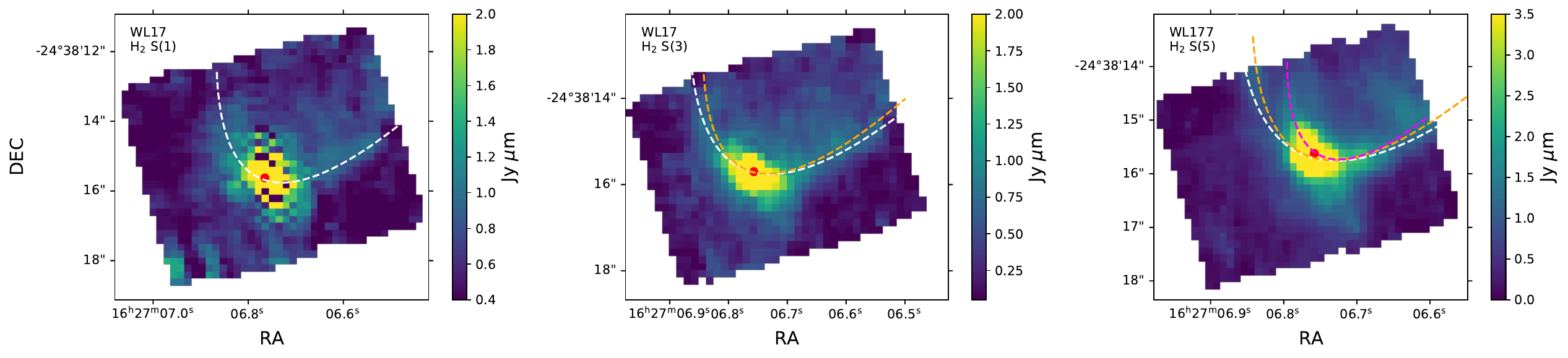}
\caption{Integrated emission of selected H$_2$ lines toward [GY92] 197 (top) and WL 17 (bottom). Dashed lines outline the outflow opening angles, in white for the S(1) line, in orange for the S(3) line, and in magenta for the S(5).
}
\label{fig:opening angle}
\end{figure*}

Figure \ref{fig:opening angle} shows the evolution of the opening angle over the different transitions of H$_2$ for the two wide-angle outflows in our sample, [GY92] 197 and WL 17. This comparison reveals very small differences in the distribution of H$_2$ emission over a large range of different transitions. 

\section{Line detections}
Table \ref{table:detections} presents a summary of all detected lines in each position for all sources in our sample.

\begin{sidewaystable*}
\caption{Line detections toward selected positions in the maps of low-mass protostars in Ophiuchus} 
\label{table:detections} 
\centering
\begin{tabular}{l l c c c| c c c c c| c c c c c c c c| c c c c c c c c| c c c c}
\hline 
Lines &transition& \multicolumn{3}{c}{GSS30-IRS 1}& \multicolumn{5}{c}{ISO-Oph 21} & \multicolumn{8}{c}{[GY92] 197} & \multicolumn{8}{c}{WL 17} & \multicolumn{4}{c}{ISO-Oph 137} \\
\hline
 & & A & B & \multicolumn{1}{c|}{C} & A & B & C & D & \multicolumn{1}{c|}{E} & A & B & C & D & E & F & G & \multicolumn{1}{c|}{H} & A & B & C & D & E & F & G & \multicolumn{1}{c|}{H} & A & B & C & D\\
\hline
H$_2$ S(1)& Iv0-J3-v0-J1& -& -& +& +& +& + &+ & +& +&+ &+ &+ &+ & +& +&+ &+ &+ & +&+ &+ &+ &+ &+ & -&+ &- & + \\
H$_2$ S(2)& Iv0-J4-v0-J2& -& -& +& +& +& +& +& +& +& +& +& +& +& +& +& +& +& +& +& +& +& +& +& +& +& +& +& +\\ 
H$_2$ S(3)& Iv0-J5-v0-J3& +& +& +& -& -& -& -& -& +& +& +& +& +& +& +& +& +& +& +& +& +& +& +& +& +& +& +& +\\ 
H$_2$ S(4)& Iv0-J6-v0-J4& +& +& +& -& -& -& -& -& +& +& +& +& +& +& +& +& +& +& +& +& +& +& +& +& +& +& +& +\\ 
H$_2$ S(5)& Iv0-J7-v0-J5& +& +& +& -& -& -& -& -& +& +& +& +& +& +& +& +& +& +& +& +& +& +& +& +& +& +& +& +\\ 
H$_2$ S(6)& Iv0-J8-v0-J6& +& +& +& -& -& -& -& -& +& +& +& +& +& +& +& +& +& +& +& +& +& +& +& +& +& +& +& +\\ 
H$_2$ S(7)& Iv0-J9-v0-J7& +& +& +& -& -& -& -& -& +& +& +& +& +& +& +& +& +& +& +& +& +& +& +& +& +& +& +& +\\ 
H$_2$ S(8)& Iv0-J10-v0-J8& +& +& +& -& -& -& -& -& +& +& +& +& +& -& -& -& +& +& +& +& +& +& +& +& -& +& -& +\\ 
$\left[\ion{Ni}{II}\right]$ 10.6& $^4$F$_{9/2}$-$^4$F$_{7/2}$& -& -& -& -& -& -& -& -& +& -& -& -& -& -& -& -& -& -& -& -& -& -& -& -& -& -& -& -\\
$\left[\ion{Ni}{II}\right]$ 6.63& $^2$D$_\mathrm{3/2}$-$^2$D$_\mathrm{5/2}$& +& +& +& -& -& -& -& -& +& -& -& -& -& -& -& -& -& -& -& -& -& -& -& -& -& -& -& +\\
$\left[\ion{Ni}{II}\right]$ 5.18& $^3$D$_\mathrm{5/2}$-$^2$D$_\mathrm{3/2}$& +& -& -& -& -& -& -& -& -& -& -& -& -& -& -& -& +& -& -& -& -& -& -& -& -& -& -& -\\
$\left[\ion{Ni}{I}\right]$ 14.8& $^3$D$_3$-$^3$D$_2$ & -& -& -& -& -& -& -& -& +& -& -& -& -& -& -& -& -& -& -& -& -& -& -& -& -& -& -& -\\
$\left[\ion{Ni}{I}\right]$ 11.3& $^3$F$_3$-$^3$F$_2$ & -& -& +& -& -& -& -& -& -& -& -& -& -& -& -& -& -& -& -& -& -& -& -& -& -& -& -& -\\
$\left[\ion{Ni}{I}\right]$ 5.89& $^3$D$_1$-$^1$D$_2$ & -& +& -& -& -& -& -& -& -& -& -& -& -& -& -& -& -& -& -& -& -& -& -& -& -& -& -& -\\
$\left[\ion{Fe}{II}\right]$ 25.9& a$^6$D$_\mathrm{7/2}$-a$^6$D$_\mathrm{9/2}$& -& +& +& -& -& -& -& -& +& -& -& -& -& -& -& -& -& -& -& -& +& -& -& -& -& +& +& +\\
$\left[\ion{Fe}{II}\right]$ 22.9& a$^4$D$_\mathrm{7/2}$-a$^4$D$_\mathrm{5/2}$& -& -& -& -& -& -& -& -& -& -& -& -& -& -& -& -& -& -& -& -& -& -& -& -& -& +& -& -\\
$\left[\ion{Fe}{II}\right]$ 17.9& a$^4$F$_\mathrm{7/2}$-a$^4$F$_\mathrm{9/2}$& -& +& +& -& -& -& -& -& +& +& +& +& +& -& -& -& -& -& -& -& -& -& -& -& -& +& +& +\\
$\left[\ion{Fe}{II}\right]$ 6.72& a$^4$F$_\mathrm{9/2}$-a$^6$D$_\mathrm{7/2}$& +& -& -& -& -& -& -& -& +& -& -& -& -& -& -& -& -& -& -& -& -& -& -& -& -& -& -& -\\
$\left[\ion{Fe}{II}\right]$ 5.34& a$^4$F$_\mathrm{9/2}$-a$^6$D$_\mathrm{9/2}$& +& +& +& -& -& -& -& -& +& +& +& -& -& -& -& -& -& -& -& -& +& -& -& -& -& -& +& +\\
$\left[\ion{Ne}{III}\right]$ 15.5& $^3$P$_2$-$^3$P$_1$& -& -& -& -& -& -& -& -& +& +& +& +& +& -& -& -& -& -& +& -& -& -& -& -& -& -& -& -\\
$\left[\ion{Ne}{II}\right]$ 12.8& $^2$P$_\mathrm{3/2}$-$^2$P$_\mathrm{1/2}$& -& -& -& -& -& -& -& -& +& +& +& +& +& +& +& +& +& +& +& -& +& +& +& +& +& +& +& -\\
$\left[\ion{S}{I}\right]$ 25.2 & $3$P$_\mathrm{1}$-$3$P$_\mathrm{2}$ & -& +& +& -& -& -& -& -& -& -& -& -& +& -& -& -& -& -& -& -& +& +& +& -& +& +& +& +\\
$\left[\ion{Co}{II}\right]$ 25.6& a$^5$F$_3$-a$^5$F$_2$& -& +& -& -& -& -& -& -& -& +& -& -& -& -& -& -& -& -& -& -& -& -& -& -& -& -& -& -\\
$\left[\ion{Co}{II}\right]$ 14.7& a$^5$F$_4$-a$^5$F$_5$& -& -& -& -& -& -& -& -& -& -& -& -& -& -& -& -& -& -& +& -& -& -& -& -& -& -& -& -\\
$\left[\ion{Ar}{II}\right]$ 6.98& $^2$P$\mathrm{1/2}$-$^2$P$_\mathrm{3/2}$& -& +& -& -& -& -& -& -& +& +& +& +& +& -& -& -& -& -& +& -& -& -& -& -& -& -& -& -\\
$\left[\ion{Ar}{III}\right]$ 21.8& $^3$P$_1$-$^3$P$_0$& -& -& -& -& -& -& -& -& -& -& -& -& -& -& -& -& -& -& +& -& +& -& +& -& -& -& -& -\\
$\left[\ion{Ar}{III}\right]$ 8.99& $^3$P$_1$-$^3$P$_2$& -& -& -& -& -& -& -& -& -& -& -& -& -& -& -& -& -& -& -& -& -& -& -& +& -& -& -& -\\
\end{tabular}
\end{sidewaystable*}

\section{Line fluxes}
Table \ref{table:ion_atom_fluxes} contains line integrated fluxes for all detected [\ion{Ne}{II}], [\ion{Ne}{III}], [\ion{Fe}{II}] and [\ion{S}{I}] transitions. Table \ref{table:H2_fluxes} contains integrated line fluxes, corrected for the local extinction values, for all H$_2$ lines detected in each apertures of all sources in our sample.

\begin{sidewaystable*}
\caption{Integrated fluxes of detected \ion{Ne}{II}, \ion{Ne}{III}, \ion{Fe}{II}, and \ion{S}{I} lines in all sources and apertures, corrected for extinction}
\label{table:ion_atom_fluxes} 
\centering
\begin{tabular}{l c c c c c c c c}
\hline \hline 
Pos. & \multicolumn{8}{c}{Integrated flux $\times10^{-18}$ { [}W m$^{-2}${]}} \\ 
\hline
 & [\ion{Ne}{II}] 12.8 $\mu$m & [\ion{Ne}{III}] 15.5 $\mu$m & [\ion{Fe}{II}] 5.34 $\mu$m & [\ion{Fe}{II}]17.9 $\mu$m& [\ion{Fe}{II}] 22.9 $\mu$m & [\ion{Fe}{II}] 24.5 $\mu$m &[\ion{Fe}{II}] 25.9 $\mu$m & [\ion{S}{I}] 25.2 $\mu$m\\
\hline 
\multicolumn{9}{c}{\textbf{GSS30-IRS 1}}\\
\hline
A & -- & -- & 15.6 $\pm$ 8.7 & -- & -- & -- & -- & -- \\
B & 14.5 $\pm$ 4.2 & -- & 8.7 $\pm$ 1.7 & 10.7 $\pm$ 3.4 & -- & -- & 26.3 $\pm$ 6.1  & --\\
C & 6.0 $\pm$ 2.5 & -- & 8.9 $\pm$ 2.3 & 9.0 $\pm$ 0.9 & -- & -- & 32.4 $\pm$ 4.9  & --\\
\hline
\multicolumn{9}{c}{\textbf{[GY92] 197}}\\
\hline
A & 28.8 $\pm$ 1.7 & 1.8 $\pm$ 0.1 & 0.8 $\pm$ 0.1 & 2.3 $\pm$ 0.2 & -- & -- & 1.2 $\pm$ 0.3  & -- \\
B & 3.9 $\pm$ 0.2 & 0.5 $\pm$ 0.04 & 0.5 $\pm$ 0.2 & 0.4 $\pm$ 0.04 & -- & 0.2 $\pm$ 0.1 & 0.4 $\pm$ 0.4  & -- \\
C & 3.5 $\pm$ 0.2 & 0.4 $\pm$ 0.05 & 0.5 $\pm$ 0.2 & 0.4 $\pm$ 0.04 & -- & --& 0.3 $\pm$ 0.2  & -- \\
D & 2.1 $\pm$ 0.2 & 0.2 $\pm$ 0.04 & -- & 0.3 $\pm$ 0.04 & -- & -- & --& -- \\
E & 10.0 $\pm$ 1.0 & 0.3 $\pm$ 0.1 & -- & 0.5 $\pm$ 0.2 & -- & -- & -- & 1.1 $\pm$ 0.4\\
F & 0.7 $\pm$ 0.1 & -- & -- & -- & -- & -- & -- & -- \\
G & 0.5 $\pm$ 0.1 & -- & -- & -- & -- & -- & -- & -- \\
H & 0.3 $\pm$ 0.06 & -- & -- & -- & -- & -- & -- & -- \\
\hline
\multicolumn{9}{c}{\textbf{WL 17}}\\
\hline
A & 1.2 $\pm$ 0.3 & -- & -- & -- & -- & -- & -- & --\\
B & 0.3 $\pm$ 0.05 & -- & -- & -- & -- & -- & -- & --\\
C & 0.1 $\pm$ 0.03 & 0.04 $\pm$ 0.02 & -- & -- & -- & -- &  -- & -- \\
D & -- & -- & -- & -- & -- & -- & -- \\
E & 1.7 $\pm$ 0.2 & -- & 0.3 $\pm$ 0.1 & -- & -- & -- & -- &-- \\
F & 0.1 $\pm$ 0.08 & -- & -- & -- & -- & -- & -- & 0.3 $\pm$ 0.1 \\
G & 0.4 $\pm$ 0.06 & -- & -- & -- & -- & -- & -- & 0.5 $\pm$ 0.2 \\
H & 0.1 $\pm$ 0.04 & -- & -- & -- & -- & -- & -- & -- \\
\hline
\multicolumn{9}{c}{\textbf{ISO-Oph 137}}\\
\hline
A & 4.3 $\pm$ 2.4 & -- & -- & -- & -- & --& -- & 7.6 $\pm$ 2.5 \\
B & 0.3 $\pm$ 0.2 & -- & -- & 0.4 $\pm$ 0.1 & 0.4 $\pm$ 0.1 & -- & -- & 2.9 $\pm$ 0.6 \\
C & 5.5 $\pm$ 2.1 & -- & 2.0 $\pm$ 0.7 & 7.0 $\pm$ 1.8 & -- & -- & 4.7 $\pm$ 1.9  & 8.0 $\pm$ 1.8 \\
D & -- & -- & 0.6 $\pm$ 0.2 & 0.9 $\pm$ 0.1 & -- & -- & 1.5 $\pm$ 0.3  & 0.2 $\pm$ 0.6 \\
\hline
\end{tabular}
\end{sidewaystable*}

\begin{table*}[htb!]
\caption{Integrated fluxes of H$_2$ lines in all sources and apertures, corrected for extinction}
\label{table:H2_fluxes} 
\centering 
\begin{tabular}{l c c c c c c c c} 
\hline \hline 
Pos. & \multicolumn{8}{c}{Integrated flux $\times10^{-18}$ { [}W m$^{-2}${]}} \\ 
\hline
 & S(1) & S(2)& S(3)& S(4)& S(5)& S(6)& S(7)& S(8) \\
\hline 
\multicolumn{9}{c}{\textbf{GSS30-IRS 1}}\\
\hline
A & -- & -- & $303.0\pm73.0$ & $50.2\pm10.1$ & $77.1\pm15.6$ & $10.9\pm6.1$ & $65.5\pm9.0$ & -- \\
B & -- & -- & \phantom{11}$6.8\pm1.9$\phantom{1} & $4.1\pm1.3$ & $19.5\pm2.2$\phantom{1} & \phantom{1}$9.0\pm1.7$ & $15.3\pm2.1$ & --  \\
C & $1.4\pm0.3$ & $1.6\pm0.5$ & \phantom{11}$2.3\pm0.6$\phantom{1} & $2.4\pm0.7$ & $3.9\pm0.7$ & \phantom{1}$2.1\pm1.4$ & \phantom{1}$4.5\pm1.2$ & --  \\
\hline
\multicolumn{9}{c}{\textbf{ISO-Oph 21}\tablefootmark{a}}\\
\hline
A & $0.6\pm0.1$& $0.9\pm0.1$ & -- & -- & -- & -- & -- & --  \\
B & $0.8\pm0.1$ & $1.1\pm0.1$ & -- & -- & -- & -- & -- & -- \\
C & $0.6\pm0.1$ & $1.1\pm0.1$ & -- & -- & -- & -- & -- & -- \\
D & $0.8\pm0.1$ & $1.1\pm0.1$ & -- & -- & -- & -- & -- & -- \\
E & $0.6\pm0.1$& $1.1\pm0.1$ & -- & -- & -- & -- & -- & -- \\
\hline
\multicolumn{9}{c}{\textbf{[GY92] 197}}\\
\hline
A & $1.7\pm0.1$ & $2.4\pm0.2$ & $6.0\pm0.5$ & $1.8\pm0.2$ & $2.8\pm0.2$ & $0.7\pm0.1$ & $1.5\pm0.2$ & $0.3\pm0.2$ \\
B & $0.4\pm0.1$ & $0.6\pm0.1$ & $1.3\pm0.2$ & $0.6\pm0.1$ & $1.3\pm0.1$ & $0.5\pm0.1$ & $0.9\pm0.2$ & $0.2\pm0.2$ \\
C & $0.9\pm0.1$ & $1.3\pm0.1$ & $3.9\pm0.4$ & $1.5\pm0.2$ & $2.1\pm0.2$ & $0.8\pm0.2$ & $1.5\pm0.2$ & $0.5\pm0.2$  \\
D & $1.1\pm0.1$ & $1.9\pm0.1$ & $4.7\pm0.4$ & $1.9\pm0.2$ & $3.2\pm0.2$ & $1.2\pm0.2$ & $2.5\pm0.3$ & $0.8\pm0.2$ \\
E & $1.3\pm0.3$ & $2.4\pm0.3$ & $5.3\pm0.5$ & $1.5\pm0.2$ & $2.3\pm0.2$ & $0.6\pm0.1$ & $1.3\pm0.2$ & -- \\
F & $0.4\pm0.1$ & $1.1\pm0.1$ & $1.7\pm0.5$ & $0.6\pm0.2$ & $1.4\pm0.1$ & $0.4\pm0.1$ & $1.0\pm0.2$ & -- \\
G & $0.7\pm0.1$ & $1.6\pm0.2$ & $2.7\pm0.5$ & $1.0\pm0.1$ & $1.5\pm0.2$ & $0.4\pm0.1$ & $0.7\pm0.1$ & $0.1\pm0.2$ \\
H & $0.6\pm0.1$ & $1.1\pm0.1$ & $2.7\pm0.4$ & $0.7\pm0.2$ & $1.2\pm0.1$ & $0.2\pm0.1$ & $0.6\pm0.2$ & $0.2\pm0.2$ \\
\hline
\multicolumn{9}{c}{\textbf{WL 17}}\\
\hline
A & $0.6\pm0.2$ & $1.7\pm0.2$& $3.5\pm0.3$& $1.3\pm0.1$& $2.3\pm0.2$& $0.5\pm0.1$& $1.4\pm0.2$& $0.4\pm0.1$ \\
B & $0.5\pm0.1$ &$0.8\pm0.1$ & $2.4\pm0.2$ & $0.9\pm0.1$ &$1.8\pm0.2$ & $0.5\pm0.1$ & $1.2\pm0.2$ & $0.5\pm0.2$ \\
C & $0.6\pm0.1$ & $1.2\pm0.1$ & $4.1\pm0.3$ & $1.5\pm0.1$ & $3.1\pm0.2$ & $0.7\pm0.1$ & $2.0\pm0.2$ & $0.5\pm0.1$ \\
D & $0.6\pm0.2$ & $1.2\pm0.2$ & $3.2\pm0.3$ & $1.1\pm0.1$ & $2.6\pm0.2$ & $0.6\pm0.1$ & $1.8\pm0.2$ & $0.2\pm0.1$ \\
E & $0.8\pm0.2$ & $1.9\pm0.2$ & $3.5\pm0.3$ & $1.3\pm0.1$ & $2.1\pm0.2$ & $0.7\pm0.1$ & $1.6\pm0.2$ & $0.5\pm0.1$ \\
F & $0.5\pm0.1$ & $0.7\pm0.1$ & $1.2\pm0.2$ & $0.4\pm0.1$ & $0.8\pm0.1$ & $0.4\pm0.2$ & $1.3\pm0.3$ & $0.5\pm0.3$ \\
G & $0.5\pm0.1$ & $0.8\pm0.1$ & $2.3\pm0.2$ & $0.9\pm0.1$ & $2.2\pm0.2$ & $0.7\pm0.1$ & $1.1\pm0.1$ & $0.3\pm0.1$ \\
H & $0.5\pm0.1$ & $1.0\pm0.1$ & $2.1\pm0.3$ & $0.6\pm0.1$ & $1.1\pm0.1$ & $0.3\pm0.1$ & $0.6\pm0.1$ & $0.3\pm0.1$  \\
\hline
\multicolumn{9}{c}{\textbf{ISO-Oph 137}}\\
\hline
A & -- & $21.3\pm7.9$& $29.6\pm5.9$ & $5.0\pm1.0$ & $10.2\pm1.1$ & $2.5\pm4.1$ & $3.7\pm2.3$ & -- \\
B & $0.5\pm0.1$ & \phantom{1}$0.9\pm0.2$ & \phantom{1}$2.1\pm0.4$ & $0.6\pm0.2$ & \phantom{1}$1.2\pm0.2$ & $0.3\pm0.2$ & $1.2\pm0.2$ & $0.5\pm0.2$ \\
C & -- & \phantom{1}$6.6\pm1.5$ & $12.0\pm2.0$ & $3.0\pm0.6$ & \phantom{1}$4.9\pm0.8$ & $1.5\pm0.6$ & $2.1\pm1.1$ & -- \\
D & $0.4\pm0.1$ & \phantom{1}$0.6\pm0.1$ & \phantom{1}$1.5\pm0.5$ & $0.4\pm0.1$ & \phantom{1}$0.9\pm0.1$ & $0.3\pm0.1$ & $0.9\pm0.3$ & $0.4\pm0.3$  \\
\hline
\end{tabular}
\tablefoot{\tablefoottext{a}{Values given for ISO-Oph 21 are uncorrected for extinction since the extinction value could not be estimated.}
}
\end{table*} 

\section{Total mid-IR luminosities}
Table \ref{table:line_luminosities} shows the total mid-IR luminosities of H$_2$, [\ion{Fe}{II}], [\ion{Ne}{II}] and [\ion{S}{I}], by summing up the intensities of all available transitions of each species in every given aperture. Values in parenthesis give the percentage of the total emission coming from each species.    

 \begin{table*}[!]
\caption{Total mid-IR line luminosities for all detected H$_2$, \ion{Fe}{II},  \ion{Ne}{II}, and \ion{S}{I} lines.}
\label{table:line_luminosities} 
\centering 
\begin{tabular}{l c c c c c} 
\hline \hline 
Pos. & \multicolumn{5}{c}{Line luminosities $\times10^{-6}$ { [}L$_\odot$ {]}}  \\ 
\hline
 & H$_2$ & [\ion{Fe}{II}]& [\ion{Ne}{II}] & [\ion{S}{I}] & Total \\
\hline 
\multicolumn{6}{c}{\textbf{GSS30-IRS 1}}\\
\hline
A & 294.04 (97.0\%)& 9.05 (3.0\%)& -- & -- & 303.09 \\
B & 31.74 (47.6\%)& 26.52 (39.8\%)& 8.41 (12.6\%)& -- & 66.67 \\
C & 10.56 (24.5\%)& 29.19 (67.5\%)& 3.48 (8.0\%)& -- & 43.23 \\
\hline
\multicolumn{6}{c}{\textbf{[GY92] 197}}\\
\hline
A & 9.98 (34.2\%)& 2.50 (8.5\%)& 16.71 (57.2\%)& -- & 29.19 \\
B & 3.37 (51.8\%)& 0.87 (13.4\%)& 2.26 (34.8\%)& -- & 6.50 \\
C & 7.25 (72.6\%)& 0.70 (7.0\%)& 2.03 (20.3\%)& -- & 9.98 \\
D & 10.04 (87.8\%)& 0.17 (1.5\%)& 1.21 (10.6\%)& -- & 11.43 \\
E & 8.53 (55.9\%)& 0.29 (1.9\%)& 5.80 (38.0\%)& 0.64 (4.2\%)& 15.26 \\
F & 3.83 (90.3\%)& -- & 0.41 (9.7\%)& -- & 4.24 \\
G & 5.05 (94.6\%)& -- & 0.29 (5.4\%)& -- & 5.34 \\
H & 4.24 (96.1\%)& -- & 0.17 (3.9\%)& -- & 4.41 \\
\hline
\multicolumn{6}{c}{\textbf{WL 17}}\\
\hline
A & 6.79 (90.7\%)& -- & 0.70 (9.3\%)& -- & 7.49 \\
B & 4.99 (96.7\%)& -- & 0.17 (3.3\%)& -- & 5.16 \\
C & 7.95 (99.3\%)& -- & 0.06 (0.7\%)& -- & 8.01 \\
D & 6.56 (100\%)& -- & -- & -- & 6.56 \\
E & 7.20 (86.1\%)& 0.17 (2.0\%)& 0.99 (11.8\%)& -- & 8.36 \\
F & 3.89 (94.4\%)& -- & 0.06 (1.5\%)& 0.17 (4.1\%)& 4.12 \\
G & 5.11 (90.8\%)& -- & 0.23 (4.1\%)& 0.29 (5.1\%)& 5.63 \\
H & 3.77 (98.4\%)& -- & 0.06 (1.6\%)& -- & 3.83 \\
\hline
\multicolumn{6}{c}{\textbf{ISO-Oph 137}}\\
\hline
A & 41.96 (85.9\%)& -- & 2.50 (5.1\%)& 4.41 (9.0\%)& 48.86 \\
B & 4.24 (64.6\%)& 0.46 (7.0\%)& 0.17 (2.6\%)& 1.68 (25.6\%)& 6.56 \\
C & 17.47 (52.5\%)& 7.95 (23.9\%)& 3.19 (9.6\%)& 4.64 (14.0\%)& 33.25 \\
D & 3.13 (62.7\%)& 1.74 (34.9\%)& -- & 0.12 (2.4\%)& 4.99 \\
\hline
\end{tabular}

\end{table*} 

\section{Additional shock model comparisons}
\label{app:shocks}
\begin{figure*}[ht!]
\centering 
\includegraphics[width=\linewidth]{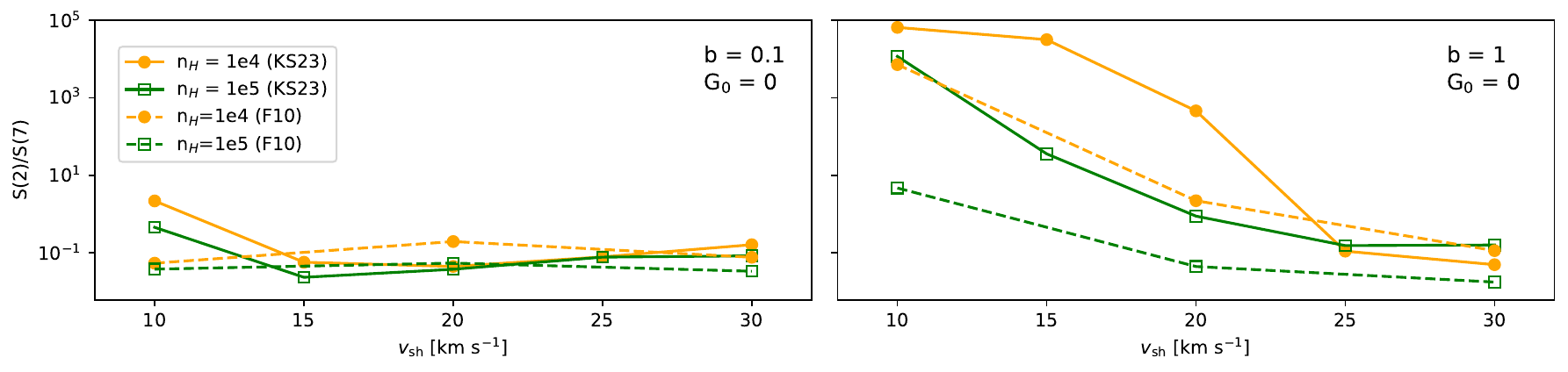}
\caption{\label{fig:KS23_vs_Flower} Ratio of the S(2) over the S(7) transition of H$_2$ against the shock velocity. Full lines show values from \cite{Kristensen23} and dashed lines from \cite{fp10}, while yellow and green lines represent log(n$_H$) = 4 and 5 respectively. Left panel shows values for b=0.1, and right for b=1.}
\end{figure*}

Figure \ref{fig:KS23_vs_Flower} shows a direct comparison of the results between the non-UV irradiated shock model from \cite{fp10} and those of \cite{Kristensen23} with G$_0$ = 0. For $J$-type shocks (b=0.1) the two models appear in relatively good agreement, with less than an order of magnitude difference between them. That is not the case in $C$-type shocks (b=1) where for certain shock velocities we see a difference up to $\sim$2 orders of magnitude. Since the external UV field is set to 0 in the \cite{Kristensen23} model the observed differences are a result of differences in the two shock models. This difference between the two iterations of shock models highlights the need for an up-to-date comparison between observations and shock models. 

\begin{figure}[ht!]
\centering 
\includegraphics[width=\linewidth]{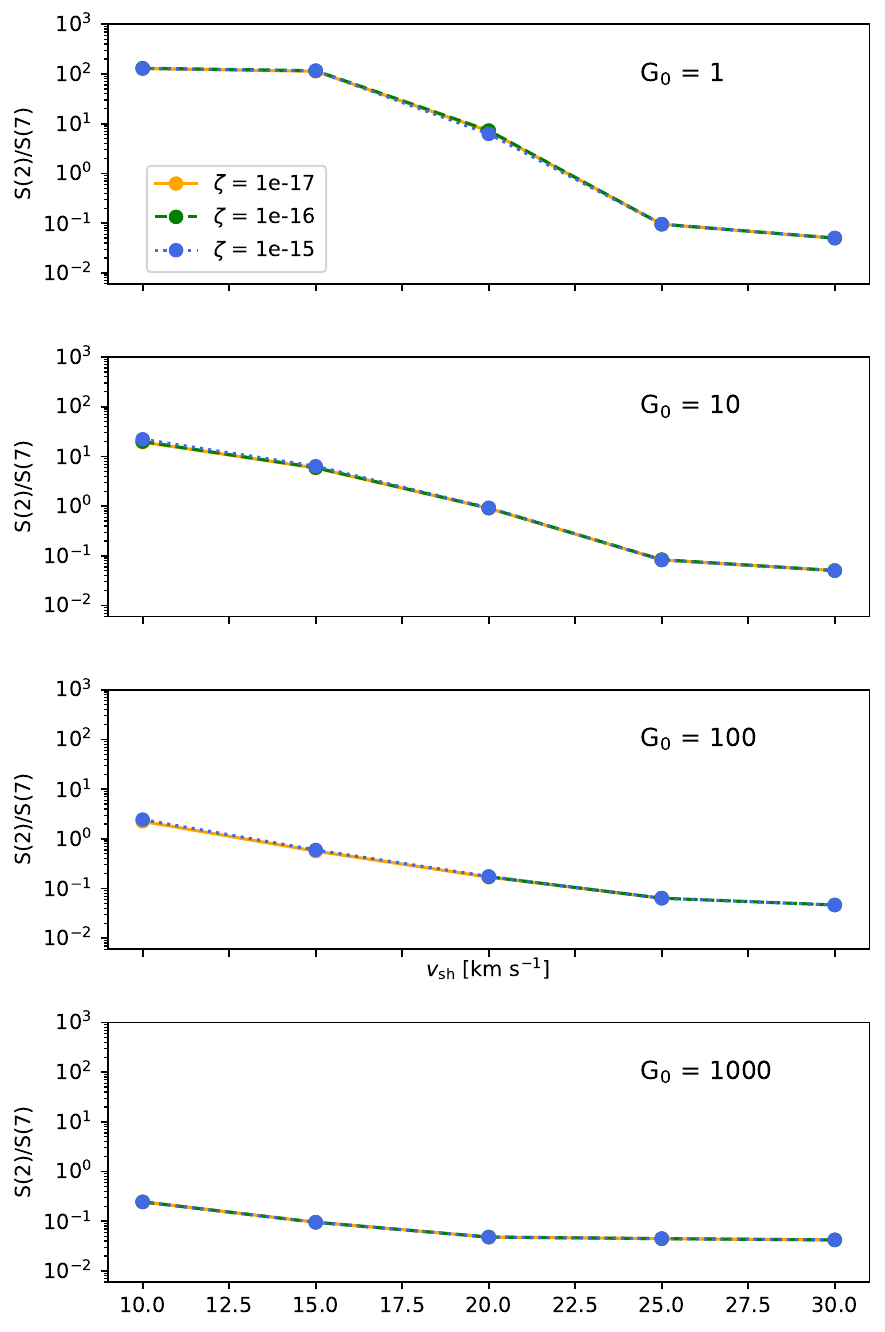} 
\caption{\label{fig:zeta_comparison} Ratio of the S(2) over the S(7) transition of H$_2$ against the shock velocity for different cosmic ray ionization rates ($\zeta$). 
The radiation field varies between panels, with $F_\mathrm{UV}$ = 1, 10, 100, and 1000 from top to bottom panel respectively. For all panels b=1, X(PAH)= $10^{-17}$, and log(n$_H$) = 4 are used.}
\end{figure}

\begin{figure}[ht!]
\centering 
\includegraphics[width=\linewidth]{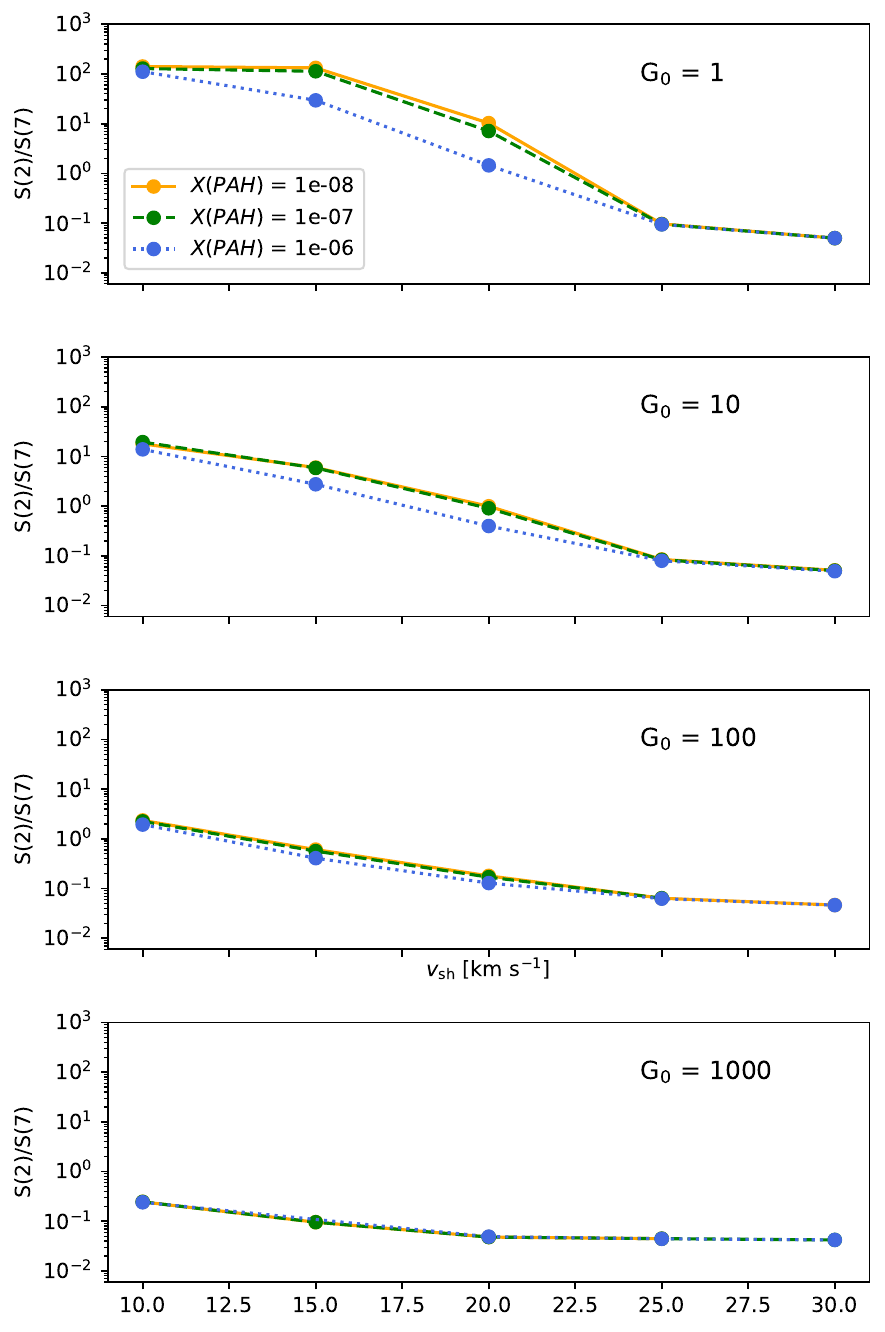} 
\caption{\label{fig:XPAH_comparison} Ratio of the S(2) over the S(7) transition of H$_2$ against the shock velocity for different PAH abundances (X(PAH)). 
The radiation field varies between panels, with $F_\mathrm{UV}$ = 1, 10, 100, and 1000 from top to bottom panel respectively. For all panels b=1, $\zeta$ = 10$^{-7}$ s$^{-1}$, and log(n$_H$) = 4 are used.}
\end{figure}

Both the cosmic ray ionization rate and the abundance of Polycyclic Aromatic Hydrocarbons (PAHs) show minimal impact on the S(2)/S(7) ratio (see Figs. \ref{fig:zeta_comparison} and \ref{fig:XPAH_comparison}).

\begin{figure}[ht!]
\centering 
\includegraphics[width=\linewidth]{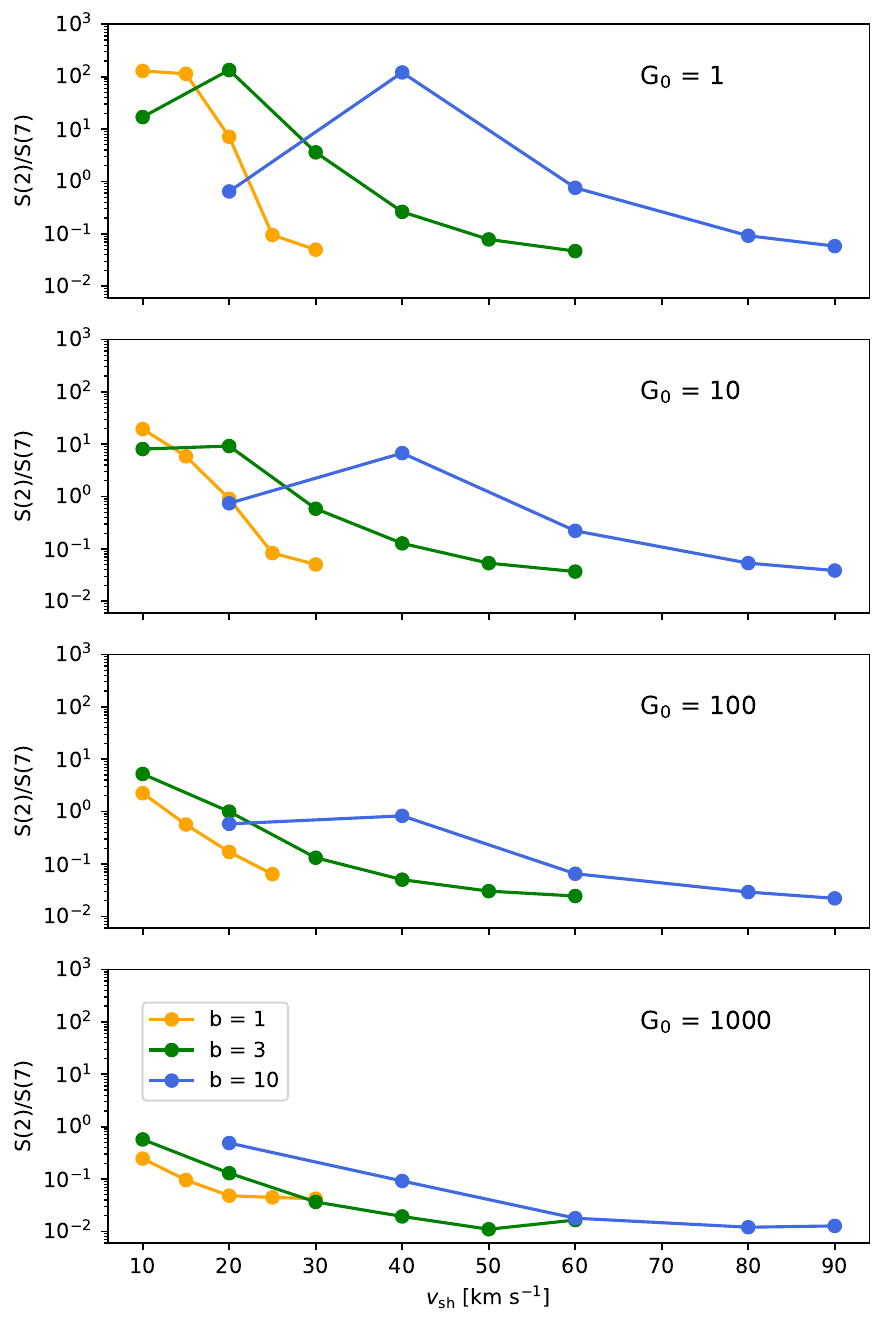} 
\caption{\label{fig:b_comparison} Ratio of the S(2) over the S(7) transition of H$_2$ against the shock velocity for different magnetic field strengths (b). 
The radiation field varies between panels, with $F_\mathrm{UV}$ = 1, 10, and 100 for top, middle and bottom panel respectively. For all panels $\zeta$ = 10$^{-7}$ s$^{-1}$, X(PAH)= $10^{-17}$, and log(n$_H$) = 4 are used.}
\end{figure}

Subsequently, we investigate the impact of even higher $b$ values. As seen in Fig. \ref{fig:b_comparison}, the shape of the S(2)/S(7) curve remains similar, but shifts towards higher velocities for stronger magnetic fields. Since a stronger magnetic field allows for $C$-type shocks to occur at higher shock velocities, this shift indicates the change in the velocity at which the shock transitions between $C$- and $J$-type. The velocity ranges presented in Figs. \ref{fig:KS23_vs_Flower} and \ref{fig:b_comparison} are restricted by the grid in \cite{Kristensen23}. Figure \ref{fig:b_comparison} also shows the impact of the external UV field, as for higher G$_0$ values the S(2)/S(7) ratio appears to become flatter irrespective of the magnetic field strength.

\begin{figure*}[ht!]
\centering 
\includegraphics[width=0.9\linewidth]{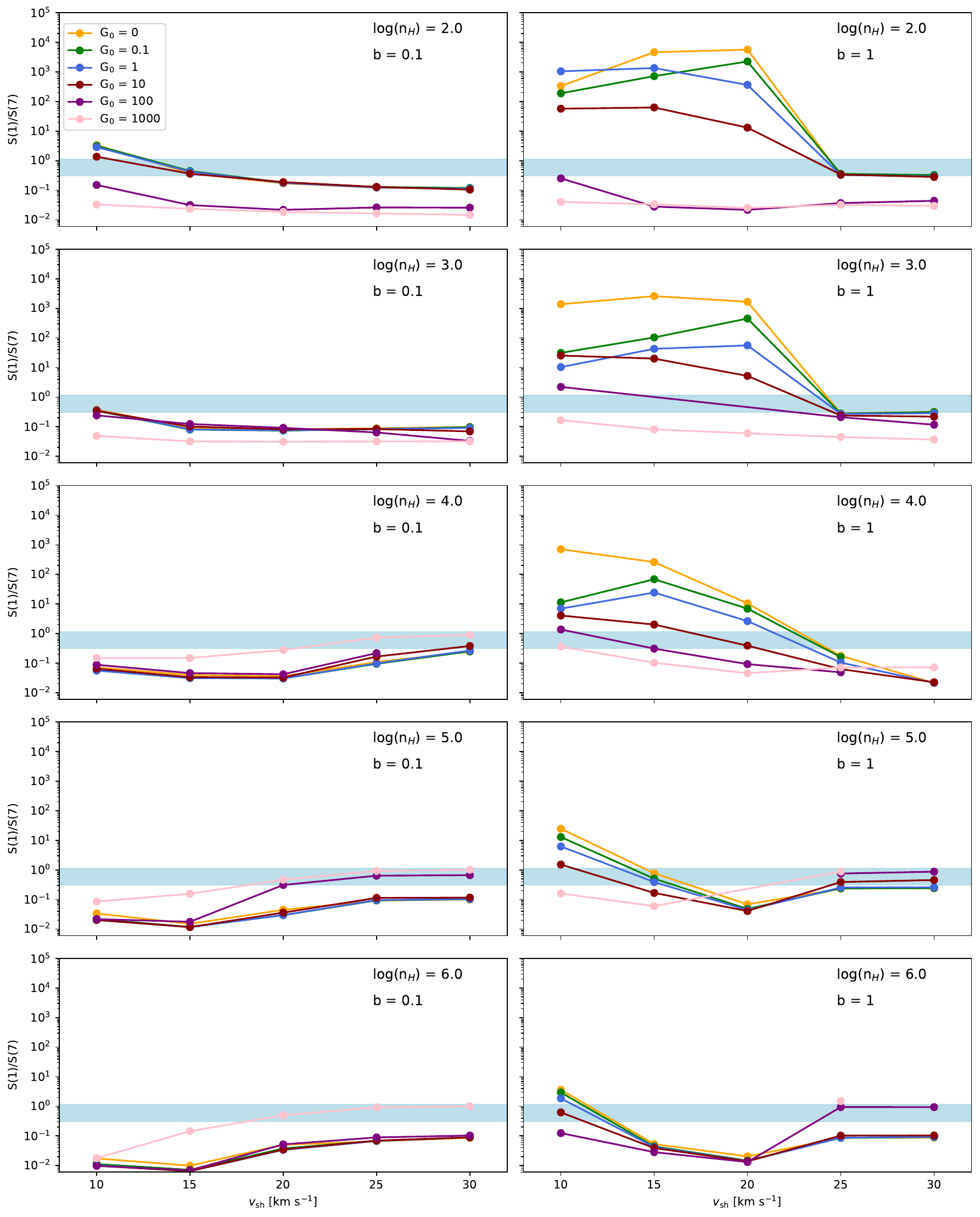} 
\caption{\label{fig:G0_comparison2} Ratio of the S(1) over the S(7) transition of H$_2$ against the shock velocity for different $F_\mathrm{UV}$ field strengths. Log(n$_H$) = 4 for the top row and 5 for bottom row, while magnetic field strength is b = 0.1 for the left column and 1 for the right. For all panels $\zeta$ = 10$^{-7}$ s$^{-1}$, and X(PAH)= $10^{-17}$ are used. The blue shaded region marks the observed range of S(2)/S(7) ratios for the sources in Ophiuchus.}
\end{figure*}

\begin{figure*}[ht!]
\centering 
\includegraphics[width=0.9\linewidth]{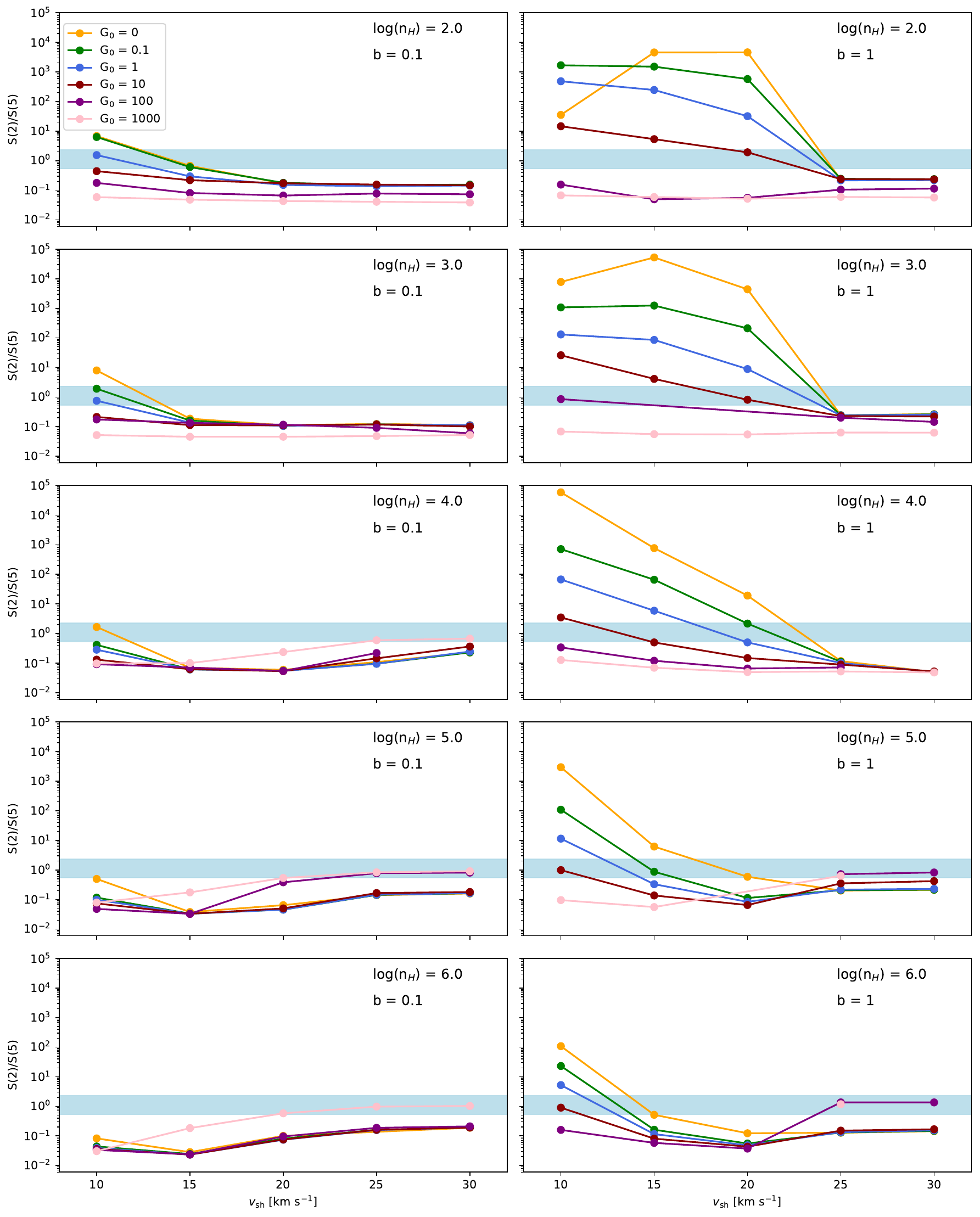} 
\caption{\label{fig:G0_comparison3} Same as Fig. \ref{fig:G0_comparison2} but for the ratio of the S(2) over the S(5) transition of H$_2$.}
\end{figure*}

Finally, Figs. \ref{fig:G0_comparison2} and \ref{fig:G0_comparison3} show comparisons of the shock model results for the S(1)/S(7) and S(2)/S(6) line ratios. Both these ratios represent comparisons of the warm to hot gas components, and show behavior similar to that of the S(2)/S(7) ratio discussed in detail in Sec. \ref{sec:shocks}, indicating that the selection of the line ratio has no significant impact on the results.

\end{document}